\documentclass[apjl]{emulateapj}
\usepackage{comment}
\usepackage{ifthen}
\usepackage{lineno}

\newcommand{\forloop}[5][1]%
{%
\setcounter{#2}{#3}%
\ifthenelse{#4}%
	{%
	#5%
	\addtocounter{#2}{#1}%
	\forloop[#1]{#2}{\value{#2}}{#4}{#5}%
	}%
	{%
	}%
}%

\newcommand{\ctbd}[1]{}

\newcommand{\lc}{light curve}
\newcommand{\lcs}{light curves}
\newcommand{\Lc}{Light curve}

\newcommand{\band}[1]{\ensuremath{#1}~band}

\newcommand{\kms}{\ensuremath{\rm km\,s^{-1}}}
\newcommand{\ms}{\ensuremath{\rm m\,s^{-1}}}

\newcommand{\gcmc}{\ensuremath{\rm g\,cm^{-3}}}
\newcommand{\ergscmsq}{\ensuremath{\rm erg\,s^{-1}\,cm^{-2}}}

\newcommand{\vsini}{\ensuremath{v \sin{i}}}
\newcommand{\feh}{\ensuremath{\rm [Fe/H]}}

\newcommand{\rsun}{\ensuremath{R_\sun}}
\newcommand{\msun}{\ensuremath{M_\sun}}
\newcommand{\lsun}{\ensuremath{L_\sun}}

\newcommand{\rstar}{\ensuremath{R_\star}}
\newcommand{\mstar}{\ensuremath{M_\star}}
\newcommand{\lstar}{\ensuremath{L_\star}}

\newcommand{\teffstar}{\ensuremath{T_{\rm eff\star}}}

\newcommand{\loggstar}{\ensuremath{\log{g_{\star}}}}

\newcommand{\rpl}{\ensuremath{R_{p}}}
\newcommand{\mpl}{\ensuremath{M_{p}}}

\newcommand{\rhopl}{\ensuremath{\rho_{p}}}

\newcommand{\arstar}{\ensuremath{a/\rstar}}
\newcommand{\zrstar}{\ensuremath{\zeta/\rstar}}

\newcommand{\rjup}{\ensuremath{R_{\rm J}}}
\newcommand{\mjup}{\ensuremath{M_{\rm J}}}

\newcommand{\reffig}[1]{Fig.~\ref{fig:#1}}
\newcommand{\refsec}[1]{\mbox{\S\ \ref{sec:#1}}}

\newcommand{\reftab}[1]{Tab.~\ref{tab:#1}}

\newcommand{\reffigl}[1]{Figure~\ref{fig:#1}}
\newcommand{\refsecl}[1]{\mbox{Section \ref{sec:#1}}}

\newcommand{\reftabl}[1]{Table~\ref{tab:#1}}

\newcommand{\flwof}{\mbox{FLWO 1.2\,m}}

\newcommand{\loopand}{\ifnum\value{planetcounter}=3 and \else\fi}
\newcommand{\loopcomma}{\ifnum\value{planetcounter}<3 ,\else. \fi}
\newcommand{\loopcommanoperiod}{\ifnum\value{planetcounter}<3 ,\else \space\fi}
\newcommand{\loopcommanospace}{\ifnum\value{planetcounter}<3 ,\else \fi}

\newcommand{\hatcurhtrxxxxxA}{HTR212-002}                              
\newcommand{\hatcurfieldxxxxxA}{212}                                   
\newcommand{\hatcurCCraxxxxxA}{\ensuremath{02^{\mathrm h}33^{\mathrm m}13.97{\mathrm s}}}                            
\newcommand{\hatcurCCdecxxxxxA}{\ensuremath{+30{\arcdeg}21{\arcmin}37.8{\arcsec}}}                           
\newcommand{\hatcurCCmagxxxxxA}{10.694}                                
\newcommand{\hatcurCCtwomassxxxxxA}{2MASS~02331396+3021377}            
\newcommand{\hatcurCCgscxxxxxA}{GSC~2324-00031}                        
\newcommand{\hatcurCCtassmvxxxxxA}{\ensuremath{10.694\pm0.063}}                             
\newcommand{\hatcurCCtassmIxxxxxA}{\ensuremath{10.084\pm0.069}}                             
\newcommand{\hatcurCCtwomassJmagxxxxxA}{\ensuremath{9.713\pm0.021}}    
\newcommand{\hatcurCCtwomassHmagxxxxxA}{\ensuremath{9.454\pm0.022}}    
\newcommand{\hatcurCCtwomassKmagxxxxxA}{\ensuremath{9.404\pm0.017}}    
\newcommand{\hatcurCCcitJmagxxxxxA}{\ensuremath{9.733\pm0.021}}        
\newcommand{\hatcurCCcitHmagxxxxxA}{\ensuremath{9.449\pm0.022}}        
\newcommand{\hatcurCCcitKmagxxxxxA}{\ensuremath{9.428\pm0.017}}        
\newcommand{\hatcurCCbbJmagxxxxxA}{\ensuremath{9.777\pm0.023}}         
\newcommand{\hatcurCCbbHmagxxxxxA}{\ensuremath{9.470\pm0.023}}         
\newcommand{\hatcurCCbbKmagxxxxxA}{\ensuremath{9.448\pm0.017}}         
\newcommand{\hatcurCCesoJmagxxxxxA}{\ensuremath{9.779\pm0.024}}        
\newcommand{\hatcurCCesoHmagxxxxxA}{\ensuremath{9.464\pm0.026}}        
\newcommand{\hatcurCCesoKmagxxxxxA}{\ensuremath{9.447\pm0.018}}        
\newcommand{\hatcurCCesoJHmagxxxxxA}{\ensuremath{0.315\pm0.033}}       
\newcommand{\hatcurCCesoJKmagxxxxxA}{\ensuremath{0.331\pm0.033}}       
\newcommand{\hatcurCCesoHKmagxxxxxA}{\ensuremath{0.017\pm0.031}}       
\newcommand{\hatcurLCdipxxxxxA}{\ensuremath{3.8}}                      
\newcommand{\hatcurLCrprstarxxxxxA}{\ensuremath{0.0890\pm0.0013}}      
\newcommand{\hatcurLCbsqxxxxxA}{\ensuremath{0.630_{-0.017}^{+0.017}}}  
\newcommand{\hatcurLCimpxxxxxA}{\ensuremath{0.794_{-0.011}^{+0.010}}}  
\newcommand{\hatcurLCzetaxxxxxA}{\ensuremath{19.05\pm0.19}}            
\newcommand{\hatcurLCdurxxxxxA}{\ensuremath{0.1287\pm0.0015}}          
\newcommand{\hatcurLCdurshortxxxxxA}{\ensuremath{0.1287}}              
\newcommand{\hatcurLCdurhrxxxxxA}{\ensuremath{3.088\pm0.037}}          
\newcommand{\hatcurLCdurhrshortxxxxxA}{\ensuremath{3.088}}             
\newcommand{\hatcurLCqxxxxxA}{\ensuremath{0.0272\pm0.0003}}            
\newcommand{\hatcurLCqshortxxxxxA}{\ensuremath{0.027}}                 
\newcommand{\hatcurLCingdurxxxxxA}{\ensuremath{0.0258\pm0.0014}}       
\newcommand{\hatcurLCPxxxxxA}{\ensuremath{4.732182\pm0.000013}}        
\newcommand{\hatcurLCPprecxxxxxA}{\ensuremath{4.7321824}}              
\newcommand{\hatcurLCPshortxxxxxA}{\ensuremath{4.7322}}                
\newcommand{\hatcurLCTxxxxxA}{\ensuremath{2455661.63669\pm0.00041}}    
\newcommand{\hatcurLCTAxxxxxA}{\ensuremath{2455415.56324\pm0.00081}}   
\newcommand{\hatcurLCTBxxxxxA}{\ensuremath{2455940.83548\pm0.00087}}   
\newcommand{\hatcurLChatnetmxxxxxA}{\ensuremath{10.7358\pm0.0000}}     
\newcommand{\hatcurLCiblendxxxxxA}{\ensuremath{0.71\pm0.05}}           
\newcommand{\hatcurSMEiteffxxxxxA}{\ensuremath{6625\pm50}}             
\newcommand{\hatcurSMEizfehxxxxxA}{\ensuremath{-0.08\pm0.08}}          
\newcommand{\hatcurSMEizfehshortxxxxxA}{\ensuremath{-0.08}}            
\newcommand{\hatcurSMEiloggxxxxxA}{\ensuremath{4.10\pm0.1}}            
\newcommand{\hatcurSMEivsinxxxxxA}{\ensuremath{14.1\pm0.5}}            
\newcommand{\hatcurSMEivmacxxxxxA}{\ensuremath{NULL}}                  
\newcommand{\hatcurSMEivmicxxxxxA}{\ensuremath{NULL}}                  
\newcommand{\hatcurSMEiiteffxxxxxA}{\ensuremath{6703\pm50}}            
\newcommand{\hatcurSMEiizfehxxxxxA}{\ensuremath{0.0\pm0.08}}           
\newcommand{\hatcurSMEiizfehshortxxxxxA}{\ensuremath{0.0}}             
\newcommand{\hatcurSMEiiloggxxxxxA}{\ensuremath{4.21\pm0.10}}          
\newcommand{\hatcurSMEiivsinxxxxxA}{\ensuremath{14.1\pm0.5}}           
\newcommand{\hatcurSMEiivmacxxxxxA}{\ensuremath{NULL}}                 
\newcommand{\hatcurSMEiivmicxxxxxA}{\ensuremath{NULL}}                 
\newcommand{\hatcurDSteffxxxxxA}{\ensuremath{NULL\pmNULL}}             
\newcommand{\hatcurDSzfehxxxxxA}{\ensuremath{NULL\pmNULL}}             
\newcommand{\hatcurDSloggxxxxxA}{\ensuremath{NULL\pmNULL}}             
\newcommand{\hatcurDSvsinixxxxxA}{\ensuremath{NULL\pmNULL}}            
\newcommand{\hatcurDSgammaxxxxxA}{\ensuremath{NULL\pmNULL}}            
\newcommand{\hatcurDSnumspecxxxxxA}{\ensuremath{0}}                    
\newcommand{\hatcurDSspanxxxxxA}{\ensuremath{0}}                       
\newcommand{\hatcurDSrvrmsxxxxxA}{\ensuremath{0.00}}                   
\newcommand{\hatcurTRESteffxxxxxA}{\ensuremath{5300\pm100}}            
\newcommand{\hatcurTRESzfehxxxxxA}{\ensuremath{NULL\pm0.1}}            
\newcommand{\hatcurTRESloggxxxxxA}{\ensuremath{4.5\pm0.2}}             
\newcommand{\hatcurTRESvsinixxxxxA}{\ensuremath{2.7\pm0.5}}            
\newcommand{\hatcurTRESgammaxxxxxA}{\ensuremath{-19.0\pm0.1}}          
\newcommand{\hatcurTRESnumspecxxxxxA}{\ensuremath{3}}                  
\newcommand{\hatcurTRESspanxxxxxA}{\ensuremath{32}}                    
\newcommand{\hatcurTRESrvrmsxxxxxA}{\ensuremath{0.19}}                 
\newcommand{\hatcurFIESteffxxxxxA}{\ensuremath{NULL\pmNULL}}           
\newcommand{\hatcurFIESzfehxxxxxA}{\ensuremath{NULL\pmNULL}}           
\newcommand{\hatcurFIESloggxxxxxA}{\ensuremath{NULL\pmNULL}}           
\newcommand{\hatcurFIESvsinixxxxxA}{\ensuremath{NULL\pmNULL}}          
\newcommand{\hatcurFIESgammaxxxxxA}{\ensuremath{NULL\pmNULL}}          
\newcommand{\hatcurFIESnumspecxxxxxA}{\ensuremath{0}}                  
\newcommand{\hatcurFIESspanxxxxxA}{\ensuremath{0}}                     
\newcommand{\hatcurFIESrvrmsxxxxxA}{\ensuremath{0.00}}                 
\newcommand{\hatcurLBizxxxxxA}{\ensuremath{0.1095}}                    
\newcommand{\hatcurLBiizxxxxxA}{\ensuremath{0.3680}}                   
\newcommand{\hatcurLBiixxxxxA}{\ensuremath{0.1592}}                    
\newcommand{\hatcurLBiiixxxxxA}{\ensuremath{0.3774}}                   
\newcommand{\hatcurLBiIxxxxxA}{\ensuremath{0.1415}}                    
\newcommand{\hatcurLBiiIxxxxxA}{\ensuremath{0.3751}}                   
\newcommand{\hatcurLBigxxxxxA}{\ensuremath{0.3734}}                    
\newcommand{\hatcurLBiigxxxxxA}{\ensuremath{0.3575}}                   
\newcommand{\hatcurLBirxxxxxA}{\ensuremath{0.2273}}                    
\newcommand{\hatcurLBiirxxxxxA}{\ensuremath{0.3906}}                   
\newcommand{\hatcurLBiRxxxxxA}{\ensuremath{0.2076}}                    
\newcommand{\hatcurLBiiRxxxxxA}{\ensuremath{0.3888}}                   
\newcommand{\hatcurLBikepxxxxxA}{\ensuremath{}}                
\newcommand{\hatcurLBiikepxxxxxA}{\ensuremath{}}               
\newcommand{\hatcurISOmxxxxxA}{\ensuremath{1.39\pm0.04}}               
\newcommand{\hatcurISOmshortxxxxxA}{\ensuremath{1.39}}                 
\newcommand{\hatcurISOmlongxxxxxA}{\ensuremath{1.387\pm0.038}}         
\newcommand{\hatcurISOrxxxxxA}{\ensuremath{1.52\pm0.04}}               
\newcommand{\hatcurISOrshortxxxxxA}{\ensuremath{1.52}}                 
\newcommand{\hatcurISOrlongxxxxxA}{\ensuremath{1.515\pm0.040}}         
\newcommand{\hatcurISOrhoxxxxxA}{\ensuremath{0.56\pm0.04}}             
\newcommand{\hatcurISOloggxxxxxA}{\ensuremath{4.22\pm0.02}}            
\newcommand{\hatcurISOlumxxxxxA}{\ensuremath{4.15\pm0.27}}             
\newcommand{\hatcurISOlumshortxxxxxA}{\ensuremath{4.15}}               
\newcommand{\hatcurISOmvxxxxxA}{\ensuremath{3.17\pm0.07}}              
\newcommand{\hatcurISOvixxxxxA}{\ensuremath{0.452\pm0.012}}            
\newcommand{\hatcurISOagexxxxxA}{\ensuremath{1.5\pm0.3}}               
\newcommand{\hatcurISOsigmaxxxxxA}{\ensuremath{0.00010\pm0.00003}}     
\newcommand{\hatcurISOMJxxxxxA}{\ensuremath{2.47\pm0.06}}              
\newcommand{\hatcurISOMHxxxxxA}{\ensuremath{2.29\pm0.06}}              
\newcommand{\hatcurISOMKxxxxxA}{\ensuremath{2.25\pm0.06}}              
\newcommand{\hatcurISOJKxxxxxA}{\ensuremath{0.22\pm0.01}}              
\newcommand{\hatcurISOspecxxxxxA}{F}                                   
\newcommand{\hatcurRVKxxxxxA}{\ensuremath{19.9\pm3.8}}                 
\newcommand{\hatcurRVrkxxxxxA}{\ensuremath{0.000\pm0.000}}             
\newcommand{\hatcurRVrhxxxxxA}{\ensuremath{0.000\pm0.000}}             
\newcommand{\hatcurRVkxxxxxA}{\ensuremath{0.000\pm0.000}}              
\newcommand{\hatcurRVhxxxxxA}{\ensuremath{0.000\pm0.000}}              
\newcommand{\hatcurRVtronexxxxxA}{\ensuremath{0.0000\pm0.0000}}        
\newcommand{\hatcurRVtrtwoxxxxxA}{\ensuremath{0.0000\pm0.0000}}        
\newcommand{\hatcurRVgammaAxxxxxA}{\ensuremath{-1.7\pm4.7}}            
\newcommand{\hatcurRVjitterAxxxxxA}{\ensuremath{22.0}}                 
\newcommand{\hatcurRVfitrmsAxxxxxA}{\ensuremath{24.1}}                 
\newcommand{\hatcurRVgammaBxxxxxA}{\ensuremath{2.0\pm4.2}}             
\newcommand{\hatcurRVjitterBxxxxxA}{\ensuremath{0.0}}                  
\newcommand{\hatcurRVfitrmsBxxxxxA}{\ensuremath{14.7}}                 
\newcommand{\hatcurRVeccenxxxxxA}{\ensuremath{0.000\pm0.000}}          
\newcommand{\hatcurRVomegaxxxxxA}{\ensuremath{0\pm0}}                  
\newcommand{\hatcurPPixxxxxA}{\ensuremath{84.8\pm0.2}}                 
\newcommand{\hatcurPPgxxxxxA}{\ensuremath{3.0\pm0.6}}                  
\newcommand{\hatcurPPloggxxxxxA}{\ensuremath{2.47_{-0.11}^{+0.07}}}    
\newcommand{\hatcurPParxxxxxA}{\ensuremath{8.73\pm0.20}}               
\newcommand{\hatcurPParelxxxxxA}{\ensuremath{0.0615\pm0.0006}}         
\newcommand{\hatcurPPrhoxxxxxA}{\ensuremath{0.11\pm0.02}}              
\newcommand{\hatcurPPmxxxxxA}{\ensuremath{0.21\pm0.04}}                
\newcommand{\hatcurPPmshortxxxxxA}{\ensuremath{0.21}}                  
\newcommand{\hatcurPPmlongxxxxxA}{\ensuremath{0.206\pm0.039}}          
\newcommand{\hatcurPPmexxxxxA}{\ensuremath{65.3\pm12.5}}               
\newcommand{\hatcurPPmeshortxxxxxA}{\ensuremath{65.3}}                 
\newcommand{\hatcurPPmelongxxxxxA}{\ensuremath{65.34\pm12.53}}         
\newcommand{\hatcurPPrxxxxxA}{\ensuremath{1.31\pm0.05}}                
\newcommand{\hatcurPPrshortxxxxxA}{\ensuremath{1.31}}                  
\newcommand{\hatcurPPrlongxxxxxA}{\ensuremath{1.313\pm0.045}}          
\newcommand{\hatcurPPrexxxxxA}{\ensuremath{14.7\pm0.5}}                
\newcommand{\hatcurPPreshortxxxxxA}{\ensuremath{14.7}}                 
\newcommand{\hatcurPPrelongxxxxxA}{\ensuremath{14.72\pm0.51}}          
\newcommand{\hatcurPPmrcorrxxxxxA}{\ensuremath{0.05}}                  
\newcommand{\hatcurPPteffxxxxxA}{\ensuremath{1605\pm22}}               
\newcommand{\hatcurPPthetaxxxxxA}{\ensuremath{0.014\pm0.003}}          
\newcommand{\hatcurPPfluxperixxxxxA}{\ensuremath{15.0\pm0.8}}         
\newcommand{\hatcurPPfluxperidimxxxxxA}{\ensuremath{8}}                
\newcommand{\hatcurPPfluxapxxxxxA}{\ensuremath{15.0\pm0.8}}           
\newcommand{\hatcurPPfluxapdimxxxxxA}{\ensuremath{8}}                  
\newcommand{\hatcurPPfluxavgxxxxxA}{\ensuremath{15.0\pm0.8}}          
\newcommand{\hatcurPPfluxavgdimxxxxxA}{\ensuremath{8}}                 
\newcommand{\hatcurXsecphasexxxxxA}{\ensuremath{0.5000\pm0.0000}}      
\newcommand{\hatcurXsecondaryxxxxxA}{\ensuremath{2455664.003\pm0.000}} 
\newcommand{\hatcurXsecdurxxxxxA}{\ensuremath{0.1287\pm0.0015}}        
\newcommand{\hatcurXsecingdurxxxxxA}{\ensuremath{0.0258\pm0.0014}}     
\newcommand{\hatcurPPphiconjxxxxxA}{\ensuremath{0.2500\pm0.0000}}      
\newcommand{\hatcurPPperixxxxxA}{\ensuremath{2455660.45\pm0.00}}       
\newcommand{\hatcurPPaequivxxxxxA}{\ensuremath{0.0302\pm0.0008}}       
\newcommand{\hatcurPPtcircxxxxxA}{\ensuremath{168.1\pm42.1}}           
\newcommand{\hatcurPPtinfallxxxxxA}{\ensuremath{15481.4_{-2799.4}^{+4997.6}}} 
\newcommand{\hatcurXdistxxxxxA}{\ensuremath{274\pm7}}                  
\newcommand{\hatcurXAvxxxxxA}{\ensuremath{0.414\pm0.077}}              
\newcommand{\hatcurXdistredxxxxxA}{\ensuremath{268\pm7}}               
\newcommand{\hatcurXEBVxxxxxA}{\ensuremath{0.133\pm0.025}}             
\newcommand{\hatcurXmvisoredxxxxxA}{\ensuremath{10.734\pm0.055}}       
\newcommand{\hatcurXmiisoredxxxxxA}{\ensuremath{10.067\pm0.022}}       
\newcommand{\hatcurXmjisoredxxxxxA}{\ensuremath{9.735\pm0.013}}        
\newcommand{\hatcurXmhisoredxxxxxA}{\ensuremath{9.513\pm0.012}}        
\newcommand{\hatcurXmkisoredxxxxxA}{\ensuremath{9.447\pm0.014}}        
\newcommand{\hatcurXviisoredxxxxxA}{\ensuremath{0.667\pm0.037}}        
\newcommand{\hatcurXvkisoredxxxxxA}{\ensuremath{1.287\pm0.061}}        
\newcommand{\hatcurXjhisoredxxxxxA}{\ensuremath{0.221\pm0.008}}        
\newcommand{\hatcurXjkisoredxxxxxA}{\ensuremath{0.288\pm0.012}}        
\newcommand{\hatcurCCpmraxxxxxA}{\ensuremath{-13.9\pm0.6}}             
\newcommand{\hatcurCCpmdecxxxxxA}{\ensuremath{-9.8\pm0.7}}             
\newcommand{\hatcurCCpmxxxxxA}{\ensuremath{17.0074\pm0.921954}}        

\newcommand{\hatcurhtrxxxxxB}{HTR213-008}                              
\newcommand{\hatcurfieldxxxxxB}{212}                                   
\newcommand{\hatcurCCraxxxxxB}{\ensuremath{02^{\mathrm h}57^{\mathrm m}53.03{\mathrm s}}}                            
\newcommand{\hatcurCCdecxxxxxB}{\ensuremath{+30{\arcdeg}37{\arcmin}32.5{\arcsec}}}                           
\newcommand{\hatcurCCmagxxxxxB}{12.16}                                 
\newcommand{\hatcurCCtwomassxxxxxB}{2MASS~02575301+3037324}            
\newcommand{\hatcurCCgscxxxxxB}{GSC~2326-00214}                        
\newcommand{\hatcurCCtassmvxxxxxB}{\ensuremath{12.16\pm0.11}}                              
\newcommand{\hatcurCCtassmIxxxxxB}{\ensuremath{11.253\pm0.071}}                              
\newcommand{\hatcurCCtwomassJmagxxxxxB}{\ensuremath{10.696\pm0.023}}   
\newcommand{\hatcurCCtwomassHmagxxxxxB}{\ensuremath{10.340\pm0.026}}   
\newcommand{\hatcurCCtwomassKmagxxxxxB}{\ensuremath{10.255\pm0.021}}   
\newcommand{\hatcurCCcitJmagxxxxxB}{\ensuremath{10.709\pm0.023}}       
\newcommand{\hatcurCCcitHmagxxxxxB}{\ensuremath{10.334\pm0.026}}       
\newcommand{\hatcurCCcitKmagxxxxxB}{\ensuremath{10.279\pm0.021}}       
\newcommand{\hatcurCCbbJmagxxxxxB}{\ensuremath{10.764\pm0.025}}        
\newcommand{\hatcurCCbbHmagxxxxxB}{\ensuremath{10.356\pm0.027}}        
\newcommand{\hatcurCCbbKmagxxxxxB}{\ensuremath{10.299\pm0.021}}        
\newcommand{\hatcurCCesoJmagxxxxxB}{\ensuremath{10.767\pm0.027}}       
\newcommand{\hatcurCCesoHmagxxxxxB}{\ensuremath{10.351\pm0.031}}       
\newcommand{\hatcurCCesoKmagxxxxxB}{\ensuremath{10.298\pm0.022}}       
\newcommand{\hatcurCCesoJHmagxxxxxB}{\ensuremath{0.415\pm0.038}}       
\newcommand{\hatcurCCesoJKmagxxxxxB}{\ensuremath{0.470\pm0.034}}       
\newcommand{\hatcurCCesoHKmagxxxxxB}{\ensuremath{0.054\pm0.038}}       
\newcommand{\hatcurLCdipxxxxxB}{\ensuremath{6.4}}                      
\newcommand{\hatcurLCrprstarxxxxxB}{\ensuremath{0.0951\pm0.0016}}      
\newcommand{\hatcurLCbsqxxxxxB}{\ensuremath{0.281_{-0.057}^{+0.040}}}  
\newcommand{\hatcurLCimpxxxxxB}{\ensuremath{0.530_{-0.060}^{+0.036}}}  
\newcommand{\hatcurLCzetaxxxxxB}{\ensuremath{16.01\pm0.08}}            
\newcommand{\hatcurLCdurxxxxxB}{\ensuremath{0.1411\pm0.0013}}          
\newcommand{\hatcurLCdurshortxxxxxB}{\ensuremath{0.1411}}              
\newcommand{\hatcurLCdurhrxxxxxB}{\ensuremath{3.386\pm0.031}}          
\newcommand{\hatcurLCdurhrshortxxxxxB}{\ensuremath{3.386}}             
\newcommand{\hatcurLCqxxxxxB}{\ensuremath{0.0320\pm0.0003}}            
\newcommand{\hatcurLCqshortxxxxxB}{\ensuremath{0.032}}                 
\newcommand{\hatcurLCingdurxxxxxB}{\ensuremath{0.0165\pm0.0013}}       
\newcommand{\hatcurLCPxxxxxB}{\ensuremath{4.408650\pm0.000008}}        
\newcommand{\hatcurLCPprecxxxxxB}{\ensuremath{4.4086501}}              
\newcommand{\hatcurLCPshortxxxxxB}{\ensuremath{4.4087}}                
\newcommand{\hatcurLCTxxxxxB}{\ensuremath{2455839.21023\pm0.00029}}    
\newcommand{\hatcurLCTAxxxxxB}{\ensuremath{2455415.97983\pm0.00078}}   
\newcommand{\hatcurLCTBxxxxxB}{\ensuremath{2455953.83514\pm0.00041}}   
\newcommand{\hatcurLChatnetmxxxxxB}{\ensuremath{11.9267\pm0.0001}}     
\newcommand{\hatcurLCiblendxxxxxB}{\ensuremath{0.70\pm0.05}}           
\newcommand{\hatcurSMEiteffxxxxxB}{\ensuremath{5880\pm50}}             
\newcommand{\hatcurSMEizfehxxxxxB}{\ensuremath{0.01\pm0.08}}           
\newcommand{\hatcurSMEizfehshortxxxxxB}{\ensuremath{0.01}}             
\newcommand{\hatcurSMEiloggxxxxxB}{\ensuremath{4.23\pm0.1}}            
\newcommand{\hatcurSMEivsinxxxxxB}{\ensuremath{2.7\pm0.5}}             
\newcommand{\hatcurSMEivmacxxxxxB}{\ensuremath{NULL}}                  
\newcommand{\hatcurSMEivmicxxxxxB}{\ensuremath{NULL}}                  
\newcommand{\hatcurSMEiiteffxxxxxB}{\ensuremath{5946\pm50}}            
\newcommand{\hatcurSMEiizfehxxxxxB}{\ensuremath{0.02\pm0.08}}          
\newcommand{\hatcurSMEiizfehshortxxxxxB}{\ensuremath{0.02}}            
\newcommand{\hatcurSMEiiloggxxxxxB}{\ensuremath{4.30\pm0.10}}          
\newcommand{\hatcurSMEiivsinxxxxxB}{\ensuremath{2.6\pm0.5}}            
\newcommand{\hatcurSMEiivmacxxxxxB}{\ensuremath{NULL}}                 
\newcommand{\hatcurSMEiivmicxxxxxB}{\ensuremath{NULL}}                 
\newcommand{\hatcurDSteffxxxxxB}{\ensuremath{NULL\pmNULL}}             
\newcommand{\hatcurDSzfehxxxxxB}{\ensuremath{NULL\pmNULL}}             
\newcommand{\hatcurDSloggxxxxxB}{\ensuremath{NULL\pmNULL}}             
\newcommand{\hatcurDSvsinixxxxxB}{\ensuremath{NULL\pmNULL}}            
\newcommand{\hatcurDSgammaxxxxxB}{\ensuremath{NULL\pmNULL}}            
\newcommand{\hatcurDSnumspecxxxxxB}{\ensuremath{0}}                    
\newcommand{\hatcurDSspanxxxxxB}{\ensuremath{0}}                       
\newcommand{\hatcurDSrvrmsxxxxxB}{\ensuremath{0.00}}                   
\newcommand{\hatcurTRESteffxxxxxB}{\ensuremath{5300\pm100}}            
\newcommand{\hatcurTRESzfehxxxxxB}{\ensuremath{NULL\pm0.1}}            
\newcommand{\hatcurTRESloggxxxxxB}{\ensuremath{4.5\pm0.2}}             
\newcommand{\hatcurTRESvsinixxxxxB}{\ensuremath{2.7\pm0.5}}            
\newcommand{\hatcurTRESgammaxxxxxB}{\ensuremath{-19.0\pm0.1}}          
\newcommand{\hatcurTRESnumspecxxxxxB}{\ensuremath{3}}                  
\newcommand{\hatcurTRESspanxxxxxB}{\ensuremath{32}}                    
\newcommand{\hatcurTRESrvrmsxxxxxB}{\ensuremath{0.19}}                 
\newcommand{\hatcurFIESteffxxxxxB}{\ensuremath{NULL\pmNULL}}           
\newcommand{\hatcurFIESzfehxxxxxB}{\ensuremath{NULL\pmNULL}}           
\newcommand{\hatcurFIESloggxxxxxB}{\ensuremath{NULL\pmNULL}}           
\newcommand{\hatcurFIESvsinixxxxxB}{\ensuremath{NULL\pmNULL}}          
\newcommand{\hatcurFIESgammaxxxxxB}{\ensuremath{NULL\pmNULL}}          
\newcommand{\hatcurFIESnumspecxxxxxB}{\ensuremath{0}}                  
\newcommand{\hatcurFIESspanxxxxxB}{\ensuremath{0}}                     
\newcommand{\hatcurFIESrvrmsxxxxxB}{\ensuremath{0.00}}                 
\newcommand{\hatcurLBizxxxxxB}{\ensuremath{0.1868}}                    
\newcommand{\hatcurLBiizxxxxxB}{\ensuremath{0.3383}}                   
\newcommand{\hatcurLBiixxxxxB}{\ensuremath{0.2434}}                    
\newcommand{\hatcurLBiiixxxxxB}{\ensuremath{0.3424}}                   
\newcommand{\hatcurLBiIxxxxxB}{\ensuremath{0.2241}}                    
\newcommand{\hatcurLBiiIxxxxxB}{\ensuremath{0.3420}}                   
\newcommand{\hatcurLBigxxxxxB}{\ensuremath{0.5066}}                    
\newcommand{\hatcurLBiigxxxxxB}{\ensuremath{0.2702}}                   
\newcommand{\hatcurLBirxxxxxB}{\ensuremath{0.3246}}                    
\newcommand{\hatcurLBiirxxxxxB}{\ensuremath{0.3445}}                   
\newcommand{\hatcurLBiRxxxxxB}{\ensuremath{0.3020}}                    
\newcommand{\hatcurLBiiRxxxxxB}{\ensuremath{0.3450}}                   
\newcommand{\hatcurLBikepxxxxxB}{\ensuremath{}}                
\newcommand{\hatcurLBiikepxxxxxB}{\ensuremath{}}               
\newcommand{\hatcurISOmxxxxxB}{\ensuremath{1.10\pm0.04}}               
\newcommand{\hatcurISOmshortxxxxxB}{\ensuremath{1.10}}                 
\newcommand{\hatcurISOmlongxxxxxB}{\ensuremath{1.099\pm0.041}}         
\newcommand{\hatcurISOrxxxxxB}{\ensuremath{1.22\pm0.05}}               
\newcommand{\hatcurISOrshortxxxxxB}{\ensuremath{1.22}}                 
\newcommand{\hatcurISOrlongxxxxxB}{\ensuremath{1.223\pm0.046}}         
\newcommand{\hatcurISOrhoxxxxxB}{\ensuremath{0.84_{-0.07}^{+0.10}}}    
\newcommand{\hatcurISOloggxxxxxB}{\ensuremath{4.30\pm0.03}}            
\newcommand{\hatcurISOlumxxxxxB}{\ensuremath{1.67\pm0.14}}             
\newcommand{\hatcurISOlumshortxxxxxB}{\ensuremath{1.67}}               
\newcommand{\hatcurISOmvxxxxxB}{\ensuremath{4.25\pm0.10}}              
\newcommand{\hatcurISOvixxxxxB}{\ensuremath{0.646\pm0.014}}            
\newcommand{\hatcurISOagexxxxxB}{\ensuremath{4.7_{-0.8}^{+1.3}}}       
\newcommand{\hatcurISOsigmaxxxxxB}{\ensuremath{0.00020\pm0.00007}}     
\newcommand{\hatcurISOMJxxxxxB}{\ensuremath{3.18\pm0.09}}              
\newcommand{\hatcurISOMHxxxxxB}{\ensuremath{2.87\pm0.09}}              
\newcommand{\hatcurISOMKxxxxxB}{\ensuremath{2.81\pm0.09}}              
\newcommand{\hatcurISOJKxxxxxB}{\ensuremath{0.37\pm0.01}}              
\newcommand{\hatcurISOspecxxxxxB}{G}                                   
\newcommand{\hatcurRVKxxxxxB}{\ensuremath{19.6\pm2.8}}                 
\newcommand{\hatcurRVrkxxxxxB}{\ensuremath{0.000\pm0.000}}             
\newcommand{\hatcurRVrhxxxxxB}{\ensuremath{0.000\pm0.000}}             
\newcommand{\hatcurRVkxxxxxB}{\ensuremath{0.000\pm0.000}}              
\newcommand{\hatcurRVhxxxxxB}{\ensuremath{0.000\pm0.000}}              
\newcommand{\hatcurRVtronexxxxxB}{\ensuremath{0.0000\pm0.0000}}        
\newcommand{\hatcurRVtrtwoxxxxxB}{\ensuremath{0.0000\pm0.0000}}        
\newcommand{\hatcurRVgammaAxxxxxB}{\ensuremath{-0.2\pm6.9}}            
\newcommand{\hatcurRVjitterAxxxxxB}{\ensuremath{20.5}}                 
\newcommand{\hatcurRVfitrmsAxxxxxB}{\ensuremath{21.8}}                 
\newcommand{\hatcurRVgammaBxxxxxB}{\ensuremath{3.1\pm2.2}}             
\newcommand{\hatcurRVjitterBxxxxxB}{\ensuremath{0.0}}                  
\newcommand{\hatcurRVfitrmsBxxxxxB}{\ensuremath{9.0}}                  
\newcommand{\hatcurRVgammaCxxxxxB}{\ensuremath{11.8\pm8.0}}            
\newcommand{\hatcurRVjitterCxxxxxB}{\ensuremath{17.5}}                 
\newcommand{\hatcurRVfitrmsCxxxxxB}{\ensuremath{21.0}}                 
\newcommand{\hatcurRVeccenxxxxxB}{\ensuremath{0.000\pm0.000}}          
\newcommand{\hatcurRVomegaxxxxxB}{\ensuremath{0\pm0}}                  
\newcommand{\hatcurPPixxxxxB}{\ensuremath{86.8_{-0.3}^{+0.5}}}         
\newcommand{\hatcurPPgxxxxxB}{\ensuremath{3.3\pm0.6}}                  
\newcommand{\hatcurPPloggxxxxxB}{\ensuremath{2.51\pm0.07}}             
\newcommand{\hatcurPParxxxxxB}{\ensuremath{9.53_{-0.27}^{+0.36}}}      
\newcommand{\hatcurPParelxxxxxB}{\ensuremath{0.0543\pm0.0007}}         
\newcommand{\hatcurPPrhoxxxxxB}{\ensuremath{0.14_{-0.02}^{+0.03}}}     
\newcommand{\hatcurPPmxxxxxB}{\ensuremath{0.17\pm0.02}}                
\newcommand{\hatcurPPmshortxxxxxB}{\ensuremath{0.17}}                  
\newcommand{\hatcurPPmlongxxxxxB}{\ensuremath{0.168\pm0.024}}          
\newcommand{\hatcurPPmexxxxxB}{\ensuremath{53.5\pm7.6}}                
\newcommand{\hatcurPPmeshortxxxxxB}{\ensuremath{53.5}}                 
\newcommand{\hatcurPPmelongxxxxxB}{\ensuremath{53.49\pm7.64}}          
\newcommand{\hatcurPPrxxxxxB}{\ensuremath{1.13\pm0.05}}                
\newcommand{\hatcurPPrshortxxxxxB}{\ensuremath{1.13}}                  
\newcommand{\hatcurPPrlongxxxxxB}{\ensuremath{1.131\pm0.054}}          
\newcommand{\hatcurPPrexxxxxB}{\ensuremath{12.7\pm0.6}}                
\newcommand{\hatcurPPreshortxxxxxB}{\ensuremath{12.7}}                 
\newcommand{\hatcurPPrelongxxxxxB}{\ensuremath{12.68\pm0.61}}          
\newcommand{\hatcurPPmrcorrxxxxxB}{\ensuremath{0.07}}                  
\newcommand{\hatcurPPteffxxxxxB}{\ensuremath{1361\pm25}}               
\newcommand{\hatcurPPthetaxxxxxB}{\ensuremath{0.015\pm0.002}}          
\newcommand{\hatcurPPfluxperixxxxxB}{\ensuremath{7.75\pm0.57}}         
\newcommand{\hatcurPPfluxperidimxxxxxB}{\ensuremath{8}}                
\newcommand{\hatcurPPfluxapxxxxxB}{\ensuremath{7.75\pm0.57}}           
\newcommand{\hatcurPPfluxapdimxxxxxB}{\ensuremath{8}}                  
\newcommand{\hatcurPPfluxavgxxxxxB}{\ensuremath{7.75\pm0.57}}          
\newcommand{\hatcurPPfluxavgdimxxxxxB}{\ensuremath{8}}                 
\newcommand{\hatcurXsecphasexxxxxB}{\ensuremath{0.5000\pm0.0000}}      
\newcommand{\hatcurXsecondaryxxxxxB}{\ensuremath{2455841.415\pm0.000}} 
\newcommand{\hatcurXsecdurxxxxxB}{\ensuremath{0.1411\pm0.0013}}        
\newcommand{\hatcurXsecingdurxxxxxB}{\ensuremath{0.0165\pm0.0013}}     
\newcommand{\hatcurPPphiconjxxxxxB}{\ensuremath{0.2500\pm0.0000}}      
\newcommand{\hatcurPPperixxxxxB}{\ensuremath{2455838.11\pm0.00}}       
\newcommand{\hatcurPPaequivxxxxxB}{\ensuremath{0.0420\pm0.0016}}       
\newcommand{\hatcurPPtcircxxxxxB}{\ensuremath{180.7_{-37.9}^{+66.0}}}  
\newcommand{\hatcurPPtinfallxxxxxB}{\ensuremath{21581.1_{-3579.9}^{+6594.0}}} 
\newcommand{\hatcurXdistxxxxxB}{\ensuremath{314\pm12}}                 
\newcommand{\hatcurXAvxxxxxB}{\ensuremath{0.500\pm0.110}}              
\newcommand{\hatcurXdistredxxxxxB}{\ensuremath{305\pm12}}              
\newcommand{\hatcurXEBVxxxxxB}{\ensuremath{0.161\pm0.036}}             
\newcommand{\hatcurXmvisoredxxxxxB}{\ensuremath{12.176\pm0.084}}       
\newcommand{\hatcurXmiisoredxxxxxB}{\ensuremath{11.269\pm0.032}}       
\newcommand{\hatcurXmjisoredxxxxxB}{\ensuremath{10.746\pm0.016}}       
\newcommand{\hatcurXmhisoredxxxxxB}{\ensuremath{10.384\pm0.015}}       
\newcommand{\hatcurXmkisoredxxxxxB}{\ensuremath{10.294\pm0.018}}       
\newcommand{\hatcurXviisoredxxxxxB}{\ensuremath{0.907\pm0.055}}        
\newcommand{\hatcurXvkisoredxxxxxB}{\ensuremath{1.882\pm0.092}}        
\newcommand{\hatcurXjhisoredxxxxxB}{\ensuremath{0.362\pm0.011}}        
\newcommand{\hatcurXjkisoredxxxxxB}{\ensuremath{0.452\pm0.018}}        
\newcommand{\hatcurCCpmraxxxxxB}{\ensuremath{18.1\pm3.0}}              
\newcommand{\hatcurCCpmdecxxxxxB}{\ensuremath{-12.5\pm3.0}}            
\newcommand{\hatcurCCpmxxxxxB}{\ensuremath{21.9968\pm4.24264}}         

\newcommand{\hatcurCCbbHmag}[1]{\ifnum#1=47 %
\hatcurCCbbHmagxxxxxA
\else
\ifnum#1=48 %
\hatcurCCbbHmagxxxxxB
\else
??????\fi
\fi
}
\newcommand{\hatcurCCbbJmag}[1]{\ifnum#1=47 %
\hatcurCCbbJmagxxxxxA
\else
\ifnum#1=48 %
\hatcurCCbbJmagxxxxxB
\else
??????\fi
\fi
}
\newcommand{\hatcurCCbbKmag}[1]{\ifnum#1=47 %
\hatcurCCbbKmagxxxxxA
\else
\ifnum#1=48 %
\hatcurCCbbKmagxxxxxB
\else
??????\fi
\fi
}
\newcommand{\hatcurCCcitHmag}[1]{\ifnum#1=47 %
\hatcurCCcitHmagxxxxxA
\else
\ifnum#1=48 %
\hatcurCCcitHmagxxxxxB
\else
??????\fi
\fi
}
\newcommand{\hatcurCCcitJmag}[1]{\ifnum#1=47 %
\hatcurCCcitJmagxxxxxA
\else
\ifnum#1=48 %
\hatcurCCcitJmagxxxxxB
\else
??????\fi
\fi
}
\newcommand{\hatcurCCcitKmag}[1]{\ifnum#1=47 %
\hatcurCCcitKmagxxxxxA
\else
\ifnum#1=48 %
\hatcurCCcitKmagxxxxxB
\else
??????\fi
\fi
}
\newcommand{\hatcurCCdec}[1]{\ifnum#1=47 %
\hatcurCCdecxxxxxA
\else
\ifnum#1=48 %
\hatcurCCdecxxxxxB
\else
??????\fi
\fi
}
\newcommand{\hatcurCCesoHKmag}[1]{\ifnum#1=47 %
\hatcurCCesoHKmagxxxxxA
\else
\ifnum#1=48 %
\hatcurCCesoHKmagxxxxxB
\else
??????\fi
\fi
}
\newcommand{\hatcurCCesoHmag}[1]{\ifnum#1=47 %
\hatcurCCesoHmagxxxxxA
\else
\ifnum#1=48 %
\hatcurCCesoHmagxxxxxB
\else
??????\fi
\fi
}
\newcommand{\hatcurCCesoJHmag}[1]{\ifnum#1=47 %
\hatcurCCesoJHmagxxxxxA
\else
\ifnum#1=48 %
\hatcurCCesoJHmagxxxxxB
\else
??????\fi
\fi
}
\newcommand{\hatcurCCesoJKmag}[1]{\ifnum#1=47 %
\hatcurCCesoJKmagxxxxxA
\else
\ifnum#1=48 %
\hatcurCCesoJKmagxxxxxB
\else
??????\fi
\fi
}
\newcommand{\hatcurCCesoJmag}[1]{\ifnum#1=47 %
\hatcurCCesoJmagxxxxxA
\else
\ifnum#1=48 %
\hatcurCCesoJmagxxxxxB
\else
??????\fi
\fi
}
\newcommand{\hatcurCCesoKmag}[1]{\ifnum#1=47 %
\hatcurCCesoKmagxxxxxA
\else
\ifnum#1=48 %
\hatcurCCesoKmagxxxxxB
\else
??????\fi
\fi
}
\newcommand{\hatcurCCgsc}[1]{\ifnum#1=47 %
\hatcurCCgscxxxxxA
\else
\ifnum#1=48 %
\hatcurCCgscxxxxxB
\else
??????\fi
\fi
}
\newcommand{\hatcurCCmag}[1]{\ifnum#1=47 %
\hatcurCCmagxxxxxA
\else
\ifnum#1=48 %
\hatcurCCmagxxxxxB
\else
??????\fi
\fi
}
\newcommand{\hatcurCCpm}[1]{\ifnum#1=47 %
\hatcurCCpmxxxxxA
\else
\ifnum#1=48 %
\hatcurCCpmxxxxxB
\else
??????\fi
\fi
}
\newcommand{\hatcurCCpmdec}[1]{\ifnum#1=47 %
\hatcurCCpmdecxxxxxA
\else
\ifnum#1=48 %
\hatcurCCpmdecxxxxxB
\else
??????\fi
\fi
}
\newcommand{\hatcurCCpmra}[1]{\ifnum#1=47 %
\hatcurCCpmraxxxxxA
\else
\ifnum#1=48 %
\hatcurCCpmraxxxxxB
\else
??????\fi
\fi
}
\newcommand{\hatcurCCra}[1]{\ifnum#1=47 %
\hatcurCCraxxxxxA
\else
\ifnum#1=48 %
\hatcurCCraxxxxxB
\else
??????\fi
\fi
}
\newcommand{\hatcurCCtassmv}[1]{\ifnum#1=47 %
\hatcurCCtassmvxxxxxA
\else
\ifnum#1=48 %
\hatcurCCtassmvxxxxxB
\else
??????\fi
\fi
}
\newcommand{\hatcurCCtassmI}[1]{\ifnum#1=47 %
\hatcurCCtassmIxxxxxA
\else
\ifnum#1=48 %
\hatcurCCtassmIxxxxxB
\else
??????\fi
\fi
}
\newcommand{\hatcurCCtwomass}[1]{\ifnum#1=47 %
\hatcurCCtwomassxxxxxA
\else
\ifnum#1=48 %
\hatcurCCtwomassxxxxxB
\else
??????\fi
\fi
}
\newcommand{\hatcurCCtwomassHmag}[1]{\ifnum#1=47 %
\hatcurCCtwomassHmagxxxxxA
\else
\ifnum#1=48 %
\hatcurCCtwomassHmagxxxxxB
\else
??????\fi
\fi
}
\newcommand{\hatcurCCtwomassJmag}[1]{\ifnum#1=47 %
\hatcurCCtwomassJmagxxxxxA
\else
\ifnum#1=48 %
\hatcurCCtwomassJmagxxxxxB
\else
??????\fi
\fi
}
\newcommand{\hatcurCCtwomassKmag}[1]{\ifnum#1=47 %
\hatcurCCtwomassKmagxxxxxA
\else
\ifnum#1=48 %
\hatcurCCtwomassKmagxxxxxB
\else
??????\fi
\fi
}
\newcommand{\hatcurDSgamma}[1]{\ifnum#1=47 %
\hatcurDSgammaxxxxxA
\else
\ifnum#1=48 %
\hatcurDSgammaxxxxxB
\else
??????\fi
\fi
}
\newcommand{\hatcurDSlogg}[1]{\ifnum#1=47 %
\hatcurDSloggxxxxxA
\else
\ifnum#1=48 %
\hatcurDSloggxxxxxB
\else
??????\fi
\fi
}
\newcommand{\hatcurDSnumspec}[1]{\ifnum#1=47 %
\hatcurDSnumspecxxxxxA
\else
\ifnum#1=48 %
\hatcurDSnumspecxxxxxB
\else
??????\fi
\fi
}
\newcommand{\hatcurDSrvrms}[1]{\ifnum#1=47 %
\hatcurDSrvrmsxxxxxA
\else
\ifnum#1=48 %
\hatcurDSrvrmsxxxxxB
\else
??????\fi
\fi
}
\newcommand{\hatcurDSspan}[1]{\ifnum#1=47 %
\hatcurDSspanxxxxxA
\else
\ifnum#1=48 %
\hatcurDSspanxxxxxB
\else
??????\fi
\fi
}
\newcommand{\hatcurDSteff}[1]{\ifnum#1=47 %
\hatcurDSteffxxxxxA
\else
\ifnum#1=48 %
\hatcurDSteffxxxxxB
\else
??????\fi
\fi
}
\newcommand{\hatcurDSvsini}[1]{\ifnum#1=47 %
\hatcurDSvsinixxxxxA
\else
\ifnum#1=48 %
\hatcurDSvsinixxxxxB
\else
??????\fi
\fi
}
\newcommand{\hatcurDSzfeh}[1]{\ifnum#1=47 %
\hatcurDSzfehxxxxxA
\else
\ifnum#1=48 %
\hatcurDSzfehxxxxxB
\else
??????\fi
\fi
}
\newcommand{\hatcurfield}[1]{\ifnum#1=47 %
\hatcurfieldxxxxxA
\else
\ifnum#1=48 %
\hatcurfieldxxxxxB
\else
??????\fi
\fi
}
\newcommand{\hatcurFIESgamma}[1]{\ifnum#1=47 %
\hatcurFIESgammaxxxxxA
\else
\ifnum#1=48 %
\hatcurFIESgammaxxxxxB
\else
??????\fi
\fi
}
\newcommand{\hatcurFIESlogg}[1]{\ifnum#1=47 %
\hatcurFIESloggxxxxxA
\else
\ifnum#1=48 %
\hatcurFIESloggxxxxxB
\else
??????\fi
\fi
}
\newcommand{\hatcurFIESnumspec}[1]{\ifnum#1=47 %
\hatcurFIESnumspecxxxxxA
\else
\ifnum#1=48 %
\hatcurFIESnumspecxxxxxB
\else
??????\fi
\fi
}
\newcommand{\hatcurFIESrvrms}[1]{\ifnum#1=47 %
\hatcurFIESrvrmsxxxxxA
\else
\ifnum#1=48 %
\hatcurFIESrvrmsxxxxxB
\else
??????\fi
\fi
}
\newcommand{\hatcurFIESspan}[1]{\ifnum#1=47 %
\hatcurFIESspanxxxxxA
\else
\ifnum#1=48 %
\hatcurFIESspanxxxxxB
\else
??????\fi
\fi
}
\newcommand{\hatcurFIESteff}[1]{\ifnum#1=47 %
\hatcurFIESteffxxxxxA
\else
\ifnum#1=48 %
\hatcurFIESteffxxxxxB
\else
??????\fi
\fi
}
\newcommand{\hatcurFIESvsini}[1]{\ifnum#1=47 %
\hatcurFIESvsinixxxxxA
\else
\ifnum#1=48 %
\hatcurFIESvsinixxxxxB
\else
??????\fi
\fi
}
\newcommand{\hatcurFIESzfeh}[1]{\ifnum#1=47 %
\hatcurFIESzfehxxxxxA
\else
\ifnum#1=48 %
\hatcurFIESzfehxxxxxB
\else
??????\fi
\fi
}
\newcommand{\hatcurhtr}[1]{\ifnum#1=47 %
\hatcurhtrxxxxxA
\else
\ifnum#1=48 %
\hatcurhtrxxxxxB
\else
??????\fi
\fi
}
\newcommand{\hatcurISOage}[1]{\ifnum#1=47 %
\hatcurISOagexxxxxA
\else
\ifnum#1=48 %
\hatcurISOagexxxxxB
\else
??????\fi
\fi
}
\newcommand{\hatcurISOJK}[1]{\ifnum#1=47 %
\hatcurISOJKxxxxxA
\else
\ifnum#1=48 %
\hatcurISOJKxxxxxB
\else
??????\fi
\fi
}
\newcommand{\hatcurISOlogg}[1]{\ifnum#1=47 %
\hatcurISOloggxxxxxA
\else
\ifnum#1=48 %
\hatcurISOloggxxxxxB
\else
??????\fi
\fi
}
\newcommand{\hatcurISOlum}[1]{\ifnum#1=47 %
\hatcurISOlumxxxxxA
\else
\ifnum#1=48 %
\hatcurISOlumxxxxxB
\else
??????\fi
\fi
}
\newcommand{\hatcurISOlumshort}[1]{\ifnum#1=47 %
\hatcurISOlumshortxxxxxA
\else
\ifnum#1=48 %
\hatcurISOlumshortxxxxxB
\else
??????\fi
\fi
}
\newcommand{\hatcurISOm}[1]{\ifnum#1=47 %
\hatcurISOmxxxxxA
\else
\ifnum#1=48 %
\hatcurISOmxxxxxB
\else
??????\fi
\fi
}
\newcommand{\hatcurISOMH}[1]{\ifnum#1=47 %
\hatcurISOMHxxxxxA
\else
\ifnum#1=48 %
\hatcurISOMHxxxxxB
\else
??????\fi
\fi
}
\newcommand{\hatcurISOMJ}[1]{\ifnum#1=47 %
\hatcurISOMJxxxxxA
\else
\ifnum#1=48 %
\hatcurISOMJxxxxxB
\else
??????\fi
\fi
}
\newcommand{\hatcurISOMK}[1]{\ifnum#1=47 %
\hatcurISOMKxxxxxA
\else
\ifnum#1=48 %
\hatcurISOMKxxxxxB
\else
??????\fi
\fi
}
\newcommand{\hatcurISOmlong}[1]{\ifnum#1=47 %
\hatcurISOmlongxxxxxA
\else
\ifnum#1=48 %
\hatcurISOmlongxxxxxB
\else
??????\fi
\fi
}
\newcommand{\hatcurISOmshort}[1]{\ifnum#1=47 %
\hatcurISOmshortxxxxxA
\else
\ifnum#1=48 %
\hatcurISOmshortxxxxxB
\else
??????\fi
\fi
}
\newcommand{\hatcurISOmv}[1]{\ifnum#1=47 %
\hatcurISOmvxxxxxA
\else
\ifnum#1=48 %
\hatcurISOmvxxxxxB
\else
??????\fi
\fi
}
\newcommand{\hatcurISOr}[1]{\ifnum#1=47 %
\hatcurISOrxxxxxA
\else
\ifnum#1=48 %
\hatcurISOrxxxxxB
\else
??????\fi
\fi
}
\newcommand{\hatcurISOrho}[1]{\ifnum#1=47 %
\hatcurISOrhoxxxxxA
\else
\ifnum#1=48 %
\hatcurISOrhoxxxxxB
\else
??????\fi
\fi
}
\newcommand{\hatcurISOrlong}[1]{\ifnum#1=47 %
\hatcurISOrlongxxxxxA
\else
\ifnum#1=48 %
\hatcurISOrlongxxxxxB
\else
??????\fi
\fi
}
\newcommand{\hatcurISOrshort}[1]{\ifnum#1=47 %
\hatcurISOrshortxxxxxA
\else
\ifnum#1=48 %
\hatcurISOrshortxxxxxB
\else
??????\fi
\fi
}
\newcommand{\hatcurISOsigma}[1]{\ifnum#1=47 %
\hatcurISOsigmaxxxxxA
\else
\ifnum#1=48 %
\hatcurISOsigmaxxxxxB
\else
??????\fi
\fi
}
\newcommand{\hatcurISOspec}[1]{\ifnum#1=47 %
\hatcurISOspecxxxxxA
\else
\ifnum#1=48 %
\hatcurISOspecxxxxxB
\else
??????\fi
\fi
}
\newcommand{\hatcurISOvi}[1]{\ifnum#1=47 %
\hatcurISOvixxxxxA
\else
\ifnum#1=48 %
\hatcurISOvixxxxxB
\else
??????\fi
\fi
}
\newcommand{\hatcurLBig}[1]{\ifnum#1=47 %
\hatcurLBigxxxxxA
\else
\ifnum#1=48 %
\hatcurLBigxxxxxB
\else
??????\fi
\fi
}
\newcommand{\hatcurLBii}[1]{\ifnum#1=47 %
\hatcurLBiixxxxxA
\else
\ifnum#1=48 %
\hatcurLBiixxxxxB
\else
??????\fi
\fi
}
\newcommand{\hatcurLBiI}[1]{\ifnum#1=47 %
\hatcurLBiIxxxxxA
\else
\ifnum#1=48 %
\hatcurLBiIxxxxxB
\else
??????\fi
\fi
}
\newcommand{\hatcurLBiig}[1]{\ifnum#1=47 %
\hatcurLBiigxxxxxA
\else
\ifnum#1=48 %
\hatcurLBiigxxxxxB
\else
??????\fi
\fi
}
\newcommand{\hatcurLBiii}[1]{\ifnum#1=47 %
\hatcurLBiiixxxxxA
\else
\ifnum#1=48 %
\hatcurLBiiixxxxxB
\else
??????\fi
\fi
}
\newcommand{\hatcurLBiiI}[1]{\ifnum#1=47 %
\hatcurLBiiIxxxxxA
\else
\ifnum#1=48 %
\hatcurLBiiIxxxxxB
\else
??????\fi
\fi
}
\newcommand{\hatcurLBiikep}[1]{\ifnum#1=47 %
\hatcurLBiikepxxxxxA
\else
\ifnum#1=48 %
\hatcurLBiikepxxxxxB
\else
??????\fi
\fi
}
\newcommand{\hatcurLBiir}[1]{\ifnum#1=47 %
\hatcurLBiirxxxxxA
\else
\ifnum#1=48 %
\hatcurLBiirxxxxxB
\else
??????\fi
\fi
}
\newcommand{\hatcurLBiiR}[1]{\ifnum#1=47 %
\hatcurLBiiRxxxxxA
\else
\ifnum#1=48 %
\hatcurLBiiRxxxxxB
\else
??????\fi
\fi
}
\newcommand{\hatcurLBiiz}[1]{\ifnum#1=47 %
\hatcurLBiizxxxxxA
\else
\ifnum#1=48 %
\hatcurLBiizxxxxxB
\else
??????\fi
\fi
}
\newcommand{\hatcurLBikep}[1]{\ifnum#1=47 %
\hatcurLBikepxxxxxA
\else
\ifnum#1=48 %
\hatcurLBikepxxxxxB
\else
??????\fi
\fi
}
\newcommand{\hatcurLBir}[1]{\ifnum#1=47 %
\hatcurLBirxxxxxA
\else
\ifnum#1=48 %
\hatcurLBirxxxxxB
\else
??????\fi
\fi
}
\newcommand{\hatcurLBiR}[1]{\ifnum#1=47 %
\hatcurLBiRxxxxxA
\else
\ifnum#1=48 %
\hatcurLBiRxxxxxB
\else
??????\fi
\fi
}
\newcommand{\hatcurLBiz}[1]{\ifnum#1=47 %
\hatcurLBizxxxxxA
\else
\ifnum#1=48 %
\hatcurLBizxxxxxB
\else
??????\fi
\fi
}
\newcommand{\hatcurLCbsq}[1]{\ifnum#1=47 %
\hatcurLCbsqxxxxxA
\else
\ifnum#1=48 %
\hatcurLCbsqxxxxxB
\else
??????\fi
\fi
}
\newcommand{\hatcurLCdip}[1]{\ifnum#1=47 %
\hatcurLCdipxxxxxA
\else
\ifnum#1=48 %
\hatcurLCdipxxxxxB
\else
??????\fi
\fi
}
\newcommand{\hatcurLCdur}[1]{\ifnum#1=47 %
\hatcurLCdurxxxxxA
\else
\ifnum#1=48 %
\hatcurLCdurxxxxxB
\else
??????\fi
\fi
}
\newcommand{\hatcurLCdurhr}[1]{\ifnum#1=47 %
\hatcurLCdurhrxxxxxA
\else
\ifnum#1=48 %
\hatcurLCdurhrxxxxxB
\else
??????\fi
\fi
}
\newcommand{\hatcurLCdurhrshort}[1]{\ifnum#1=47 %
\hatcurLCdurhrshortxxxxxA
\else
\ifnum#1=48 %
\hatcurLCdurhrshortxxxxxB
\else
??????\fi
\fi
}
\newcommand{\hatcurLCdurshort}[1]{\ifnum#1=47 %
\hatcurLCdurshortxxxxxA
\else
\ifnum#1=48 %
\hatcurLCdurshortxxxxxB
\else
??????\fi
\fi
}
\newcommand{\hatcurLChatnetm}[1]{\ifnum#1=47 %
\hatcurLChatnetmxxxxxA
\else
\ifnum#1=48 %
\hatcurLChatnetmxxxxxB
\else
??????\fi
\fi
}
\newcommand{\hatcurLCiblend}[1]{\ifnum#1=47 %
\hatcurLCiblendxxxxxA
\else
\ifnum#1=48 %
\hatcurLCiblendxxxxxB
\else
??????\fi
\fi
}
\newcommand{\hatcurLCimp}[1]{\ifnum#1=47 %
\hatcurLCimpxxxxxA
\else
\ifnum#1=48 %
\hatcurLCimpxxxxxB
\else
??????\fi
\fi
}
\newcommand{\hatcurLCingdur}[1]{\ifnum#1=47 %
\hatcurLCingdurxxxxxA
\else
\ifnum#1=48 %
\hatcurLCingdurxxxxxB
\else
??????\fi
\fi
}
\newcommand{\hatcurLCP}[1]{\ifnum#1=47 %
\hatcurLCPxxxxxA
\else
\ifnum#1=48 %
\hatcurLCPxxxxxB
\else
??????\fi
\fi
}
\newcommand{\hatcurLCPprec}[1]{\ifnum#1=47 %
\hatcurLCPprecxxxxxA
\else
\ifnum#1=48 %
\hatcurLCPprecxxxxxB
\else
??????\fi
\fi
}
\newcommand{\hatcurLCPshort}[1]{\ifnum#1=47 %
\hatcurLCPshortxxxxxA
\else
\ifnum#1=48 %
\hatcurLCPshortxxxxxB
\else
??????\fi
\fi
}
\newcommand{\hatcurLCq}[1]{\ifnum#1=47 %
\hatcurLCqxxxxxA
\else
\ifnum#1=48 %
\hatcurLCqxxxxxB
\else
??????\fi
\fi
}
\newcommand{\hatcurLCqshort}[1]{\ifnum#1=47 %
\hatcurLCqshortxxxxxA
\else
\ifnum#1=48 %
\hatcurLCqshortxxxxxB
\else
??????\fi
\fi
}
\newcommand{\hatcurLCrprstar}[1]{\ifnum#1=47 %
\hatcurLCrprstarxxxxxA
\else
\ifnum#1=48 %
\hatcurLCrprstarxxxxxB
\else
??????\fi
\fi
}
\newcommand{\hatcurLCT}[1]{\ifnum#1=47 %
\hatcurLCTxxxxxA
\else
\ifnum#1=48 %
\hatcurLCTxxxxxB
\else
??????\fi
\fi
}
\newcommand{\hatcurLCTA}[1]{\ifnum#1=47 %
\hatcurLCTAxxxxxA
\else
\ifnum#1=48 %
\hatcurLCTAxxxxxB
\else
??????\fi
\fi
}
\newcommand{\hatcurLCTB}[1]{\ifnum#1=47 %
\hatcurLCTBxxxxxA
\else
\ifnum#1=48 %
\hatcurLCTBxxxxxB
\else
??????\fi
\fi
}
\newcommand{\hatcurLCzeta}[1]{\ifnum#1=47 %
\hatcurLCzetaxxxxxA
\else
\ifnum#1=48 %
\hatcurLCzetaxxxxxB
\else
??????\fi
\fi
}
\newcommand{\hatcurPPaequiv}[1]{\ifnum#1=47 %
\hatcurPPaequivxxxxxA
\else
\ifnum#1=48 %
\hatcurPPaequivxxxxxB
\else
??????\fi
\fi
}
\newcommand{\hatcurPPar}[1]{\ifnum#1=47 %
\hatcurPParxxxxxA
\else
\ifnum#1=48 %
\hatcurPParxxxxxB
\else
??????\fi
\fi
}
\newcommand{\hatcurPParel}[1]{\ifnum#1=47 %
\hatcurPParelxxxxxA
\else
\ifnum#1=48 %
\hatcurPParelxxxxxB
\else
??????\fi
\fi
}
\newcommand{\hatcurPPfluxap}[1]{\ifnum#1=47 %
\hatcurPPfluxapxxxxxA
\else
\ifnum#1=48 %
\hatcurPPfluxapxxxxxB
\else
??????\fi
\fi
}
\newcommand{\hatcurPPfluxapdim}[1]{\ifnum#1=47 %
\hatcurPPfluxapdimxxxxxA
\else
\ifnum#1=48 %
\hatcurPPfluxapdimxxxxxB
\else
??????\fi
\fi
}
\newcommand{\hatcurPPfluxavg}[1]{\ifnum#1=47 %
\hatcurPPfluxavgxxxxxA
\else
\ifnum#1=48 %
\hatcurPPfluxavgxxxxxB
\else
??????\fi
\fi
}
\newcommand{\hatcurPPfluxavgdim}[1]{\ifnum#1=47 %
\hatcurPPfluxavgdimxxxxxA
\else
\ifnum#1=48 %
\hatcurPPfluxavgdimxxxxxB
\else
??????\fi
\fi
}
\newcommand{\hatcurPPfluxperi}[1]{\ifnum#1=47 %
\hatcurPPfluxperixxxxxA
\else
\ifnum#1=48 %
\hatcurPPfluxperixxxxxB
\else
??????\fi
\fi
}
\newcommand{\hatcurPPfluxperidim}[1]{\ifnum#1=47 %
\hatcurPPfluxperidimxxxxxA
\else
\ifnum#1=48 %
\hatcurPPfluxperidimxxxxxB
\else
??????\fi
\fi
}
\newcommand{\hatcurPPg}[1]{\ifnum#1=47 %
\hatcurPPgxxxxxA
\else
\ifnum#1=48 %
\hatcurPPgxxxxxB
\else
??????\fi
\fi
}
\newcommand{\hatcurPPi}[1]{\ifnum#1=47 %
\hatcurPPixxxxxA
\else
\ifnum#1=48 %
\hatcurPPixxxxxB
\else
??????\fi
\fi
}
\newcommand{\hatcurPPlogg}[1]{\ifnum#1=47 %
\hatcurPPloggxxxxxA
\else
\ifnum#1=48 %
\hatcurPPloggxxxxxB
\else
??????\fi
\fi
}
\newcommand{\hatcurPPm}[1]{\ifnum#1=47 %
\hatcurPPmxxxxxA
\else
\ifnum#1=48 %
\hatcurPPmxxxxxB
\else
??????\fi
\fi
}
\newcommand{\hatcurPPme}[1]{\ifnum#1=47 %
\hatcurPPmexxxxxA
\else
\ifnum#1=48 %
\hatcurPPmexxxxxB
\else
??????\fi
\fi
}
\newcommand{\hatcurPPmelong}[1]{\ifnum#1=47 %
\hatcurPPmelongxxxxxA
\else
\ifnum#1=48 %
\hatcurPPmelongxxxxxB
\else
??????\fi
\fi
}
\newcommand{\hatcurPPmeshort}[1]{\ifnum#1=47 %
\hatcurPPmeshortxxxxxA
\else
\ifnum#1=48 %
\hatcurPPmeshortxxxxxB
\else
??????\fi
\fi
}
\newcommand{\hatcurPPmlong}[1]{\ifnum#1=47 %
\hatcurPPmlongxxxxxA
\else
\ifnum#1=48 %
\hatcurPPmlongxxxxxB
\else
??????\fi
\fi
}
\newcommand{\hatcurPPmrcorr}[1]{\ifnum#1=47 %
\hatcurPPmrcorrxxxxxA
\else
\ifnum#1=48 %
\hatcurPPmrcorrxxxxxB
\else
??????\fi
\fi
}
\newcommand{\hatcurPPmshort}[1]{\ifnum#1=47 %
\hatcurPPmshortxxxxxA
\else
\ifnum#1=48 %
\hatcurPPmshortxxxxxB
\else
??????\fi
\fi
}
\newcommand{\hatcurPPperi}[1]{\ifnum#1=47 %
\hatcurPPperixxxxxA
\else
\ifnum#1=48 %
\hatcurPPperixxxxxB
\else
??????\fi
\fi
}
\newcommand{\hatcurPPphiconj}[1]{\ifnum#1=47 %
\hatcurPPphiconjxxxxxA
\else
\ifnum#1=48 %
\hatcurPPphiconjxxxxxB
\else
??????\fi
\fi
}
\newcommand{\hatcurPPr}[1]{\ifnum#1=47 %
\hatcurPPrxxxxxA
\else
\ifnum#1=48 %
\hatcurPPrxxxxxB
\else
??????\fi
\fi
}
\newcommand{\hatcurPPre}[1]{\ifnum#1=47 %
\hatcurPPrexxxxxA
\else
\ifnum#1=48 %
\hatcurPPrexxxxxB
\else
??????\fi
\fi
}
\newcommand{\hatcurPPrelong}[1]{\ifnum#1=47 %
\hatcurPPrelongxxxxxA
\else
\ifnum#1=48 %
\hatcurPPrelongxxxxxB
\else
??????\fi
\fi
}
\newcommand{\hatcurPPreshort}[1]{\ifnum#1=47 %
\hatcurPPreshortxxxxxA
\else
\ifnum#1=48 %
\hatcurPPreshortxxxxxB
\else
??????\fi
\fi
}
\newcommand{\hatcurPPrho}[1]{\ifnum#1=47 %
\hatcurPPrhoxxxxxA
\else
\ifnum#1=48 %
\hatcurPPrhoxxxxxB
\else
??????\fi
\fi
}
\newcommand{\hatcurPPrlong}[1]{\ifnum#1=47 %
\hatcurPPrlongxxxxxA
\else
\ifnum#1=48 %
\hatcurPPrlongxxxxxB
\else
??????\fi
\fi
}
\newcommand{\hatcurPPrshort}[1]{\ifnum#1=47 %
\hatcurPPrshortxxxxxA
\else
\ifnum#1=48 %
\hatcurPPrshortxxxxxB
\else
??????\fi
\fi
}
\newcommand{\hatcurPPtcirc}[1]{\ifnum#1=47 %
\hatcurPPtcircxxxxxA
\else
\ifnum#1=48 %
\hatcurPPtcircxxxxxB
\else
??????\fi
\fi
}
\newcommand{\hatcurPPteff}[1]{\ifnum#1=47 %
\hatcurPPteffxxxxxA
\else
\ifnum#1=48 %
\hatcurPPteffxxxxxB
\else
??????\fi
\fi
}
\newcommand{\hatcurPPtheta}[1]{\ifnum#1=47 %
\hatcurPPthetaxxxxxA
\else
\ifnum#1=48 %
\hatcurPPthetaxxxxxB
\else
??????\fi
\fi
}
\newcommand{\hatcurPPtinfall}[1]{\ifnum#1=47 %
\hatcurPPtinfallxxxxxA
\else
\ifnum#1=48 %
\hatcurPPtinfallxxxxxB
\else
??????\fi
\fi
}
\newcommand{\hatcurRVeccen}[1]{\ifnum#1=47 %
\hatcurRVeccenxxxxxA
\else
\ifnum#1=48 %
\hatcurRVeccenxxxxxB
\else
??????\fi
\fi
}
\newcommand{\hatcurRVfitrmsA}[1]{\ifnum#1=47 %
\hatcurRVfitrmsAxxxxxA
\else
\ifnum#1=48 %
\hatcurRVfitrmsAxxxxxB
\else
??????\fi
\fi
}
\newcommand{\hatcurRVfitrmsB}[1]{\ifnum#1=47 %
\hatcurRVfitrmsBxxxxxA
\else
\ifnum#1=48 %
\hatcurRVfitrmsBxxxxxB
\else
??????\fi
\fi
}
\newcommand{\hatcurRVfitrmsC}[1]{\ifnum#1=48 %
\hatcurRVfitrmsCxxxxxB
\else
??????\fi
}
\newcommand{\hatcurRVgammaA}[1]{\ifnum#1=47 %
\hatcurRVgammaAxxxxxA
\else
\ifnum#1=48 %
\hatcurRVgammaAxxxxxB
\else
??????\fi
\fi
}
\newcommand{\hatcurRVgammaB}[1]{\ifnum#1=47 %
\hatcurRVgammaBxxxxxA
\else
\ifnum#1=48 %
\hatcurRVgammaBxxxxxB
\else
??????\fi
\fi
}
\newcommand{\hatcurRVgammaC}[1]{\ifnum#1=48 %
\hatcurRVgammaCxxxxxB
\else
??????\fi
}
\newcommand{\hatcurRVh}[1]{\ifnum#1=47 %
\hatcurRVhxxxxxA
\else
\ifnum#1=48 %
\hatcurRVhxxxxxB
\else
??????\fi
\fi
}
\newcommand{\hatcurRVjitterA}[1]{\ifnum#1=47 %
\hatcurRVjitterAxxxxxA
\else
\ifnum#1=48 %
\hatcurRVjitterAxxxxxB
\else
??????\fi
\fi
}
\newcommand{\hatcurRVjitterB}[1]{\ifnum#1=47 %
\hatcurRVjitterBxxxxxA
\else
\ifnum#1=48 %
\hatcurRVjitterBxxxxxB
\else
??????\fi
\fi
}
\newcommand{\hatcurRVjitterC}[1]{\ifnum#1=48 %
\hatcurRVjitterCxxxxxB
\else
??????\fi
}
\newcommand{\hatcurRVk}[1]{\ifnum#1=47 %
\hatcurRVkxxxxxA
\else
\ifnum#1=48 %
\hatcurRVkxxxxxB
\else
??????\fi
\fi
}
\newcommand{\hatcurRVK}[1]{\ifnum#1=47 %
\hatcurRVKxxxxxA
\else
\ifnum#1=48 %
\hatcurRVKxxxxxB
\else
??????\fi
\fi
}
\newcommand{\hatcurRVomega}[1]{\ifnum#1=47 %
\hatcurRVomegaxxxxxA
\else
\ifnum#1=48 %
\hatcurRVomegaxxxxxB
\else
??????\fi
\fi
}
\newcommand{\hatcurRVrh}[1]{\ifnum#1=47 %
\hatcurRVrhxxxxxA
\else
\ifnum#1=48 %
\hatcurRVrhxxxxxB
\else
??????\fi
\fi
}
\newcommand{\hatcurRVrk}[1]{\ifnum#1=47 %
\hatcurRVrkxxxxxA
\else
\ifnum#1=48 %
\hatcurRVrkxxxxxB
\else
??????\fi
\fi
}
\newcommand{\hatcurRVtrone}[1]{\ifnum#1=47 %
\hatcurRVtronexxxxxA
\else
\ifnum#1=48 %
\hatcurRVtronexxxxxB
\else
??????\fi
\fi
}
\newcommand{\hatcurRVtrtwo}[1]{\ifnum#1=47 %
\hatcurRVtrtwoxxxxxA
\else
\ifnum#1=48 %
\hatcurRVtrtwoxxxxxB
\else
??????\fi
\fi
}
\newcommand{\hatcurSMEiilogg}[1]{\ifnum#1=47 %
\hatcurSMEiiloggxxxxxA
\else
\ifnum#1=48 %
\hatcurSMEiiloggxxxxxB
\else
??????\fi
\fi
}
\newcommand{\hatcurSMEiiteff}[1]{\ifnum#1=47 %
\hatcurSMEiiteffxxxxxA
\else
\ifnum#1=48 %
\hatcurSMEiiteffxxxxxB
\else
??????\fi
\fi
}
\newcommand{\hatcurSMEiivmac}[1]{\ifnum#1=47 %
\hatcurSMEiivmacxxxxxA
\else
\ifnum#1=48 %
\hatcurSMEiivmacxxxxxB
\else
??????\fi
\fi
}
\newcommand{\hatcurSMEiivmic}[1]{\ifnum#1=47 %
\hatcurSMEiivmicxxxxxA
\else
\ifnum#1=48 %
\hatcurSMEiivmicxxxxxB
\else
??????\fi
\fi
}
\newcommand{\hatcurSMEiivsin}[1]{\ifnum#1=47 %
\hatcurSMEiivsinxxxxxA
\else
\ifnum#1=48 %
\hatcurSMEiivsinxxxxxB
\else
??????\fi
\fi
}
\newcommand{\hatcurSMEiizfeh}[1]{\ifnum#1=47 %
\hatcurSMEiizfehxxxxxA
\else
\ifnum#1=48 %
\hatcurSMEiizfehxxxxxB
\else
??????\fi
\fi
}
\newcommand{\hatcurSMEiizfehshort}[1]{\ifnum#1=47 %
\hatcurSMEiizfehshortxxxxxA
\else
\ifnum#1=48 %
\hatcurSMEiizfehshortxxxxxB
\else
??????\fi
\fi
}
\newcommand{\hatcurSMEilogg}[1]{\ifnum#1=47 %
\hatcurSMEiloggxxxxxA
\else
\ifnum#1=48 %
\hatcurSMEiloggxxxxxB
\else
??????\fi
\fi
}
\newcommand{\hatcurSMEiteff}[1]{\ifnum#1=47 %
\hatcurSMEiteffxxxxxA
\else
\ifnum#1=48 %
\hatcurSMEiteffxxxxxB
\else
??????\fi
\fi
}
\newcommand{\hatcurSMEivmac}[1]{\ifnum#1=47 %
\hatcurSMEivmacxxxxxA
\else
\ifnum#1=48 %
\hatcurSMEivmacxxxxxB
\else
??????\fi
\fi
}
\newcommand{\hatcurSMEivmic}[1]{\ifnum#1=47 %
\hatcurSMEivmicxxxxxA
\else
\ifnum#1=48 %
\hatcurSMEivmicxxxxxB
\else
??????\fi
\fi
}
\newcommand{\hatcurSMEivsin}[1]{\ifnum#1=47 %
\hatcurSMEivsinxxxxxA
\else
\ifnum#1=48 %
\hatcurSMEivsinxxxxxB
\else
??????\fi
\fi
}
\newcommand{\hatcurSMEizfeh}[1]{\ifnum#1=47 %
\hatcurSMEizfehxxxxxA
\else
\ifnum#1=48 %
\hatcurSMEizfehxxxxxB
\else
??????\fi
\fi
}
\newcommand{\hatcurSMEizfehshort}[1]{\ifnum#1=47 %
\hatcurSMEizfehshortxxxxxA
\else
\ifnum#1=48 %
\hatcurSMEizfehshortxxxxxB
\else
??????\fi
\fi
}
\newcommand{\hatcurTRESgamma}[1]{\ifnum#1=47 %
\hatcurTRESgammaxxxxxA
\else
\ifnum#1=48 %
\hatcurTRESgammaxxxxxB
\else
??????\fi
\fi
}
\newcommand{\hatcurTRESlogg}[1]{\ifnum#1=47 %
\hatcurTRESloggxxxxxA
\else
\ifnum#1=48 %
\hatcurTRESloggxxxxxB
\else
??????\fi
\fi
}
\newcommand{\hatcurTRESnumspec}[1]{\ifnum#1=47 %
\hatcurTRESnumspecxxxxxA
\else
\ifnum#1=48 %
\hatcurTRESnumspecxxxxxB
\else
??????\fi
\fi
}
\newcommand{\hatcurTRESrvrms}[1]{\ifnum#1=47 %
\hatcurTRESrvrmsxxxxxA
\else
\ifnum#1=48 %
\hatcurTRESrvrmsxxxxxB
\else
??????\fi
\fi
}
\newcommand{\hatcurTRESspan}[1]{\ifnum#1=47 %
\hatcurTRESspanxxxxxA
\else
\ifnum#1=48 %
\hatcurTRESspanxxxxxB
\else
??????\fi
\fi
}
\newcommand{\hatcurTRESteff}[1]{\ifnum#1=47 %
\hatcurTRESteffxxxxxA
\else
\ifnum#1=48 %
\hatcurTRESteffxxxxxB
\else
??????\fi
\fi
}
\newcommand{\hatcurTRESvsini}[1]{\ifnum#1=47 %
\hatcurTRESvsinixxxxxA
\else
\ifnum#1=48 %
\hatcurTRESvsinixxxxxB
\else
??????\fi
\fi
}
\newcommand{\hatcurTRESzfeh}[1]{\ifnum#1=47 %
\hatcurTRESzfehxxxxxA
\else
\ifnum#1=48 %
\hatcurTRESzfehxxxxxB
\else
??????\fi
\fi
}
\newcommand{\hatcurXAv}[1]{\ifnum#1=47 %
\hatcurXAvxxxxxA
\else
\ifnum#1=48 %
\hatcurXAvxxxxxB
\else
??????\fi
\fi
}
\newcommand{\hatcurXdist}[1]{\ifnum#1=47 %
\hatcurXdistxxxxxA
\else
\ifnum#1=48 %
\hatcurXdistxxxxxB
\else
??????\fi
\fi
}
\newcommand{\hatcurXdistred}[1]{\ifnum#1=47 %
\hatcurXdistredxxxxxA
\else
\ifnum#1=48 %
\hatcurXdistredxxxxxB
\else
??????\fi
\fi
}
\newcommand{\hatcurXEBV}[1]{\ifnum#1=47 %
\hatcurXEBVxxxxxA
\else
\ifnum#1=48 %
\hatcurXEBVxxxxxB
\else
??????\fi
\fi
}
\newcommand{\hatcurXjhisored}[1]{\ifnum#1=47 %
\hatcurXjhisoredxxxxxA
\else
\ifnum#1=48 %
\hatcurXjhisoredxxxxxB
\else
??????\fi
\fi
}
\newcommand{\hatcurXjkisored}[1]{\ifnum#1=47 %
\hatcurXjkisoredxxxxxA
\else
\ifnum#1=48 %
\hatcurXjkisoredxxxxxB
\else
??????\fi
\fi
}
\newcommand{\hatcurXmhisored}[1]{\ifnum#1=47 %
\hatcurXmhisoredxxxxxA
\else
\ifnum#1=48 %
\hatcurXmhisoredxxxxxB
\else
??????\fi
\fi
}
\newcommand{\hatcurXmiisored}[1]{\ifnum#1=47 %
\hatcurXmiisoredxxxxxA
\else
\ifnum#1=48 %
\hatcurXmiisoredxxxxxB
\else
??????\fi
\fi
}
\newcommand{\hatcurXmjisored}[1]{\ifnum#1=47 %
\hatcurXmjisoredxxxxxA
\else
\ifnum#1=48 %
\hatcurXmjisoredxxxxxB
\else
??????\fi
\fi
}
\newcommand{\hatcurXmkisored}[1]{\ifnum#1=47 %
\hatcurXmkisoredxxxxxA
\else
\ifnum#1=48 %
\hatcurXmkisoredxxxxxB
\else
??????\fi
\fi
}
\newcommand{\hatcurXmvisored}[1]{\ifnum#1=47 %
\hatcurXmvisoredxxxxxA
\else
\ifnum#1=48 %
\hatcurXmvisoredxxxxxB
\else
??????\fi
\fi
}
\newcommand{\hatcurXsecdur}[1]{\ifnum#1=47 %
\hatcurXsecdurxxxxxA
\else
\ifnum#1=48 %
\hatcurXsecdurxxxxxB
\else
??????\fi
\fi
}
\newcommand{\hatcurXsecingdur}[1]{\ifnum#1=47 %
\hatcurXsecingdurxxxxxA
\else
\ifnum#1=48 %
\hatcurXsecingdurxxxxxB
\else
??????\fi
\fi
}
\newcommand{\hatcurXsecondary}[1]{\ifnum#1=47 %
\hatcurXsecondaryxxxxxA
\else
\ifnum#1=48 %
\hatcurXsecondaryxxxxxB
\else
??????\fi
\fi
}
\newcommand{\hatcurXsecphase}[1]{\ifnum#1=47 %
\hatcurXsecphasexxxxxA
\else
\ifnum#1=48 %
\hatcurXsecphasexxxxxB
\else
??????\fi
\fi
}
\newcommand{\hatcurXviisored}[1]{\ifnum#1=47 %
\hatcurXviisoredxxxxxA
\else
\ifnum#1=48 %
\hatcurXviisoredxxxxxB
\else
??????\fi
\fi
}
\newcommand{\hatcurXvkisored}[1]{\ifnum#1=47 %
\hatcurXvkisoredxxxxxA
\else
\ifnum#1=48 %
\hatcurXvkisoredxxxxxB
\else
??????\fi
\fi
}

\newcommand{\hatcurhtreccenxxxxxA}{HTR212-002}                         
\newcommand{\hatcurfieldeccenxxxxxA}{212}                              
\newcommand{\hatcurCCraeccenxxxxxA}{\ensuremath{02^{\mathrm h}33^{\mathrm m}13.97{\mathrm s}}}                       
\newcommand{\hatcurCCdececcenxxxxxA}{\ensuremath{+30{\arcdeg}21{\arcmin}37.8{\arcsec}}}                      
\newcommand{\hatcurCCmageccenxxxxxA}{10.694}                           
\newcommand{\hatcurCCtwomasseccenxxxxxA}{2MASS~02331396+3021377}       
\newcommand{\hatcurCCgsceccenxxxxxA}{GSC~2324-00031}                   
\newcommand{\hatcurCCtassmveccenxxxxxA}{\ensuremath{10.694\pm0.063}}                        
\newcommand{\hatcurCCtassmIeccenxxxxxA}{\ensuremath{10.084\pm0.069}}                        
\newcommand{\hatcurCCtwomassJmageccenxxxxxA}{\ensuremath{9.713\pm0.021}} 
\newcommand{\hatcurCCtwomassHmageccenxxxxxA}{\ensuremath{9.454\pm0.022}} 
\newcommand{\hatcurCCtwomassKmageccenxxxxxA}{\ensuremath{9.404\pm0.017}} 
\newcommand{\hatcurCCcitJmageccenxxxxxA}{\ensuremath{9.733\pm0.021}}   
\newcommand{\hatcurCCcitHmageccenxxxxxA}{\ensuremath{9.449\pm0.022}}   
\newcommand{\hatcurCCcitKmageccenxxxxxA}{\ensuremath{9.428\pm0.017}}   
\newcommand{\hatcurCCbbJmageccenxxxxxA}{\ensuremath{9.777\pm0.023}}    
\newcommand{\hatcurCCbbHmageccenxxxxxA}{\ensuremath{9.470\pm0.023}}    
\newcommand{\hatcurCCbbKmageccenxxxxxA}{\ensuremath{9.448\pm0.017}}    
\newcommand{\hatcurCCesoJmageccenxxxxxA}{\ensuremath{9.779\pm0.024}}   
\newcommand{\hatcurCCesoHmageccenxxxxxA}{\ensuremath{9.464\pm0.026}}   
\newcommand{\hatcurCCesoKmageccenxxxxxA}{\ensuremath{9.447\pm0.018}}   
\newcommand{\hatcurCCesoJHmageccenxxxxxA}{\ensuremath{0.315\pm0.033}}  
\newcommand{\hatcurCCesoJKmageccenxxxxxA}{\ensuremath{0.331\pm0.033}}  
\newcommand{\hatcurCCesoHKmageccenxxxxxA}{\ensuremath{0.017\pm0.031}}  
\newcommand{\hatcurLCdipeccenxxxxxA}{\ensuremath{3.8}}                 
\newcommand{\hatcurLCrprstareccenxxxxxA}{\ensuremath{0.0890\pm0.0016}} 
\newcommand{\hatcurLCbsqeccenxxxxxA}{\ensuremath{0.630_{-0.019}^{+0.019}}} 
\newcommand{\hatcurLCimpeccenxxxxxA}{\ensuremath{0.794_{-0.012}^{+0.012}}} 
\newcommand{\hatcurLCzetaeccenxxxxxA}{\ensuremath{19.05\pm0.19}}       
\newcommand{\hatcurLCdureccenxxxxxA}{\ensuremath{0.1287\pm0.0018}}     
\newcommand{\hatcurLCdurshorteccenxxxxxA}{\ensuremath{0.1287}}         
\newcommand{\hatcurLCdurhreccenxxxxxA}{\ensuremath{3.089\pm0.043}}     
\newcommand{\hatcurLCdurhrshorteccenxxxxxA}{\ensuremath{3.089}}        
\newcommand{\hatcurLCqeccenxxxxxA}{\ensuremath{0.0272\pm0.0004}}       
\newcommand{\hatcurLCqshorteccenxxxxxA}{\ensuremath{0.027}}            
\newcommand{\hatcurLCingdureccenxxxxxA}{\ensuremath{0.0258\pm0.0017}}  
\newcommand{\hatcurLCPeccenxxxxxA}{\ensuremath{4.732179\pm0.000012}}   
\newcommand{\hatcurLCPprececcenxxxxxA}{\ensuremath{4.7321794}}         
\newcommand{\hatcurLCPshorteccenxxxxxA}{\ensuremath{4.7322}}           
\newcommand{\hatcurLCTeccenxxxxxA}{\ensuremath{2455690.02948\pm0.00085}} 
\newcommand{\hatcurLCTAeccenxxxxxA}{\ensuremath{2455415.56305\pm0.00110}} 
\newcommand{\hatcurLCTBeccenxxxxxA}{\ensuremath{2455940.83494\pm0.00106}} 
\newcommand{\hatcurLChatnetmeccenxxxxxA}{\ensuremath{10.7358\pm0.0001}} 
\newcommand{\hatcurLCiblendeccenxxxxxA}{\ensuremath{0.70\pm0.05}}      
\newcommand{\hatcurSMEiteffeccenxxxxxA}{\ensuremath{6625\pm50}}        
\newcommand{\hatcurSMEizfeheccenxxxxxA}{\ensuremath{-0.08\pm0.08}}     
\newcommand{\hatcurSMEizfehshorteccenxxxxxA}{\ensuremath{-0.08}}       
\newcommand{\hatcurSMEiloggeccenxxxxxA}{\ensuremath{4.10\pm0.1}}       
\newcommand{\hatcurSMEivsineccenxxxxxA}{\ensuremath{14.1\pm0.5}}       
\newcommand{\hatcurSMEivmaceccenxxxxxA}{\ensuremath{NULL}}             
\newcommand{\hatcurSMEivmiceccenxxxxxA}{\ensuremath{NULL}}             
\newcommand{\hatcurSMEiiteffeccenxxxxxA}{\ensuremath{6703\pm50}}       
\newcommand{\hatcurSMEiizfeheccenxxxxxA}{\ensuremath{0.0\pm0.08}}      
\newcommand{\hatcurSMEiizfehshorteccenxxxxxA}{\ensuremath{0.0}}        
\newcommand{\hatcurSMEiiloggeccenxxxxxA}{\ensuremath{4.21\pm0.10}}     
\newcommand{\hatcurSMEiivsineccenxxxxxA}{\ensuremath{14.1\pm0.5}}      
\newcommand{\hatcurSMEiivmaceccenxxxxxA}{\ensuremath{NULL}}            
\newcommand{\hatcurSMEiivmiceccenxxxxxA}{\ensuremath{NULL}}            
\newcommand{\hatcurDSteffeccenxxxxxA}{\ensuremath{NULL\pmNULL}}        
\newcommand{\hatcurDSzfeheccenxxxxxA}{\ensuremath{NULL\pmNULL}}        
\newcommand{\hatcurDSloggeccenxxxxxA}{\ensuremath{NULL\pmNULL}}        
\newcommand{\hatcurDSvsinieccenxxxxxA}{\ensuremath{NULL\pmNULL}}       
\newcommand{\hatcurDSgammaeccenxxxxxA}{\ensuremath{NULL\pmNULL}}       
\newcommand{\hatcurDSnumspececcenxxxxxA}{\ensuremath{0}}               
\newcommand{\hatcurDSspaneccenxxxxxA}{\ensuremath{0}}                  
\newcommand{\hatcurDSrvrmseccenxxxxxA}{\ensuremath{0.00}}              
\newcommand{\hatcurTRESteffeccenxxxxxA}{\ensuremath{5300\pm100}}       
\newcommand{\hatcurTRESzfeheccenxxxxxA}{\ensuremath{NULL\pm0.1}}       
\newcommand{\hatcurTRESloggeccenxxxxxA}{\ensuremath{4.5\pm0.2}}        
\newcommand{\hatcurTRESvsinieccenxxxxxA}{\ensuremath{2.7\pm0.5}}       
\newcommand{\hatcurTRESgammaeccenxxxxxA}{\ensuremath{-19.0\pm0.1}}     
\newcommand{\hatcurTRESnumspececcenxxxxxA}{\ensuremath{3}}             
\newcommand{\hatcurTRESspaneccenxxxxxA}{\ensuremath{32}}               
\newcommand{\hatcurTRESrvrmseccenxxxxxA}{\ensuremath{0.19}}            
\newcommand{\hatcurFIESteffeccenxxxxxA}{\ensuremath{NULL\pmNULL}}      
\newcommand{\hatcurFIESzfeheccenxxxxxA}{\ensuremath{NULL\pmNULL}}      
\newcommand{\hatcurFIESloggeccenxxxxxA}{\ensuremath{NULL\pmNULL}}      
\newcommand{\hatcurFIESvsinieccenxxxxxA}{\ensuremath{NULL\pmNULL}}     
\newcommand{\hatcurFIESgammaeccenxxxxxA}{\ensuremath{NULL\pmNULL}}     
\newcommand{\hatcurFIESnumspececcenxxxxxA}{\ensuremath{0}}             
\newcommand{\hatcurFIESspaneccenxxxxxA}{\ensuremath{0}}                
\newcommand{\hatcurFIESrvrmseccenxxxxxA}{\ensuremath{0.00}}            
\newcommand{\hatcurLBizeccenxxxxxA}{\ensuremath{0.1095}}               
\newcommand{\hatcurLBiizeccenxxxxxA}{\ensuremath{0.3680}}              
\newcommand{\hatcurLBiieccenxxxxxA}{\ensuremath{0.1592}}               
\newcommand{\hatcurLBiiieccenxxxxxA}{\ensuremath{0.3774}}              
\newcommand{\hatcurLBiIeccenxxxxxA}{\ensuremath{0.1415}}               
\newcommand{\hatcurLBiiIeccenxxxxxA}{\ensuremath{0.3751}}              
\newcommand{\hatcurLBigeccenxxxxxA}{\ensuremath{0.3734}}               
\newcommand{\hatcurLBiigeccenxxxxxA}{\ensuremath{0.3575}}              
\newcommand{\hatcurLBireccenxxxxxA}{\ensuremath{0.2273}}               
\newcommand{\hatcurLBiireccenxxxxxA}{\ensuremath{0.3906}}              
\newcommand{\hatcurLBiReccenxxxxxA}{\ensuremath{0.2076}}               
\newcommand{\hatcurLBiiReccenxxxxxA}{\ensuremath{0.3888}}              
\newcommand{\hatcurLBikepeccenxxxxxA}{\ensuremath{}}           
\newcommand{\hatcurLBiikepeccenxxxxxA}{\ensuremath{}}          
\newcommand{\hatcurISOmeccenxxxxxA}{\ensuremath{1.40_{-0.05}^{+0.10}}} 
\newcommand{\hatcurISOmshorteccenxxxxxA}{\ensuremath{1.40}}            
\newcommand{\hatcurISOmlongeccenxxxxxA}{\ensuremath{1.402_{-0.049}^{+0.099}}} 
\newcommand{\hatcurISOreccenxxxxxA}{\ensuremath{1.54_{-0.12}^{+0.32}}} 
\newcommand{\hatcurISOrshorteccenxxxxxA}{\ensuremath{1.54}}            
\newcommand{\hatcurISOrlongeccenxxxxxA}{\ensuremath{1.545_{-0.117}^{+0.316}}} 
\newcommand{\hatcurISOrhoeccenxxxxxA}{\ensuremath{0.53\pm0.16}}        
\newcommand{\hatcurISOloggeccenxxxxxA}{\ensuremath{4.20\pm0.10}}       
\newcommand{\hatcurISOlumeccenxxxxxA}{\ensuremath{4.32_{-0.65}^{+2.11}}} 
\newcommand{\hatcurISOlumshorteccenxxxxxA}{\ensuremath{4.32}}          
\newcommand{\hatcurISOmveccenxxxxxA}{\ensuremath{3.13\pm0.30}}         
\newcommand{\hatcurISOvieccenxxxxxA}{\ensuremath{0.451\pm0.012}}       
\newcommand{\hatcurISOageeccenxxxxxA}{\ensuremath{1.6_{-0.6}^{+0.3}}}  
\newcommand{\hatcurISOsigmaeccenxxxxxA}{\ensuremath{0.00010\pm0.00010}} 
\newcommand{\hatcurISOMJeccenxxxxxA}{\ensuremath{2.43\pm0.30}}         
\newcommand{\hatcurISOMHeccenxxxxxA}{\ensuremath{2.25\pm0.29}}         
\newcommand{\hatcurISOMKeccenxxxxxA}{\ensuremath{2.21\pm0.30}}         
\newcommand{\hatcurISOJKeccenxxxxxA}{\ensuremath{0.22\pm0.01}}         
\newcommand{\hatcurISOspececcenxxxxxA}{F}                              
\newcommand{\hatcurRVKeccenxxxxxA}{\ensuremath{20.0\pm4.5}}            
\newcommand{\hatcurRVrkeccenxxxxxA}{\ensuremath{-0.116\pm0.210}}       
\newcommand{\hatcurRVrheccenxxxxxA}{\ensuremath{0.058\pm0.251}}        
\newcommand{\hatcurRVkeccenxxxxxA}{\ensuremath{-0.030_{-0.113}^{+0.067}}} 
\newcommand{\hatcurRVheccenxxxxxA}{\ensuremath{0.010_{-0.071}^{+0.149}}} 
\newcommand{\hatcurRVtroneeccenxxxxxA}{\ensuremath{0.0000\pm0.0000}}   
\newcommand{\hatcurRVtrtwoeccenxxxxxA}{\ensuremath{0.0000\pm0.0000}}   
\newcommand{\hatcurRVgammaAeccenxxxxxA}{\ensuremath{-1.9\pm5.1}}       
\newcommand{\hatcurRVjitterAeccenxxxxxA}{\ensuremath{22.0}}            
\newcommand{\hatcurRVfitrmsAeccenxxxxxA}{\ensuremath{24.1}}            
\newcommand{\hatcurRVgammaBeccenxxxxxA}{\ensuremath{1.9\pm4.7}}        
\newcommand{\hatcurRVjitterBeccenxxxxxA}{\ensuremath{0.0}}             
\newcommand{\hatcurRVfitrmsBeccenxxxxxA}{\ensuremath{14.7}}            
\newcommand{\hatcurRVecceneccenxxxxxA}{\ensuremath{0.101\pm0.101}}     
\newcommand{\hatcurRVomegaeccenxxxxxA}{\ensuremath{161\pm85}}          
\newcommand{\hatcurPPieccenxxxxxA}{\ensuremath{84.6_{-2.5}^{+0.7}}}    
\newcommand{\hatcurPPgeccenxxxxxA}{\ensuremath{2.7\pm0.8}}             
\newcommand{\hatcurPPloggeccenxxxxxA}{\ensuremath{2.44_{-0.17}^{+0.11}}} 
\newcommand{\hatcurPPareccenxxxxxA}{\ensuremath{8.57_{-1.15}^{+0.68}}} 
\newcommand{\hatcurPPareleccenxxxxxA}{\ensuremath{0.0617_{-0.0007}^{+0.0014}}} 
\newcommand{\hatcurPPrhoeccenxxxxxA}{\ensuremath{0.10\pm0.04}}         
\newcommand{\hatcurPPmeccenxxxxxA}{\ensuremath{0.21\pm0.05}}           
\newcommand{\hatcurPPmshorteccenxxxxxA}{\ensuremath{0.21}}             
\newcommand{\hatcurPPmlongeccenxxxxxA}{\ensuremath{0.206\pm0.048}}     
\newcommand{\hatcurPPmeeccenxxxxxA}{\ensuremath{65.6\pm15.3}}          
\newcommand{\hatcurPPmeshorteccenxxxxxA}{\ensuremath{65.6}}            
\newcommand{\hatcurPPmelongeccenxxxxxA}{\ensuremath{65.63\pm15.26}}    
\newcommand{\hatcurPPreccenxxxxxA}{\ensuremath{1.34_{-0.11}^{+0.27}}}  
\newcommand{\hatcurPPrshorteccenxxxxxA}{\ensuremath{1.34}}             
\newcommand{\hatcurPPrlongeccenxxxxxA}{\ensuremath{1.342_{-0.109}^{+0.274}}} 
\newcommand{\hatcurPPreeccenxxxxxA}{\ensuremath{15.0_{-1.2}^{+3.1}}}   
\newcommand{\hatcurPPreshorteccenxxxxxA}{\ensuremath{15.0}}            
\newcommand{\hatcurPPrelongeccenxxxxxA}{\ensuremath{15.04_{-1.22}^{+3.07}}} 
\newcommand{\hatcurPPmrcorreccenxxxxxA}{\ensuremath{0.22}}             
\newcommand{\hatcurPPteffeccenxxxxxA}{\ensuremath{1621_{-60}^{+145}}}  
\newcommand{\hatcurPPthetaeccenxxxxxA}{\ensuremath{0.013\pm0.003}}     
\newcommand{\hatcurPPfluxperieccenxxxxxA}{\ensuremath{18.1_{-2.4}^{+23.4}}} 
\newcommand{\hatcurPPfluxperidimeccenxxxxxA}{\ensuremath{8}}           
\newcommand{\hatcurPPfluxapeccenxxxxxA}{\ensuremath{13.7_{-2.7}^{+1.7}}} 
\newcommand{\hatcurPPfluxapdimeccenxxxxxA}{\ensuremath{8}}             
\newcommand{\hatcurPPfluxavgeccenxxxxxA}{\ensuremath{15.6_{-2.1}^{+7.1}}} 
\newcommand{\hatcurPPfluxavgdimeccenxxxxxA}{\ensuremath{8}}            
\newcommand{\hatcurXsecphaseeccenxxxxxA}{\ensuremath{0.4806\pm0.0593}} 
\newcommand{\hatcurXsecondaryeccenxxxxxA}{\ensuremath{2455692.30\pm0.28}} 
\newcommand{\hatcurXsecdureccenxxxxxA}{\ensuremath{0.128\pm0.021}}   
\newcommand{\hatcurXsecingdureccenxxxxxA}{\ensuremath{0.027\pm0.028}} 
\newcommand{\hatcurPPphiconjeccenxxxxxA}{\ensuremath{-0.07\pm0.24}} 
\newcommand{\hatcurPPperieccenxxxxxA}{\ensuremath{2455690.4\pm1.2}}  
\newcommand{\hatcurPPaequiveccenxxxxxA}{\ensuremath{0.0296_{-0.0040}^{+0.0025}}} 
\newcommand{\hatcurPPtcirceccenxxxxxA}{\ensuremath{136.4\pm79.0}}      
\newcommand{\hatcurPPtinfalleccenxxxxxA}{\ensuremath{14014.6_{-6380.5}^{+10345.4}}} 
\newcommand{\hatcurXdisteccenxxxxxA}{\ensuremath{280_{-21}^{+57}}}     
\newcommand{\hatcurXAveccenxxxxxA}{\ensuremath{0.414\pm0.079}}         
\newcommand{\hatcurXdistredeccenxxxxxA}{\ensuremath{273_{-20}^{+55}}}  
\newcommand{\hatcurXEBVeccenxxxxxA}{\ensuremath{0.134\pm0.025}}        
\newcommand{\hatcurXmvisoredeccenxxxxxA}{\ensuremath{10.735\pm0.055}}  
\newcommand{\hatcurXmiisoredeccenxxxxxA}{\ensuremath{10.067\pm0.022}}  
\newcommand{\hatcurXmjisoredeccenxxxxxA}{\ensuremath{9.735\pm0.013}}   
\newcommand{\hatcurXmhisoredeccenxxxxxA}{\ensuremath{9.513\pm0.013}}   
\newcommand{\hatcurXmkisoredeccenxxxxxA}{\ensuremath{9.447\pm0.014}}   
\newcommand{\hatcurXviisoredeccenxxxxxA}{\ensuremath{0.667\pm0.037}}   
\newcommand{\hatcurXvkisoredeccenxxxxxA}{\ensuremath{1.287\pm0.061}}   
\newcommand{\hatcurXjhisoredeccenxxxxxA}{\ensuremath{0.221\pm0.008}}   
\newcommand{\hatcurXjkisoredeccenxxxxxA}{\ensuremath{0.288\pm0.012}}   
\newcommand{\hatcurCCpmraeccenxxxxxA}{\ensuremath{-13.9\pm0.6}}        
\newcommand{\hatcurCCpmdececcenxxxxxA}{\ensuremath{-9.8\pm0.7}}        
\newcommand{\hatcurCCpmeccenxxxxxA}{\ensuremath{17.0074\pm0.921954}}   

\newcommand{\hatcurhtreccenxxxxxB}{HTR213-008}                         
\newcommand{\hatcurfieldeccenxxxxxB}{212}                              
\newcommand{\hatcurCCraeccenxxxxxB}{\ensuremath{02^{\mathrm h}57^{\mathrm m}53.03{\mathrm s}}}                       
\newcommand{\hatcurCCdececcenxxxxxB}{\ensuremath{+30{\arcdeg}37{\arcmin}32.5{\arcsec}}}                      
\newcommand{\hatcurCCmageccenxxxxxB}{12.16}                            
\newcommand{\hatcurCCtwomasseccenxxxxxB}{2MASS~02575301+3037324}       
\newcommand{\hatcurCCgsceccenxxxxxB}{GSC~2326-00214}                   
\newcommand{\hatcurCCtassmveccenxxxxxB}{\ensuremath{12.16\pm0.11}}                         
\newcommand{\hatcurCCtassmIeccenxxxxxB}{\ensuremath{11.253\pm0.071}}                         
\newcommand{\hatcurCCtwomassJmageccenxxxxxB}{\ensuremath{10.696\pm0.023}} 
\newcommand{\hatcurCCtwomassHmageccenxxxxxB}{\ensuremath{10.340\pm0.026}} 
\newcommand{\hatcurCCtwomassKmageccenxxxxxB}{\ensuremath{10.255\pm0.021}} 
\newcommand{\hatcurCCcitJmageccenxxxxxB}{\ensuremath{10.709\pm0.023}}  
\newcommand{\hatcurCCcitHmageccenxxxxxB}{\ensuremath{10.334\pm0.026}}  
\newcommand{\hatcurCCcitKmageccenxxxxxB}{\ensuremath{10.279\pm0.021}}  
\newcommand{\hatcurCCbbJmageccenxxxxxB}{\ensuremath{10.764\pm0.025}}   
\newcommand{\hatcurCCbbHmageccenxxxxxB}{\ensuremath{10.356\pm0.027}}   
\newcommand{\hatcurCCbbKmageccenxxxxxB}{\ensuremath{10.299\pm0.021}}   
\newcommand{\hatcurCCesoJmageccenxxxxxB}{\ensuremath{10.767\pm0.027}}  
\newcommand{\hatcurCCesoHmageccenxxxxxB}{\ensuremath{10.351\pm0.031}}  
\newcommand{\hatcurCCesoKmageccenxxxxxB}{\ensuremath{10.298\pm0.022}}  
\newcommand{\hatcurCCesoJHmageccenxxxxxB}{\ensuremath{0.415\pm0.038}}  
\newcommand{\hatcurCCesoJKmageccenxxxxxB}{\ensuremath{0.470\pm0.034}}  
\newcommand{\hatcurCCesoHKmageccenxxxxxB}{\ensuremath{0.054\pm0.038}}  
\newcommand{\hatcurLCdipeccenxxxxxB}{\ensuremath{6.4}}                 
\newcommand{\hatcurLCrprstareccenxxxxxB}{\ensuremath{0.0951\pm0.0016}} 
\newcommand{\hatcurLCbsqeccenxxxxxB}{\ensuremath{0.275_{-0.055}^{+0.042}}} 
\newcommand{\hatcurLCimpeccenxxxxxB}{\ensuremath{0.524_{-0.058}^{+0.038}}} 
\newcommand{\hatcurLCzetaeccenxxxxxB}{\ensuremath{16.07\pm0.11}}       
\newcommand{\hatcurLCdureccenxxxxxB}{\ensuremath{0.1404\pm0.0014}}     
\newcommand{\hatcurLCdurshorteccenxxxxxB}{\ensuremath{0.1404}}         
\newcommand{\hatcurLCdurhreccenxxxxxB}{\ensuremath{3.369\pm0.033}}     
\newcommand{\hatcurLCdurhrshorteccenxxxxxB}{\ensuremath{3.369}}        
\newcommand{\hatcurLCqeccenxxxxxB}{\ensuremath{0.0318\pm0.0003}}       
\newcommand{\hatcurLCqshorteccenxxxxxB}{\ensuremath{0.032}}            
\newcommand{\hatcurLCingdureccenxxxxxB}{\ensuremath{0.0163\pm0.0012}}  
\newcommand{\hatcurLCPeccenxxxxxB}{\ensuremath{4.408650\pm0.000009}}   
\newcommand{\hatcurLCPprececcenxxxxxB}{\ensuremath{4.4086497}}         
\newcommand{\hatcurLCPshorteccenxxxxxB}{\ensuremath{4.4086}}           
\newcommand{\hatcurLCTeccenxxxxxB}{\ensuremath{2455817.16702\pm0.00038}} 
\newcommand{\hatcurLCTAeccenxxxxxB}{\ensuremath{2455415.97989\pm0.00084}} 
\newcommand{\hatcurLCTBeccenxxxxxB}{\ensuremath{2455953.83515\pm0.00048}} 
\newcommand{\hatcurLChatnetmeccenxxxxxB}{\ensuremath{11.9267\pm0.0001}} 
\newcommand{\hatcurLCiblendeccenxxxxxB}{\ensuremath{0.70\pm0.05}}      
\newcommand{\hatcurSMEiteffeccenxxxxxB}{\ensuremath{5880\pm50}}        
\newcommand{\hatcurSMEizfeheccenxxxxxB}{\ensuremath{0.01\pm0.08}}      
\newcommand{\hatcurSMEizfehshorteccenxxxxxB}{\ensuremath{0.01}}        
\newcommand{\hatcurSMEiloggeccenxxxxxB}{\ensuremath{4.23\pm0.1}}       
\newcommand{\hatcurSMEivsineccenxxxxxB}{\ensuremath{2.7\pm0.5}}        
\newcommand{\hatcurSMEivmaceccenxxxxxB}{\ensuremath{NULL}}             
\newcommand{\hatcurSMEivmiceccenxxxxxB}{\ensuremath{NULL}}             
\newcommand{\hatcurSMEiiteffeccenxxxxxB}{\ensuremath{5837\pm50}}       
\newcommand{\hatcurSMEiizfeheccenxxxxxB}{\ensuremath{-0.04\pm0.08}}    
\newcommand{\hatcurSMEiizfehshorteccenxxxxxB}{\ensuremath{-0.04}}      
\newcommand{\hatcurSMEiiloggeccenxxxxxB}{\ensuremath{4.11\pm0.10}}     
\newcommand{\hatcurSMEiivsineccenxxxxxB}{\ensuremath{2.7\pm0.5}}       
\newcommand{\hatcurSMEiivmaceccenxxxxxB}{\ensuremath{4.35}}            
\newcommand{\hatcurSMEiivmiceccenxxxxxB}{\ensuremath{0.85}}            
\newcommand{\hatcurDSteffeccenxxxxxB}{\ensuremath{NULL\pmNULL}}        
\newcommand{\hatcurDSzfeheccenxxxxxB}{\ensuremath{NULL\pmNULL}}        
\newcommand{\hatcurDSloggeccenxxxxxB}{\ensuremath{NULL\pmNULL}}        
\newcommand{\hatcurDSvsinieccenxxxxxB}{\ensuremath{NULL\pmNULL}}       
\newcommand{\hatcurDSgammaeccenxxxxxB}{\ensuremath{NULL\pmNULL}}       
\newcommand{\hatcurDSnumspececcenxxxxxB}{\ensuremath{0}}               
\newcommand{\hatcurDSspaneccenxxxxxB}{\ensuremath{0}}                  
\newcommand{\hatcurDSrvrmseccenxxxxxB}{\ensuremath{0.00}}              
\newcommand{\hatcurTRESteffeccenxxxxxB}{\ensuremath{5300\pm100}}       
\newcommand{\hatcurTRESzfeheccenxxxxxB}{\ensuremath{NULL\pm0.1}}       
\newcommand{\hatcurTRESloggeccenxxxxxB}{\ensuremath{4.5\pm0.2}}        
\newcommand{\hatcurTRESvsinieccenxxxxxB}{\ensuremath{2.7\pm0.5}}       
\newcommand{\hatcurTRESgammaeccenxxxxxB}{\ensuremath{-19.0\pm0.1}}     
\newcommand{\hatcurTRESnumspececcenxxxxxB}{\ensuremath{3}}             
\newcommand{\hatcurTRESspaneccenxxxxxB}{\ensuremath{32}}               
\newcommand{\hatcurTRESrvrmseccenxxxxxB}{\ensuremath{0.19}}            
\newcommand{\hatcurFIESteffeccenxxxxxB}{\ensuremath{NULL\pmNULL}}      
\newcommand{\hatcurFIESzfeheccenxxxxxB}{\ensuremath{NULL\pmNULL}}      
\newcommand{\hatcurFIESloggeccenxxxxxB}{\ensuremath{NULL\pmNULL}}      
\newcommand{\hatcurFIESvsinieccenxxxxxB}{\ensuremath{NULL\pmNULL}}     
\newcommand{\hatcurFIESgammaeccenxxxxxB}{\ensuremath{NULL\pmNULL}}     
\newcommand{\hatcurFIESnumspececcenxxxxxB}{\ensuremath{0}}             
\newcommand{\hatcurFIESspaneccenxxxxxB}{\ensuremath{0}}                
\newcommand{\hatcurFIESrvrmseccenxxxxxB}{\ensuremath{0.00}}            
\newcommand{\hatcurLBizeccenxxxxxB}{\ensuremath{0.1990}}               
\newcommand{\hatcurLBiizeccenxxxxxB}{\ensuremath{0.3302}}              
\newcommand{\hatcurLBiieccenxxxxxB}{\ensuremath{0.2568}}               
\newcommand{\hatcurLBiiieccenxxxxxB}{\ensuremath{0.3337}}              
\newcommand{\hatcurLBiIeccenxxxxxB}{\ensuremath{0.2372}}               
\newcommand{\hatcurLBiiIeccenxxxxxB}{\ensuremath{0.3333}}              
\newcommand{\hatcurLBigeccenxxxxxB}{\ensuremath{0.5260}}               
\newcommand{\hatcurLBiigeccenxxxxxB}{\ensuremath{0.2556}}              
\newcommand{\hatcurLBireccenxxxxxB}{\ensuremath{0.3402}}               
\newcommand{\hatcurLBiireccenxxxxxB}{\ensuremath{0.3341}}              
\newcommand{\hatcurLBiReccenxxxxxB}{\ensuremath{0.3171}}               
\newcommand{\hatcurLBiiReccenxxxxxB}{\ensuremath{0.3350}}              
\newcommand{\hatcurLBikepeccenxxxxxB}{\ensuremath{}}           
\newcommand{\hatcurLBiikepeccenxxxxxB}{\ensuremath{}}          
\newcommand{\hatcurISOmeccenxxxxxB}{\ensuremath{1.08_{-0.06}^{+0.09}}} 
\newcommand{\hatcurISOmshorteccenxxxxxB}{\ensuremath{1.08}}            
\newcommand{\hatcurISOmlongeccenxxxxxB}{\ensuremath{1.083_{-0.060}^{+0.087}}} 
\newcommand{\hatcurISOreccenxxxxxB}{\ensuremath{1.52\pm0.27}}          
\newcommand{\hatcurISOrshorteccenxxxxxB}{\ensuremath{1.52}}            
\newcommand{\hatcurISOrlongeccenxxxxxB}{\ensuremath{1.524\pm0.270}}    
\newcommand{\hatcurISOrhoeccenxxxxxB}{\ensuremath{0.43_{-0.14}^{+0.30}}} 
\newcommand{\hatcurISOloggeccenxxxxxB}{\ensuremath{4.10\pm0.13}}       
\newcommand{\hatcurISOlumeccenxxxxxB}{\ensuremath{2.41_{-0.68}^{+1.12}}} 
\newcommand{\hatcurISOlumshorteccenxxxxxB}{\ensuremath{2.41}}          
\newcommand{\hatcurISOmveccenxxxxxB}{\ensuremath{3.87\pm0.38}}         
\newcommand{\hatcurISOvieccenxxxxxB}{\ensuremath{0.677\pm0.015}}       
\newcommand{\hatcurISOageeccenxxxxxB}{\ensuremath{6.9\pm1.3}}          
\newcommand{\hatcurISOsigmaeccenxxxxxB}{\ensuremath{0.00010\pm0.00004}} 
\newcommand{\hatcurISOMJeccenxxxxxB}{\ensuremath{2.75\pm0.38}}         
\newcommand{\hatcurISOMHeccenxxxxxB}{\ensuremath{2.42\pm0.38}}         
\newcommand{\hatcurISOMKeccenxxxxxB}{\ensuremath{2.37\pm0.38}}         
\newcommand{\hatcurISOJKeccenxxxxxB}{\ensuremath{0.39\pm0.01}}         
\newcommand{\hatcurISOspececcenxxxxxB}{G}                              
\newcommand{\hatcurRVKeccenxxxxxB}{\ensuremath{20.9\pm2.8}}            
\newcommand{\hatcurRVrkeccenxxxxxB}{\ensuremath{0.014\pm0.155}}        
\newcommand{\hatcurRVrheccenxxxxxB}{\ensuremath{0.464_{-0.294}^{+0.124}}} 
\newcommand{\hatcurRVkeccenxxxxxB}{\ensuremath{0.004\pm0.076}}         
\newcommand{\hatcurRVheccenxxxxxB}{\ensuremath{0.227\pm0.145}}         
\newcommand{\hatcurRVtroneeccenxxxxxB}{\ensuremath{0.0000\pm0.0000}}   
\newcommand{\hatcurRVtrtwoeccenxxxxxB}{\ensuremath{0.0000\pm0.0000}}   
\newcommand{\hatcurRVgammaAeccenxxxxxB}{\ensuremath{-1.2\pm6.5}}       
\newcommand{\hatcurRVjitterAeccenxxxxxB}{\ensuremath{20.5}}            
\newcommand{\hatcurRVfitrmsAeccenxxxxxB}{\ensuremath{21.8}}            
\newcommand{\hatcurRVgammaBeccenxxxxxB}{\ensuremath{3.9\pm1.8}}        
\newcommand{\hatcurRVjitterBeccenxxxxxB}{\ensuremath{0.0}}             
\newcommand{\hatcurRVfitrmsBeccenxxxxxB}{\ensuremath{9.0}}             
\newcommand{\hatcurRVgammaCeccenxxxxxB}{\ensuremath{12.2\pm8.0}}       
\newcommand{\hatcurRVjitterCeccenxxxxxB}{\ensuremath{17.5}}            
\newcommand{\hatcurRVfitrmsCeccenxxxxxB}{\ensuremath{21.0}}            
\newcommand{\hatcurRVecceneccenxxxxxB}{\ensuremath{0.242\pm0.136}}     
\newcommand{\hatcurRVomegaeccenxxxxxB}{\ensuremath{90\pm60}}           
\newcommand{\hatcurPPieccenxxxxxB}{\ensuremath{84.9_{-2.5}^{+1.5}}}    
\newcommand{\hatcurPPgeccenxxxxxB}{\ensuremath{2.1_{-0.6}^{+1.0}}}     
\newcommand{\hatcurPPloggeccenxxxxxB}{\ensuremath{2.33\pm0.15}}        
\newcommand{\hatcurPPareccenxxxxxB}{\ensuremath{7.61_{-1.01}^{+1.37}}} 
\newcommand{\hatcurPPareleccenxxxxxB}{\ensuremath{0.0540_{-0.0010}^{+0.0014}}} 
\newcommand{\hatcurPPrhoeccenxxxxxB}{\ensuremath{0.08_{-0.03}^{+0.06}}} 
\newcommand{\hatcurPPmeccenxxxxxB}{\ensuremath{0.17\pm0.02}}           
\newcommand{\hatcurPPmshorteccenxxxxxB}{\ensuremath{0.17}}             
\newcommand{\hatcurPPmlongeccenxxxxxB}{\ensuremath{0.173\pm0.023}}     
\newcommand{\hatcurPPmeeccenxxxxxB}{\ensuremath{54.9\pm7.4}}           
\newcommand{\hatcurPPmeshorteccenxxxxxB}{\ensuremath{54.9}}            
\newcommand{\hatcurPPmelongeccenxxxxxB}{\ensuremath{54.95\pm7.38}}     
\newcommand{\hatcurPPreccenxxxxxB}{\ensuremath{1.41\pm0.25}}           
\newcommand{\hatcurPPrshorteccenxxxxxB}{\ensuremath{1.41}}             
\newcommand{\hatcurPPrlongeccenxxxxxB}{\ensuremath{1.407\pm0.253}}     
\newcommand{\hatcurPPreeccenxxxxxB}{\ensuremath{15.8\pm2.8}}           
\newcommand{\hatcurPPreshorteccenxxxxxB}{\ensuremath{15.8}}            
\newcommand{\hatcurPPrelongeccenxxxxxB}{\ensuremath{15.77\pm2.84}}     
\newcommand{\hatcurPPmrcorreccenxxxxxB}{\ensuremath{0.29}}             
\newcommand{\hatcurPPteffeccenxxxxxB}{\ensuremath{1508\pm131}}         
\newcommand{\hatcurPPthetaeccenxxxxxB}{\ensuremath{0.012_{-0.002}^{+0.003}}} 
\newcommand{\hatcurPPfluxperieccenxxxxxB}{\ensuremath{19.7_{-8.8}^{+25.7}}} 
\newcommand{\hatcurPPfluxperidimeccenxxxxxB}{\ensuremath{8}}           
\newcommand{\hatcurPPfluxapeccenxxxxxB}{\ensuremath{7.46\pm0.88}}      
\newcommand{\hatcurPPfluxapdimeccenxxxxxB}{\ensuremath{8}}             
\newcommand{\hatcurPPfluxavgeccenxxxxxB}{\ensuremath{11.7_{-3.2}^{+5.4}}} 
\newcommand{\hatcurPPfluxavgdimeccenxxxxxB}{\ensuremath{8}}            
\newcommand{\hatcurXsecphaseeccenxxxxxB}{\ensuremath{0.5028\pm0.0508}} 
\newcommand{\hatcurXsecondaryeccenxxxxxB}{\ensuremath{2455819.38\pm0.22}} 
\newcommand{\hatcurXsecdureccenxxxxxB}{\ensuremath{0.166\pm0.030}}   
\newcommand{\hatcurXsecingdureccenxxxxxB}{\ensuremath{0.039\pm0.040}} 
\newcommand{\hatcurPPphiconjeccenxxxxxB}{\ensuremath{0.0024_{-0.0848}^{+0.1289}}} 
\newcommand{\hatcurPPperieccenxxxxxB}{\ensuremath{2455817.16\pm0.48}}  
\newcommand{\hatcurPPaequiveccenxxxxxB}{\ensuremath{0.0347_{-0.0046}^{+0.0063}}} 
\newcommand{\hatcurPPtcirceccenxxxxxB}{\ensuremath{41.0_{-26.0}^{+111.6}}} 
\newcommand{\hatcurPPtinfalleccenxxxxxB}{\ensuremath{6783.1_{-3114.8}^{+10736.9}}} 
\newcommand{\hatcurXdisteccenxxxxxB}{\ensuremath{385\pm68}}            
\newcommand{\hatcurXAveccenxxxxxB}{\ensuremath{0.419\pm0.111}}         
\newcommand{\hatcurXdistredeccenxxxxxB}{\ensuremath{376\pm67}}         
\newcommand{\hatcurXEBVeccenxxxxxB}{\ensuremath{0.135\pm0.036}}        
\newcommand{\hatcurXmvisoredeccenxxxxxB}{\ensuremath{12.166\pm0.084}}  
\newcommand{\hatcurXmiisoredeccenxxxxxB}{\ensuremath{11.270\pm0.032}}  
\newcommand{\hatcurXmjisoredeccenxxxxxB}{\ensuremath{10.750\pm0.017}}  
\newcommand{\hatcurXmhisoredeccenxxxxxB}{\ensuremath{10.380\pm0.015}}  
\newcommand{\hatcurXmkisoredeccenxxxxxB}{\ensuremath{10.293\pm0.018}}  
\newcommand{\hatcurXviisoredeccenxxxxxB}{\ensuremath{0.895\pm0.055}}   
\newcommand{\hatcurXvkisoredeccenxxxxxB}{\ensuremath{1.872\pm0.092}}   
\newcommand{\hatcurXjhisoredeccenxxxxxB}{\ensuremath{0.370\pm0.011}}   
\newcommand{\hatcurXjkisoredeccenxxxxxB}{\ensuremath{0.457\pm0.018}}   
\newcommand{\hatcurCCpmraeccenxxxxxB}{\ensuremath{18.1\pm3.0}}         
\newcommand{\hatcurCCpmdececcenxxxxxB}{\ensuremath{-12.5\pm3.0}}       
\newcommand{\hatcurCCpmeccenxxxxxB}{\ensuremath{21.9968\pm4.24264}}    

\newcommand{\hatcurCCbbHmageccen}[1]{\ifnum#1=47 %
\hatcurCCbbHmageccenxxxxxA
\else
\ifnum#1=48 %
\hatcurCCbbHmageccenxxxxxB
\else
??????\fi
\fi
}
\newcommand{\hatcurCCbbJmageccen}[1]{\ifnum#1=47 %
\hatcurCCbbJmageccenxxxxxA
\else
\ifnum#1=48 %
\hatcurCCbbJmageccenxxxxxB
\else
??????\fi
\fi
}
\newcommand{\hatcurCCbbKmageccen}[1]{\ifnum#1=47 %
\hatcurCCbbKmageccenxxxxxA
\else
\ifnum#1=48 %
\hatcurCCbbKmageccenxxxxxB
\else
??????\fi
\fi
}
\newcommand{\hatcurCCcitHmageccen}[1]{\ifnum#1=47 %
\hatcurCCcitHmageccenxxxxxA
\else
\ifnum#1=48 %
\hatcurCCcitHmageccenxxxxxB
\else
??????\fi
\fi
}
\newcommand{\hatcurCCcitJmageccen}[1]{\ifnum#1=47 %
\hatcurCCcitJmageccenxxxxxA
\else
\ifnum#1=48 %
\hatcurCCcitJmageccenxxxxxB
\else
??????\fi
\fi
}
\newcommand{\hatcurCCcitKmageccen}[1]{\ifnum#1=47 %
\hatcurCCcitKmageccenxxxxxA
\else
\ifnum#1=48 %
\hatcurCCcitKmageccenxxxxxB
\else
??????\fi
\fi
}
\newcommand{\hatcurCCdececcen}[1]{\ifnum#1=47 %
\hatcurCCdececcenxxxxxA
\else
\ifnum#1=48 %
\hatcurCCdececcenxxxxxB
\else
??????\fi
\fi
}
\newcommand{\hatcurCCesoHKmageccen}[1]{\ifnum#1=47 %
\hatcurCCesoHKmageccenxxxxxA
\else
\ifnum#1=48 %
\hatcurCCesoHKmageccenxxxxxB
\else
??????\fi
\fi
}
\newcommand{\hatcurCCesoHmageccen}[1]{\ifnum#1=47 %
\hatcurCCesoHmageccenxxxxxA
\else
\ifnum#1=48 %
\hatcurCCesoHmageccenxxxxxB
\else
??????\fi
\fi
}
\newcommand{\hatcurCCesoJHmageccen}[1]{\ifnum#1=47 %
\hatcurCCesoJHmageccenxxxxxA
\else
\ifnum#1=48 %
\hatcurCCesoJHmageccenxxxxxB
\else
??????\fi
\fi
}
\newcommand{\hatcurCCesoJKmageccen}[1]{\ifnum#1=47 %
\hatcurCCesoJKmageccenxxxxxA
\else
\ifnum#1=48 %
\hatcurCCesoJKmageccenxxxxxB
\else
??????\fi
\fi
}
\newcommand{\hatcurCCesoJmageccen}[1]{\ifnum#1=47 %
\hatcurCCesoJmageccenxxxxxA
\else
\ifnum#1=48 %
\hatcurCCesoJmageccenxxxxxB
\else
??????\fi
\fi
}
\newcommand{\hatcurCCesoKmageccen}[1]{\ifnum#1=47 %
\hatcurCCesoKmageccenxxxxxA
\else
\ifnum#1=48 %
\hatcurCCesoKmageccenxxxxxB
\else
??????\fi
\fi
}
\newcommand{\hatcurCCgsceccen}[1]{\ifnum#1=47 %
\hatcurCCgsceccenxxxxxA
\else
\ifnum#1=48 %
\hatcurCCgsceccenxxxxxB
\else
??????\fi
\fi
}
\newcommand{\hatcurCCmageccen}[1]{\ifnum#1=47 %
\hatcurCCmageccenxxxxxA
\else
\ifnum#1=48 %
\hatcurCCmageccenxxxxxB
\else
??????\fi
\fi
}
\newcommand{\hatcurCCpmeccen}[1]{\ifnum#1=47 %
\hatcurCCpmeccenxxxxxA
\else
\ifnum#1=48 %
\hatcurCCpmeccenxxxxxB
\else
??????\fi
\fi
}
\newcommand{\hatcurCCpmdececcen}[1]{\ifnum#1=47 %
\hatcurCCpmdececcenxxxxxA
\else
\ifnum#1=48 %
\hatcurCCpmdececcenxxxxxB
\else
??????\fi
\fi
}
\newcommand{\hatcurCCpmraeccen}[1]{\ifnum#1=47 %
\hatcurCCpmraeccenxxxxxA
\else
\ifnum#1=48 %
\hatcurCCpmraeccenxxxxxB
\else
??????\fi
\fi
}
\newcommand{\hatcurCCraeccen}[1]{\ifnum#1=47 %
\hatcurCCraeccenxxxxxA
\else
\ifnum#1=48 %
\hatcurCCraeccenxxxxxB
\else
??????\fi
\fi
}
\newcommand{\hatcurCCtassmveccen}[1]{\ifnum#1=47 %
\hatcurCCtassmveccenxxxxxA
\else
\ifnum#1=48 %
\hatcurCCtassmveccenxxxxxB
\else
??????\fi
\fi
}
\newcommand{\hatcurCCtassmIeccen}[1]{\ifnum#1=47 %
\hatcurCCtassmIeccenxxxxxA
\else
\ifnum#1=48 %
\hatcurCCtassmIeccenxxxxxB
\else
??????\fi
\fi
}
\newcommand{\hatcurCCtwomasseccen}[1]{\ifnum#1=47 %
\hatcurCCtwomasseccenxxxxxA
\else
\ifnum#1=48 %
\hatcurCCtwomasseccenxxxxxB
\else
??????\fi
\fi
}
\newcommand{\hatcurCCtwomassHmageccen}[1]{\ifnum#1=47 %
\hatcurCCtwomassHmageccenxxxxxA
\else
\ifnum#1=48 %
\hatcurCCtwomassHmageccenxxxxxB
\else
??????\fi
\fi
}
\newcommand{\hatcurCCtwomassJmageccen}[1]{\ifnum#1=47 %
\hatcurCCtwomassJmageccenxxxxxA
\else
\ifnum#1=48 %
\hatcurCCtwomassJmageccenxxxxxB
\else
??????\fi
\fi
}
\newcommand{\hatcurCCtwomassKmageccen}[1]{\ifnum#1=47 %
\hatcurCCtwomassKmageccenxxxxxA
\else
\ifnum#1=48 %
\hatcurCCtwomassKmageccenxxxxxB
\else
??????\fi
\fi
}
\newcommand{\hatcurDSgammaeccen}[1]{\ifnum#1=47 %
\hatcurDSgammaeccenxxxxxA
\else
\ifnum#1=48 %
\hatcurDSgammaeccenxxxxxB
\else
??????\fi
\fi
}
\newcommand{\hatcurDSloggeccen}[1]{\ifnum#1=47 %
\hatcurDSloggeccenxxxxxA
\else
\ifnum#1=48 %
\hatcurDSloggeccenxxxxxB
\else
??????\fi
\fi
}
\newcommand{\hatcurDSnumspececcen}[1]{\ifnum#1=47 %
\hatcurDSnumspececcenxxxxxA
\else
\ifnum#1=48 %
\hatcurDSnumspececcenxxxxxB
\else
??????\fi
\fi
}
\newcommand{\hatcurDSrvrmseccen}[1]{\ifnum#1=47 %
\hatcurDSrvrmseccenxxxxxA
\else
\ifnum#1=48 %
\hatcurDSrvrmseccenxxxxxB
\else
??????\fi
\fi
}
\newcommand{\hatcurDSspaneccen}[1]{\ifnum#1=47 %
\hatcurDSspaneccenxxxxxA
\else
\ifnum#1=48 %
\hatcurDSspaneccenxxxxxB
\else
??????\fi
\fi
}
\newcommand{\hatcurDSteffeccen}[1]{\ifnum#1=47 %
\hatcurDSteffeccenxxxxxA
\else
\ifnum#1=48 %
\hatcurDSteffeccenxxxxxB
\else
??????\fi
\fi
}
\newcommand{\hatcurDSvsinieccen}[1]{\ifnum#1=47 %
\hatcurDSvsinieccenxxxxxA
\else
\ifnum#1=48 %
\hatcurDSvsinieccenxxxxxB
\else
??????\fi
\fi
}
\newcommand{\hatcurDSzfeheccen}[1]{\ifnum#1=47 %
\hatcurDSzfeheccenxxxxxA
\else
\ifnum#1=48 %
\hatcurDSzfeheccenxxxxxB
\else
??????\fi
\fi
}
\newcommand{\hatcurfieldeccen}[1]{\ifnum#1=47 %
\hatcurfieldeccenxxxxxA
\else
\ifnum#1=48 %
\hatcurfieldeccenxxxxxB
\else
??????\fi
\fi
}
\newcommand{\hatcurFIESgammaeccen}[1]{\ifnum#1=47 %
\hatcurFIESgammaeccenxxxxxA
\else
\ifnum#1=48 %
\hatcurFIESgammaeccenxxxxxB
\else
??????\fi
\fi
}
\newcommand{\hatcurFIESloggeccen}[1]{\ifnum#1=47 %
\hatcurFIESloggeccenxxxxxA
\else
\ifnum#1=48 %
\hatcurFIESloggeccenxxxxxB
\else
??????\fi
\fi
}
\newcommand{\hatcurFIESnumspececcen}[1]{\ifnum#1=47 %
\hatcurFIESnumspececcenxxxxxA
\else
\ifnum#1=48 %
\hatcurFIESnumspececcenxxxxxB
\else
??????\fi
\fi
}
\newcommand{\hatcurFIESrvrmseccen}[1]{\ifnum#1=47 %
\hatcurFIESrvrmseccenxxxxxA
\else
\ifnum#1=48 %
\hatcurFIESrvrmseccenxxxxxB
\else
??????\fi
\fi
}
\newcommand{\hatcurFIESspaneccen}[1]{\ifnum#1=47 %
\hatcurFIESspaneccenxxxxxA
\else
\ifnum#1=48 %
\hatcurFIESspaneccenxxxxxB
\else
??????\fi
\fi
}
\newcommand{\hatcurFIESteffeccen}[1]{\ifnum#1=47 %
\hatcurFIESteffeccenxxxxxA
\else
\ifnum#1=48 %
\hatcurFIESteffeccenxxxxxB
\else
??????\fi
\fi
}
\newcommand{\hatcurFIESvsinieccen}[1]{\ifnum#1=47 %
\hatcurFIESvsinieccenxxxxxA
\else
\ifnum#1=48 %
\hatcurFIESvsinieccenxxxxxB
\else
??????\fi
\fi
}
\newcommand{\hatcurFIESzfeheccen}[1]{\ifnum#1=47 %
\hatcurFIESzfeheccenxxxxxA
\else
\ifnum#1=48 %
\hatcurFIESzfeheccenxxxxxB
\else
??????\fi
\fi
}
\newcommand{\hatcurhtreccen}[1]{\ifnum#1=47 %
\hatcurhtreccenxxxxxA
\else
\ifnum#1=48 %
\hatcurhtreccenxxxxxB
\else
??????\fi
\fi
}
\newcommand{\hatcurISOageeccen}[1]{\ifnum#1=47 %
\hatcurISOageeccenxxxxxA
\else
\ifnum#1=48 %
\hatcurISOageeccenxxxxxB
\else
??????\fi
\fi
}
\newcommand{\hatcurISOJKeccen}[1]{\ifnum#1=47 %
\hatcurISOJKeccenxxxxxA
\else
\ifnum#1=48 %
\hatcurISOJKeccenxxxxxB
\else
??????\fi
\fi
}
\newcommand{\hatcurISOloggeccen}[1]{\ifnum#1=47 %
\hatcurISOloggeccenxxxxxA
\else
\ifnum#1=48 %
\hatcurISOloggeccenxxxxxB
\else
??????\fi
\fi
}
\newcommand{\hatcurISOlumeccen}[1]{\ifnum#1=47 %
\hatcurISOlumeccenxxxxxA
\else
\ifnum#1=48 %
\hatcurISOlumeccenxxxxxB
\else
??????\fi
\fi
}
\newcommand{\hatcurISOlumshorteccen}[1]{\ifnum#1=47 %
\hatcurISOlumshorteccenxxxxxA
\else
\ifnum#1=48 %
\hatcurISOlumshorteccenxxxxxB
\else
??????\fi
\fi
}
\newcommand{\hatcurISOmeccen}[1]{\ifnum#1=47 %
\hatcurISOmeccenxxxxxA
\else
\ifnum#1=48 %
\hatcurISOmeccenxxxxxB
\else
??????\fi
\fi
}
\newcommand{\hatcurISOMHeccen}[1]{\ifnum#1=47 %
\hatcurISOMHeccenxxxxxA
\else
\ifnum#1=48 %
\hatcurISOMHeccenxxxxxB
\else
??????\fi
\fi
}
\newcommand{\hatcurISOMJeccen}[1]{\ifnum#1=47 %
\hatcurISOMJeccenxxxxxA
\else
\ifnum#1=48 %
\hatcurISOMJeccenxxxxxB
\else
??????\fi
\fi
}
\newcommand{\hatcurISOMKeccen}[1]{\ifnum#1=47 %
\hatcurISOMKeccenxxxxxA
\else
\ifnum#1=48 %
\hatcurISOMKeccenxxxxxB
\else
??????\fi
\fi
}
\newcommand{\hatcurISOmlongeccen}[1]{\ifnum#1=47 %
\hatcurISOmlongeccenxxxxxA
\else
\ifnum#1=48 %
\hatcurISOmlongeccenxxxxxB
\else
??????\fi
\fi
}
\newcommand{\hatcurISOmshorteccen}[1]{\ifnum#1=47 %
\hatcurISOmshorteccenxxxxxA
\else
\ifnum#1=48 %
\hatcurISOmshorteccenxxxxxB
\else
??????\fi
\fi
}
\newcommand{\hatcurISOmveccen}[1]{\ifnum#1=47 %
\hatcurISOmveccenxxxxxA
\else
\ifnum#1=48 %
\hatcurISOmveccenxxxxxB
\else
??????\fi
\fi
}
\newcommand{\hatcurISOreccen}[1]{\ifnum#1=47 %
\hatcurISOreccenxxxxxA
\else
\ifnum#1=48 %
\hatcurISOreccenxxxxxB
\else
??????\fi
\fi
}
\newcommand{\hatcurISOrhoeccen}[1]{\ifnum#1=47 %
\hatcurISOrhoeccenxxxxxA
\else
\ifnum#1=48 %
\hatcurISOrhoeccenxxxxxB
\else
??????\fi
\fi
}
\newcommand{\hatcurISOrlongeccen}[1]{\ifnum#1=47 %
\hatcurISOrlongeccenxxxxxA
\else
\ifnum#1=48 %
\hatcurISOrlongeccenxxxxxB
\else
??????\fi
\fi
}
\newcommand{\hatcurISOrshorteccen}[1]{\ifnum#1=47 %
\hatcurISOrshorteccenxxxxxA
\else
\ifnum#1=48 %
\hatcurISOrshorteccenxxxxxB
\else
??????\fi
\fi
}
\newcommand{\hatcurISOsigmaeccen}[1]{\ifnum#1=47 %
\hatcurISOsigmaeccenxxxxxA
\else
\ifnum#1=48 %
\hatcurISOsigmaeccenxxxxxB
\else
??????\fi
\fi
}
\newcommand{\hatcurISOspececcen}[1]{\ifnum#1=47 %
\hatcurISOspececcenxxxxxA
\else
\ifnum#1=48 %
\hatcurISOspececcenxxxxxB
\else
??????\fi
\fi
}
\newcommand{\hatcurISOvieccen}[1]{\ifnum#1=47 %
\hatcurISOvieccenxxxxxA
\else
\ifnum#1=48 %
\hatcurISOvieccenxxxxxB
\else
??????\fi
\fi
}
\newcommand{\hatcurLBigeccen}[1]{\ifnum#1=47 %
\hatcurLBigeccenxxxxxA
\else
\ifnum#1=48 %
\hatcurLBigeccenxxxxxB
\else
??????\fi
\fi
}
\newcommand{\hatcurLBiieccen}[1]{\ifnum#1=47 %
\hatcurLBiieccenxxxxxA
\else
\ifnum#1=48 %
\hatcurLBiieccenxxxxxB
\else
??????\fi
\fi
}
\newcommand{\hatcurLBiIeccen}[1]{\ifnum#1=47 %
\hatcurLBiIeccenxxxxxA
\else
\ifnum#1=48 %
\hatcurLBiIeccenxxxxxB
\else
??????\fi
\fi
}
\newcommand{\hatcurLBiigeccen}[1]{\ifnum#1=47 %
\hatcurLBiigeccenxxxxxA
\else
\ifnum#1=48 %
\hatcurLBiigeccenxxxxxB
\else
??????\fi
\fi
}
\newcommand{\hatcurLBiiieccen}[1]{\ifnum#1=47 %
\hatcurLBiiieccenxxxxxA
\else
\ifnum#1=48 %
\hatcurLBiiieccenxxxxxB
\else
??????\fi
\fi
}
\newcommand{\hatcurLBiiIeccen}[1]{\ifnum#1=47 %
\hatcurLBiiIeccenxxxxxA
\else
\ifnum#1=48 %
\hatcurLBiiIeccenxxxxxB
\else
??????\fi
\fi
}
\newcommand{\hatcurLBiikepeccen}[1]{\ifnum#1=47 %
\hatcurLBiikepeccenxxxxxA
\else
\ifnum#1=48 %
\hatcurLBiikepeccenxxxxxB
\else
??????\fi
\fi
}
\newcommand{\hatcurLBiireccen}[1]{\ifnum#1=47 %
\hatcurLBiireccenxxxxxA
\else
\ifnum#1=48 %
\hatcurLBiireccenxxxxxB
\else
??????\fi
\fi
}
\newcommand{\hatcurLBiiReccen}[1]{\ifnum#1=47 %
\hatcurLBiiReccenxxxxxA
\else
\ifnum#1=48 %
\hatcurLBiiReccenxxxxxB
\else
??????\fi
\fi
}
\newcommand{\hatcurLBiizeccen}[1]{\ifnum#1=47 %
\hatcurLBiizeccenxxxxxA
\else
\ifnum#1=48 %
\hatcurLBiizeccenxxxxxB
\else
??????\fi
\fi
}
\newcommand{\hatcurLBikepeccen}[1]{\ifnum#1=47 %
\hatcurLBikepeccenxxxxxA
\else
\ifnum#1=48 %
\hatcurLBikepeccenxxxxxB
\else
??????\fi
\fi
}
\newcommand{\hatcurLBireccen}[1]{\ifnum#1=47 %
\hatcurLBireccenxxxxxA
\else
\ifnum#1=48 %
\hatcurLBireccenxxxxxB
\else
??????\fi
\fi
}
\newcommand{\hatcurLBiReccen}[1]{\ifnum#1=47 %
\hatcurLBiReccenxxxxxA
\else
\ifnum#1=48 %
\hatcurLBiReccenxxxxxB
\else
??????\fi
\fi
}
\newcommand{\hatcurLBizeccen}[1]{\ifnum#1=47 %
\hatcurLBizeccenxxxxxA
\else
\ifnum#1=48 %
\hatcurLBizeccenxxxxxB
\else
??????\fi
\fi
}
\newcommand{\hatcurLCbsqeccen}[1]{\ifnum#1=47 %
\hatcurLCbsqeccenxxxxxA
\else
\ifnum#1=48 %
\hatcurLCbsqeccenxxxxxB
\else
??????\fi
\fi
}
\newcommand{\hatcurLCdipeccen}[1]{\ifnum#1=47 %
\hatcurLCdipeccenxxxxxA
\else
\ifnum#1=48 %
\hatcurLCdipeccenxxxxxB
\else
??????\fi
\fi
}
\newcommand{\hatcurLCdureccen}[1]{\ifnum#1=47 %
\hatcurLCdureccenxxxxxA
\else
\ifnum#1=48 %
\hatcurLCdureccenxxxxxB
\else
??????\fi
\fi
}
\newcommand{\hatcurLCdurhreccen}[1]{\ifnum#1=47 %
\hatcurLCdurhreccenxxxxxA
\else
\ifnum#1=48 %
\hatcurLCdurhreccenxxxxxB
\else
??????\fi
\fi
}
\newcommand{\hatcurLCdurhrshorteccen}[1]{\ifnum#1=47 %
\hatcurLCdurhrshorteccenxxxxxA
\else
\ifnum#1=48 %
\hatcurLCdurhrshorteccenxxxxxB
\else
??????\fi
\fi
}
\newcommand{\hatcurLCdurshorteccen}[1]{\ifnum#1=47 %
\hatcurLCdurshorteccenxxxxxA
\else
\ifnum#1=48 %
\hatcurLCdurshorteccenxxxxxB
\else
??????\fi
\fi
}
\newcommand{\hatcurLChatnetmeccen}[1]{\ifnum#1=47 %
\hatcurLChatnetmeccenxxxxxA
\else
\ifnum#1=48 %
\hatcurLChatnetmeccenxxxxxB
\else
??????\fi
\fi
}
\newcommand{\hatcurLCiblendeccen}[1]{\ifnum#1=47 %
\hatcurLCiblendeccenxxxxxA
\else
\ifnum#1=48 %
\hatcurLCiblendeccenxxxxxB
\else
??????\fi
\fi
}
\newcommand{\hatcurLCimpeccen}[1]{\ifnum#1=47 %
\hatcurLCimpeccenxxxxxA
\else
\ifnum#1=48 %
\hatcurLCimpeccenxxxxxB
\else
??????\fi
\fi
}
\newcommand{\hatcurLCingdureccen}[1]{\ifnum#1=47 %
\hatcurLCingdureccenxxxxxA
\else
\ifnum#1=48 %
\hatcurLCingdureccenxxxxxB
\else
??????\fi
\fi
}
\newcommand{\hatcurLCPeccen}[1]{\ifnum#1=47 %
\hatcurLCPeccenxxxxxA
\else
\ifnum#1=48 %
\hatcurLCPeccenxxxxxB
\else
??????\fi
\fi
}
\newcommand{\hatcurLCPprececcen}[1]{\ifnum#1=47 %
\hatcurLCPprececcenxxxxxA
\else
\ifnum#1=48 %
\hatcurLCPprececcenxxxxxB
\else
??????\fi
\fi
}
\newcommand{\hatcurLCPshorteccen}[1]{\ifnum#1=47 %
\hatcurLCPshorteccenxxxxxA
\else
\ifnum#1=48 %
\hatcurLCPshorteccenxxxxxB
\else
??????\fi
\fi
}
\newcommand{\hatcurLCqeccen}[1]{\ifnum#1=47 %
\hatcurLCqeccenxxxxxA
\else
\ifnum#1=48 %
\hatcurLCqeccenxxxxxB
\else
??????\fi
\fi
}
\newcommand{\hatcurLCqshorteccen}[1]{\ifnum#1=47 %
\hatcurLCqshorteccenxxxxxA
\else
\ifnum#1=48 %
\hatcurLCqshorteccenxxxxxB
\else
??????\fi
\fi
}
\newcommand{\hatcurLCrprstareccen}[1]{\ifnum#1=47 %
\hatcurLCrprstareccenxxxxxA
\else
\ifnum#1=48 %
\hatcurLCrprstareccenxxxxxB
\else
??????\fi
\fi
}
\newcommand{\hatcurLCTeccen}[1]{\ifnum#1=47 %
\hatcurLCTeccenxxxxxA
\else
\ifnum#1=48 %
\hatcurLCTeccenxxxxxB
\else
??????\fi
\fi
}
\newcommand{\hatcurLCTAeccen}[1]{\ifnum#1=47 %
\hatcurLCTAeccenxxxxxA
\else
\ifnum#1=48 %
\hatcurLCTAeccenxxxxxB
\else
??????\fi
\fi
}
\newcommand{\hatcurLCTBeccen}[1]{\ifnum#1=47 %
\hatcurLCTBeccenxxxxxA
\else
\ifnum#1=48 %
\hatcurLCTBeccenxxxxxB
\else
??????\fi
\fi
}
\newcommand{\hatcurLCzetaeccen}[1]{\ifnum#1=47 %
\hatcurLCzetaeccenxxxxxA
\else
\ifnum#1=48 %
\hatcurLCzetaeccenxxxxxB
\else
??????\fi
\fi
}
\newcommand{\hatcurPPaequiveccen}[1]{\ifnum#1=47 %
\hatcurPPaequiveccenxxxxxA
\else
\ifnum#1=48 %
\hatcurPPaequiveccenxxxxxB
\else
??????\fi
\fi
}
\newcommand{\hatcurPPareccen}[1]{\ifnum#1=47 %
\hatcurPPareccenxxxxxA
\else
\ifnum#1=48 %
\hatcurPPareccenxxxxxB
\else
??????\fi
\fi
}
\newcommand{\hatcurPPareleccen}[1]{\ifnum#1=47 %
\hatcurPPareleccenxxxxxA
\else
\ifnum#1=48 %
\hatcurPPareleccenxxxxxB
\else
??????\fi
\fi
}
\newcommand{\hatcurPPfluxapeccen}[1]{\ifnum#1=47 %
\hatcurPPfluxapeccenxxxxxA
\else
\ifnum#1=48 %
\hatcurPPfluxapeccenxxxxxB
\else
??????\fi
\fi
}
\newcommand{\hatcurPPfluxapdimeccen}[1]{\ifnum#1=47 %
\hatcurPPfluxapdimeccenxxxxxA
\else
\ifnum#1=48 %
\hatcurPPfluxapdimeccenxxxxxB
\else
??????\fi
\fi
}
\newcommand{\hatcurPPfluxavgeccen}[1]{\ifnum#1=47 %
\hatcurPPfluxavgeccenxxxxxA
\else
\ifnum#1=48 %
\hatcurPPfluxavgeccenxxxxxB
\else
??????\fi
\fi
}
\newcommand{\hatcurPPfluxavgdimeccen}[1]{\ifnum#1=47 %
\hatcurPPfluxavgdimeccenxxxxxA
\else
\ifnum#1=48 %
\hatcurPPfluxavgdimeccenxxxxxB
\else
??????\fi
\fi
}
\newcommand{\hatcurPPfluxperieccen}[1]{\ifnum#1=47 %
\hatcurPPfluxperieccenxxxxxA
\else
\ifnum#1=48 %
\hatcurPPfluxperieccenxxxxxB
\else
??????\fi
\fi
}
\newcommand{\hatcurPPfluxperidimeccen}[1]{\ifnum#1=47 %
\hatcurPPfluxperidimeccenxxxxxA
\else
\ifnum#1=48 %
\hatcurPPfluxperidimeccenxxxxxB
\else
??????\fi
\fi
}
\newcommand{\hatcurPPgeccen}[1]{\ifnum#1=47 %
\hatcurPPgeccenxxxxxA
\else
\ifnum#1=48 %
\hatcurPPgeccenxxxxxB
\else
??????\fi
\fi
}
\newcommand{\hatcurPPieccen}[1]{\ifnum#1=47 %
\hatcurPPieccenxxxxxA
\else
\ifnum#1=48 %
\hatcurPPieccenxxxxxB
\else
??????\fi
\fi
}
\newcommand{\hatcurPPloggeccen}[1]{\ifnum#1=47 %
\hatcurPPloggeccenxxxxxA
\else
\ifnum#1=48 %
\hatcurPPloggeccenxxxxxB
\else
??????\fi
\fi
}
\newcommand{\hatcurPPmeccen}[1]{\ifnum#1=47 %
\hatcurPPmeccenxxxxxA
\else
\ifnum#1=48 %
\hatcurPPmeccenxxxxxB
\else
??????\fi
\fi
}
\newcommand{\hatcurPPmeeccen}[1]{\ifnum#1=47 %
\hatcurPPmeeccenxxxxxA
\else
\ifnum#1=48 %
\hatcurPPmeeccenxxxxxB
\else
??????\fi
\fi
}
\newcommand{\hatcurPPmelongeccen}[1]{\ifnum#1=47 %
\hatcurPPmelongeccenxxxxxA
\else
\ifnum#1=48 %
\hatcurPPmelongeccenxxxxxB
\else
??????\fi
\fi
}
\newcommand{\hatcurPPmeshorteccen}[1]{\ifnum#1=47 %
\hatcurPPmeshorteccenxxxxxA
\else
\ifnum#1=48 %
\hatcurPPmeshorteccenxxxxxB
\else
??????\fi
\fi
}
\newcommand{\hatcurPPmlongeccen}[1]{\ifnum#1=47 %
\hatcurPPmlongeccenxxxxxA
\else
\ifnum#1=48 %
\hatcurPPmlongeccenxxxxxB
\else
??????\fi
\fi
}
\newcommand{\hatcurPPmrcorreccen}[1]{\ifnum#1=47 %
\hatcurPPmrcorreccenxxxxxA
\else
\ifnum#1=48 %
\hatcurPPmrcorreccenxxxxxB
\else
??????\fi
\fi
}
\newcommand{\hatcurPPmshorteccen}[1]{\ifnum#1=47 %
\hatcurPPmshorteccenxxxxxA
\else
\ifnum#1=48 %
\hatcurPPmshorteccenxxxxxB
\else
??????\fi
\fi
}
\newcommand{\hatcurPPperieccen}[1]{\ifnum#1=47 %
\hatcurPPperieccenxxxxxA
\else
\ifnum#1=48 %
\hatcurPPperieccenxxxxxB
\else
??????\fi
\fi
}
\newcommand{\hatcurPPphiconjeccen}[1]{\ifnum#1=47 %
\hatcurPPphiconjeccenxxxxxA
\else
\ifnum#1=48 %
\hatcurPPphiconjeccenxxxxxB
\else
??????\fi
\fi
}
\newcommand{\hatcurPPreccen}[1]{\ifnum#1=47 %
\hatcurPPreccenxxxxxA
\else
\ifnum#1=48 %
\hatcurPPreccenxxxxxB
\else
??????\fi
\fi
}
\newcommand{\hatcurPPreeccen}[1]{\ifnum#1=47 %
\hatcurPPreeccenxxxxxA
\else
\ifnum#1=48 %
\hatcurPPreeccenxxxxxB
\else
??????\fi
\fi
}
\newcommand{\hatcurPPrelongeccen}[1]{\ifnum#1=47 %
\hatcurPPrelongeccenxxxxxA
\else
\ifnum#1=48 %
\hatcurPPrelongeccenxxxxxB
\else
??????\fi
\fi
}
\newcommand{\hatcurPPreshorteccen}[1]{\ifnum#1=47 %
\hatcurPPreshorteccenxxxxxA
\else
\ifnum#1=48 %
\hatcurPPreshorteccenxxxxxB
\else
??????\fi
\fi
}
\newcommand{\hatcurPPrhoeccen}[1]{\ifnum#1=47 %
\hatcurPPrhoeccenxxxxxA
\else
\ifnum#1=48 %
\hatcurPPrhoeccenxxxxxB
\else
??????\fi
\fi
}
\newcommand{\hatcurPPrlongeccen}[1]{\ifnum#1=47 %
\hatcurPPrlongeccenxxxxxA
\else
\ifnum#1=48 %
\hatcurPPrlongeccenxxxxxB
\else
??????\fi
\fi
}
\newcommand{\hatcurPPrshorteccen}[1]{\ifnum#1=47 %
\hatcurPPrshorteccenxxxxxA
\else
\ifnum#1=48 %
\hatcurPPrshorteccenxxxxxB
\else
??????\fi
\fi
}
\newcommand{\hatcurPPtcirceccen}[1]{\ifnum#1=47 %
\hatcurPPtcirceccenxxxxxA
\else
\ifnum#1=48 %
\hatcurPPtcirceccenxxxxxB
\else
??????\fi
\fi
}
\newcommand{\hatcurPPteffeccen}[1]{\ifnum#1=47 %
\hatcurPPteffeccenxxxxxA
\else
\ifnum#1=48 %
\hatcurPPteffeccenxxxxxB
\else
??????\fi
\fi
}
\newcommand{\hatcurPPthetaeccen}[1]{\ifnum#1=47 %
\hatcurPPthetaeccenxxxxxA
\else
\ifnum#1=48 %
\hatcurPPthetaeccenxxxxxB
\else
??????\fi
\fi
}
\newcommand{\hatcurPPtinfalleccen}[1]{\ifnum#1=47 %
\hatcurPPtinfalleccenxxxxxA
\else
\ifnum#1=48 %
\hatcurPPtinfalleccenxxxxxB
\else
??????\fi
\fi
}
\newcommand{\hatcurRVecceneccen}[1]{\ifnum#1=47 %
\hatcurRVecceneccenxxxxxA
\else
\ifnum#1=48 %
\hatcurRVecceneccenxxxxxB
\else
??????\fi
\fi
}
\newcommand{\hatcurRVfitrmsAeccen}[1]{\ifnum#1=47 %
\hatcurRVfitrmsAeccenxxxxxA
\else
\ifnum#1=48 %
\hatcurRVfitrmsAeccenxxxxxB
\else
??????\fi
\fi
}
\newcommand{\hatcurRVfitrmsBeccen}[1]{\ifnum#1=47 %
\hatcurRVfitrmsBeccenxxxxxA
\else
\ifnum#1=48 %
\hatcurRVfitrmsBeccenxxxxxB
\else
??????\fi
\fi
}
\newcommand{\hatcurRVfitrmsCeccen}[1]{\ifnum#1=48 %
\hatcurRVfitrmsCeccenxxxxxB
\else
??????\fi
}
\newcommand{\hatcurRVgammaAeccen}[1]{\ifnum#1=47 %
\hatcurRVgammaAeccenxxxxxA
\else
\ifnum#1=48 %
\hatcurRVgammaAeccenxxxxxB
\else
??????\fi
\fi
}
\newcommand{\hatcurRVgammaBeccen}[1]{\ifnum#1=47 %
\hatcurRVgammaBeccenxxxxxA
\else
\ifnum#1=48 %
\hatcurRVgammaBeccenxxxxxB
\else
??????\fi
\fi
}
\newcommand{\hatcurRVgammaCeccen}[1]{\ifnum#1=48 %
\hatcurRVgammaCeccenxxxxxB
\else
??????\fi
}
\newcommand{\hatcurRVheccen}[1]{\ifnum#1=47 %
\hatcurRVheccenxxxxxA
\else
\ifnum#1=48 %
\hatcurRVheccenxxxxxB
\else
??????\fi
\fi
}
\newcommand{\hatcurRVjitterAeccen}[1]{\ifnum#1=47 %
\hatcurRVjitterAeccenxxxxxA
\else
\ifnum#1=48 %
\hatcurRVjitterAeccenxxxxxB
\else
??????\fi
\fi
}
\newcommand{\hatcurRVjitterBeccen}[1]{\ifnum#1=47 %
\hatcurRVjitterBeccenxxxxxA
\else
\ifnum#1=48 %
\hatcurRVjitterBeccenxxxxxB
\else
??????\fi
\fi
}
\newcommand{\hatcurRVjitterCeccen}[1]{\ifnum#1=48 %
\hatcurRVjitterCeccenxxxxxB
\else
??????\fi
}
\newcommand{\hatcurRVkeccen}[1]{\ifnum#1=47 %
\hatcurRVkeccenxxxxxA
\else
\ifnum#1=48 %
\hatcurRVkeccenxxxxxB
\else
??????\fi
\fi
}
\newcommand{\hatcurRVKeccen}[1]{\ifnum#1=47 %
\hatcurRVKeccenxxxxxA
\else
\ifnum#1=48 %
\hatcurRVKeccenxxxxxB
\else
??????\fi
\fi
}
\newcommand{\hatcurRVomegaeccen}[1]{\ifnum#1=47 %
\hatcurRVomegaeccenxxxxxA
\else
\ifnum#1=48 %
\hatcurRVomegaeccenxxxxxB
\else
??????\fi
\fi
}
\newcommand{\hatcurRVrheccen}[1]{\ifnum#1=47 %
\hatcurRVrheccenxxxxxA
\else
\ifnum#1=48 %
\hatcurRVrheccenxxxxxB
\else
??????\fi
\fi
}
\newcommand{\hatcurRVrkeccen}[1]{\ifnum#1=47 %
\hatcurRVrkeccenxxxxxA
\else
\ifnum#1=48 %
\hatcurRVrkeccenxxxxxB
\else
??????\fi
\fi
}
\newcommand{\hatcurRVtroneeccen}[1]{\ifnum#1=47 %
\hatcurRVtroneeccenxxxxxA
\else
\ifnum#1=48 %
\hatcurRVtroneeccenxxxxxB
\else
??????\fi
\fi
}
\newcommand{\hatcurRVtrtwoeccen}[1]{\ifnum#1=47 %
\hatcurRVtrtwoeccenxxxxxA
\else
\ifnum#1=48 %
\hatcurRVtrtwoeccenxxxxxB
\else
??????\fi
\fi
}
\newcommand{\hatcurSMEiiloggeccen}[1]{\ifnum#1=47 %
\hatcurSMEiiloggeccenxxxxxA
\else
\ifnum#1=48 %
\hatcurSMEiiloggeccenxxxxxB
\else
??????\fi
\fi
}
\newcommand{\hatcurSMEiiteffeccen}[1]{\ifnum#1=47 %
\hatcurSMEiiteffeccenxxxxxA
\else
\ifnum#1=48 %
\hatcurSMEiiteffeccenxxxxxB
\else
??????\fi
\fi
}
\newcommand{\hatcurSMEiivmaceccen}[1]{\ifnum#1=47 %
\hatcurSMEiivmaceccenxxxxxA
\else
\ifnum#1=48 %
\hatcurSMEiivmaceccenxxxxxB
\else
??????\fi
\fi
}
\newcommand{\hatcurSMEiivmiceccen}[1]{\ifnum#1=47 %
\hatcurSMEiivmiceccenxxxxxA
\else
\ifnum#1=48 %
\hatcurSMEiivmiceccenxxxxxB
\else
??????\fi
\fi
}
\newcommand{\hatcurSMEiivsineccen}[1]{\ifnum#1=47 %
\hatcurSMEiivsineccenxxxxxA
\else
\ifnum#1=48 %
\hatcurSMEiivsineccenxxxxxB
\else
??????\fi
\fi
}
\newcommand{\hatcurSMEiizfeheccen}[1]{\ifnum#1=47 %
\hatcurSMEiizfeheccenxxxxxA
\else
\ifnum#1=48 %
\hatcurSMEiizfeheccenxxxxxB
\else
??????\fi
\fi
}
\newcommand{\hatcurSMEiizfehshorteccen}[1]{\ifnum#1=47 %
\hatcurSMEiizfehshorteccenxxxxxA
\else
\ifnum#1=48 %
\hatcurSMEiizfehshorteccenxxxxxB
\else
??????\fi
\fi
}
\newcommand{\hatcurSMEiloggeccen}[1]{\ifnum#1=47 %
\hatcurSMEiloggeccenxxxxxA
\else
\ifnum#1=48 %
\hatcurSMEiloggeccenxxxxxB
\else
??????\fi
\fi
}
\newcommand{\hatcurSMEiteffeccen}[1]{\ifnum#1=47 %
\hatcurSMEiteffeccenxxxxxA
\else
\ifnum#1=48 %
\hatcurSMEiteffeccenxxxxxB
\else
??????\fi
\fi
}
\newcommand{\hatcurSMEivmaceccen}[1]{\ifnum#1=47 %
\hatcurSMEivmaceccenxxxxxA
\else
\ifnum#1=48 %
\hatcurSMEivmaceccenxxxxxB
\else
??????\fi
\fi
}
\newcommand{\hatcurSMEivmiceccen}[1]{\ifnum#1=47 %
\hatcurSMEivmiceccenxxxxxA
\else
\ifnum#1=48 %
\hatcurSMEivmiceccenxxxxxB
\else
??????\fi
\fi
}
\newcommand{\hatcurSMEivsineccen}[1]{\ifnum#1=47 %
\hatcurSMEivsineccenxxxxxA
\else
\ifnum#1=48 %
\hatcurSMEivsineccenxxxxxB
\else
??????\fi
\fi
}
\newcommand{\hatcurSMEizfeheccen}[1]{\ifnum#1=47 %
\hatcurSMEizfeheccenxxxxxA
\else
\ifnum#1=48 %
\hatcurSMEizfeheccenxxxxxB
\else
??????\fi
\fi
}
\newcommand{\hatcurSMEizfehshorteccen}[1]{\ifnum#1=47 %
\hatcurSMEizfehshorteccenxxxxxA
\else
\ifnum#1=48 %
\hatcurSMEizfehshorteccenxxxxxB
\else
??????\fi
\fi
}
\newcommand{\hatcurTRESgammaeccen}[1]{\ifnum#1=47 %
\hatcurTRESgammaeccenxxxxxA
\else
\ifnum#1=48 %
\hatcurTRESgammaeccenxxxxxB
\else
??????\fi
\fi
}
\newcommand{\hatcurTRESloggeccen}[1]{\ifnum#1=47 %
\hatcurTRESloggeccenxxxxxA
\else
\ifnum#1=48 %
\hatcurTRESloggeccenxxxxxB
\else
??????\fi
\fi
}
\newcommand{\hatcurTRESnumspececcen}[1]{\ifnum#1=47 %
\hatcurTRESnumspececcenxxxxxA
\else
\ifnum#1=48 %
\hatcurTRESnumspececcenxxxxxB
\else
??????\fi
\fi
}
\newcommand{\hatcurTRESrvrmseccen}[1]{\ifnum#1=47 %
\hatcurTRESrvrmseccenxxxxxA
\else
\ifnum#1=48 %
\hatcurTRESrvrmseccenxxxxxB
\else
??????\fi
\fi
}
\newcommand{\hatcurTRESspaneccen}[1]{\ifnum#1=47 %
\hatcurTRESspaneccenxxxxxA
\else
\ifnum#1=48 %
\hatcurTRESspaneccenxxxxxB
\else
??????\fi
\fi
}
\newcommand{\hatcurTRESteffeccen}[1]{\ifnum#1=47 %
\hatcurTRESteffeccenxxxxxA
\else
\ifnum#1=48 %
\hatcurTRESteffeccenxxxxxB
\else
??????\fi
\fi
}
\newcommand{\hatcurTRESvsinieccen}[1]{\ifnum#1=47 %
\hatcurTRESvsinieccenxxxxxA
\else
\ifnum#1=48 %
\hatcurTRESvsinieccenxxxxxB
\else
??????\fi
\fi
}
\newcommand{\hatcurTRESzfeheccen}[1]{\ifnum#1=47 %
\hatcurTRESzfeheccenxxxxxA
\else
\ifnum#1=48 %
\hatcurTRESzfeheccenxxxxxB
\else
??????\fi
\fi
}
\newcommand{\hatcurXAveccen}[1]{\ifnum#1=47 %
\hatcurXAveccenxxxxxA
\else
\ifnum#1=48 %
\hatcurXAveccenxxxxxB
\else
??????\fi
\fi
}
\newcommand{\hatcurXdisteccen}[1]{\ifnum#1=47 %
\hatcurXdisteccenxxxxxA
\else
\ifnum#1=48 %
\hatcurXdisteccenxxxxxB
\else
??????\fi
\fi
}
\newcommand{\hatcurXdistredeccen}[1]{\ifnum#1=47 %
\hatcurXdistredeccenxxxxxA
\else
\ifnum#1=48 %
\hatcurXdistredeccenxxxxxB
\else
??????\fi
\fi
}
\newcommand{\hatcurXEBVeccen}[1]{\ifnum#1=47 %
\hatcurXEBVeccenxxxxxA
\else
\ifnum#1=48 %
\hatcurXEBVeccenxxxxxB
\else
??????\fi
\fi
}
\newcommand{\hatcurXjhisoredeccen}[1]{\ifnum#1=47 %
\hatcurXjhisoredeccenxxxxxA
\else
\ifnum#1=48 %
\hatcurXjhisoredeccenxxxxxB
\else
??????\fi
\fi
}
\newcommand{\hatcurXjkisoredeccen}[1]{\ifnum#1=47 %
\hatcurXjkisoredeccenxxxxxA
\else
\ifnum#1=48 %
\hatcurXjkisoredeccenxxxxxB
\else
??????\fi
\fi
}
\newcommand{\hatcurXmhisoredeccen}[1]{\ifnum#1=47 %
\hatcurXmhisoredeccenxxxxxA
\else
\ifnum#1=48 %
\hatcurXmhisoredeccenxxxxxB
\else
??????\fi
\fi
}
\newcommand{\hatcurXmiisoredeccen}[1]{\ifnum#1=47 %
\hatcurXmiisoredeccenxxxxxA
\else
\ifnum#1=48 %
\hatcurXmiisoredeccenxxxxxB
\else
??????\fi
\fi
}
\newcommand{\hatcurXmjisoredeccen}[1]{\ifnum#1=47 %
\hatcurXmjisoredeccenxxxxxA
\else
\ifnum#1=48 %
\hatcurXmjisoredeccenxxxxxB
\else
??????\fi
\fi
}
\newcommand{\hatcurXmkisoredeccen}[1]{\ifnum#1=47 %
\hatcurXmkisoredeccenxxxxxA
\else
\ifnum#1=48 %
\hatcurXmkisoredeccenxxxxxB
\else
??????\fi
\fi
}
\newcommand{\hatcurXmvisoredeccen}[1]{\ifnum#1=47 %
\hatcurXmvisoredeccenxxxxxA
\else
\ifnum#1=48 %
\hatcurXmvisoredeccenxxxxxB
\else
??????\fi
\fi
}
\newcommand{\hatcurXsecdureccen}[1]{\ifnum#1=47 %
\hatcurXsecdureccenxxxxxA
\else
\ifnum#1=48 %
\hatcurXsecdureccenxxxxxB
\else
??????\fi
\fi
}
\newcommand{\hatcurXsecingdureccen}[1]{\ifnum#1=47 %
\hatcurXsecingdureccenxxxxxA
\else
\ifnum#1=48 %
\hatcurXsecingdureccenxxxxxB
\else
??????\fi
\fi
}
\newcommand{\hatcurXsecondaryeccen}[1]{\ifnum#1=47 %
\hatcurXsecondaryeccenxxxxxA
\else
\ifnum#1=48 %
\hatcurXsecondaryeccenxxxxxB
\else
??????\fi
\fi
}
\newcommand{\hatcurXsecphaseeccen}[1]{\ifnum#1=47 %
\hatcurXsecphaseeccenxxxxxA
\else
\ifnum#1=48 %
\hatcurXsecphaseeccenxxxxxB
\else
??????\fi
\fi
}
\newcommand{\hatcurXviisoredeccen}[1]{\ifnum#1=47 %
\hatcurXviisoredeccenxxxxxA
\else
\ifnum#1=48 %
\hatcurXviisoredeccenxxxxxB
\else
??????\fi
\fi
}
\newcommand{\hatcurXvkisoredeccen}[1]{\ifnum#1=47 %
\hatcurXvkisoredeccenxxxxxA
\else
\ifnum#1=48 %
\hatcurXvkisoredeccenxxxxxB
\else
??????\fi
\fi
}

\newcommand{\hatcurxxxxxA}{HAT-P-47}
\newcommand{\hatcurbxxxxxA}{HAT-P-47b}
\newcommand{\hatcurcxxxxxA}{HAT-P-47c}

\newcommand{\hatcurplanetnumxxxxxA}{47}
\newcommand{\hatcurplanetnumeccenxxxxxA}{47}

\newcommand{\hatcurRVgammaabsxxxxxA}{\hatcurDSgamma{\hatcurplanetnumxxxxxA}}                           

\newcommand{\hatcurRVgammarelxxxxxA}{\hatcurRVgamma{\hatcurplanetnumxxxxxA}}                           

\newcommand{\hatcurCCtassvixxxxxA}{\ensuremath{0.58\pm0.15}}                  
\newcommand{\hatcurCCtassvieccenxxxxxA}{\ensuremath{0.58\pm0.15}}                  

\newcommand{\hatcurSMEversionxxxxxA}{ii}                                       
\newcommand{\hatcurSMEversioneccenxxxxxA}{ii}                                       

\newcommand{\hatcurisoshortxxxxxA}{YY}
\newcommand{\hatcurisofullxxxxxA}{Yonsei-Yale (YY)}
\newcommand{\hatcurisocitexxxxxA}{yi:2001}

\newcommand{\hatcurlumindxxxxxA}{\arstar}

\newcommand{\hatcurjhkfilsetxxxxxA}{ESO}

\newcommand{\hatcurSMEteffxxxxxA}{\ifthenelse{\equal{\hatcurSMEversionxxxxxA}{i}}{\hatcurSMEiteff{\hatcurplanetnumxxxxxA}}{\hatcurSMEiiteff{\hatcurplanetnumxxxxxA}}}
\newcommand{\hatcurSMEzfehxxxxxA}{\ifthenelse{\equal{\hatcurSMEversionxxxxxA}{i}}{\hatcurSMEizfeh{\hatcurplanetnumxxxxxA}}{\hatcurSMEiizfeh{\hatcurplanetnumxxxxxA}}}
\newcommand{\hatcurSMEzfehshortxxxxxA}{\ifthenelse{\equal{\hatcurSMEversionxxxxxA}{i}}{\hatcurSMEizfehshort{\hatcurplanetnumxxxxxA}}{\hatcurSMEiizfehshort{\hatcurplanetnumxxxxxA}}}
\newcommand{\hatcurSMEloggxxxxxA}{\ifthenelse{\equal{\hatcurSMEversionxxxxxA}{i}}{\hatcurSMEilogg{\hatcurplanetnumxxxxxA}}{\hatcurSMEiilogg{\hatcurplanetnumxxxxxA}}}
\newcommand{\hatcurSMEvsinxxxxxA}{\ifthenelse{\equal{\hatcurSMEversionxxxxxA}{i}}{\hatcurSMEivsin{\hatcurplanetnumxxxxxA}}{\hatcurSMEiivsin{\hatcurplanetnumxxxxxA}}}
\newcommand{\hatcurSMEvmacxxxxxA}{\ifthenelse{\equal{\hatcurSMEversionxxxxxA}{i}}{\hatcurSMEivmac{\hatcurplanetnumxxxxxA}}{\hatcurSMEiivmac{\hatcurplanetnumxxxxxA}}}
\newcommand{\hatcurSMEvmicxxxxxA}{\ifthenelse{\equal{\hatcurSMEversionxxxxxA}{i}}{\hatcurSMEivmic{\hatcurplanetnumxxxxxA}}{\hatcurSMEiivmic{\hatcurplanetnumxxxxxA}}}

\newcommand{\hatcurSMEteffeccenxxxxxA}{\ifthenelse{\equal{\hatcurSMEversioneccenxxxxxA}{i}}{\hatcurSMEiteff{\hatcurplanetnumeccenxxxxxA}}{\hatcurSMEiiteff{\hatcurplanetnumeccenxxxxxA}}}
\newcommand{\hatcurSMEzfeheccenxxxxxA}{\ifthenelse{\equal{\hatcurSMEversioneccenxxxxxA}{i}}{\hatcurSMEizfeh{\hatcurplanetnumeccenxxxxxA}}{\hatcurSMEiizfeh{\hatcurplanetnumeccenxxxxxA}}}
\newcommand{\hatcurSMEzfehshorteccenxxxxxA}{\ifthenelse{\equal{\hatcurSMEversioneccenxxxxxA}{i}}{\hatcurSMEizfehshort{\hatcurplanetnumeccenxxxxxA}}{\hatcurSMEiizfehshort{\hatcurplanetnumeccenxxxxxA}}}
\newcommand{\hatcurSMEloggeccenxxxxxA}{\ifthenelse{\equal{\hatcurSMEversioneccenxxxxxA}{i}}{\hatcurSMEilogg{\hatcurplanetnumeccenxxxxxA}}{\hatcurSMEiilogg{\hatcurplanetnumeccenxxxxxA}}}
\newcommand{\hatcurSMEvsineccenxxxxxA}{\ifthenelse{\equal{\hatcurSMEversioneccenxxxxxA}{i}}{\hatcurSMEivsin{\hatcurplanetnumeccenxxxxxA}}{\hatcurSMEiivsin{\hatcurplanetnumeccenxxxxxA}}}
\newcommand{\hatcurSMEvmaceccenxxxxxA}{\ifthenelse{\equal{\hatcurSMEversioneccenxxxxxA}{i}}{\hatcurSMEivmac{\hatcurplanetnumeccenxxxxxA}}{\hatcurSMEiivmac{\hatcurplanetnumeccenxxxxxA}}}
\newcommand{\hatcurSMEvmiceccenxxxxxA}{\ifthenelse{\equal{\hatcurSMEversioneccenxxxxxA}{i}}{\hatcurSMEivmic{\hatcurplanetnumeccenxxxxxA}}{\hatcurSMEiivmic{\hatcurplanetnumeccenxxxxxA}}}

\newcommand{\hatcurxxxxxB}{HAT-P-48}
\newcommand{\hatcurbxxxxxB}{HAT-P-48b}
\newcommand{\hatcurcxxxxxB}{HAT-P-48c}

\newcommand{\hatcurplanetnumxxxxxB}{48}
\newcommand{\hatcurplanetnumeccenxxxxxB}{48}

\newcommand{\hatcurRVgammaabsxxxxxB}{\hatcurDSgamma{\hatcurplanetnumxxxxxB}}                           

\newcommand{\hatcurRVgammarelxxxxxB}{\hatcurRVgamma{\hatcurplanetnumxxxxxB}}                           

\newcommand{\hatcurCCtassvixxxxxB}{\ensuremath{0.58\pm0.15}}                  
\newcommand{\hatcurCCtassvieccenxxxxxB}{\ensuremath{0.58\pm0.15}}                  

\newcommand{\hatcurSMEversionxxxxxB}{ii}                                       
\newcommand{\hatcurSMEversioneccenxxxxxB}{ii}                                       

\newcommand{\hatcurisoshortxxxxxB}{YY}
\newcommand{\hatcurisofullxxxxxB}{Yonsei-Yale (YY)}
\newcommand{\hatcurisocitexxxxxB}{yi:2001}

\newcommand{\hatcurlumindxxxxxB}{\arstar}

\newcommand{\hatcurjhkfilsetxxxxxB}{ESO}

\newcommand{\hatcurSMEteffxxxxxB}{\ifthenelse{\equal{\hatcurSMEversionxxxxxB}{i}}{\hatcurSMEiteff{\hatcurplanetnumxxxxxB}}{\hatcurSMEiiteff{\hatcurplanetnumxxxxxB}}}
\newcommand{\hatcurSMEzfehxxxxxB}{\ifthenelse{\equal{\hatcurSMEversionxxxxxB}{i}}{\hatcurSMEizfeh{\hatcurplanetnumxxxxxB}}{\hatcurSMEiizfeh{\hatcurplanetnumxxxxxB}}}
\newcommand{\hatcurSMEzfehshortxxxxxB}{\ifthenelse{\equal{\hatcurSMEversionxxxxxB}{i}}{\hatcurSMEizfehshort{\hatcurplanetnumxxxxxB}}{\hatcurSMEiizfehshort{\hatcurplanetnumxxxxxB}}}
\newcommand{\hatcurSMEloggxxxxxB}{\ifthenelse{\equal{\hatcurSMEversionxxxxxB}{i}}{\hatcurSMEilogg{\hatcurplanetnumxxxxxB}}{\hatcurSMEiilogg{\hatcurplanetnumxxxxxB}}}
\newcommand{\hatcurSMEvsinxxxxxB}{\ifthenelse{\equal{\hatcurSMEversionxxxxxB}{i}}{\hatcurSMEivsin{\hatcurplanetnumxxxxxB}}{\hatcurSMEiivsin{\hatcurplanetnumxxxxxB}}}
\newcommand{\hatcurSMEvmacxxxxxB}{\ifthenelse{\equal{\hatcurSMEversionxxxxxB}{i}}{\hatcurSMEivmac{\hatcurplanetnumxxxxxB}}{\hatcurSMEiivmac{\hatcurplanetnumxxxxxB}}}
\newcommand{\hatcurSMEvmicxxxxxB}{\ifthenelse{\equal{\hatcurSMEversionxxxxxB}{i}}{\hatcurSMEivmic{\hatcurplanetnumxxxxxB}}{\hatcurSMEiivmic{\hatcurplanetnumxxxxxB}}}

\newcommand{\hatcurSMEteffeccenxxxxxB}{\ifthenelse{\equal{\hatcurSMEversioneccenxxxxxB}{i}}{\hatcurSMEiteff{\hatcurplanetnumeccenxxxxxB}}{\hatcurSMEiiteff{\hatcurplanetnumeccenxxxxxB}}}
\newcommand{\hatcurSMEzfeheccenxxxxxB}{\ifthenelse{\equal{\hatcurSMEversioneccenxxxxxB}{i}}{\hatcurSMEizfeh{\hatcurplanetnumeccenxxxxxB}}{\hatcurSMEiizfeh{\hatcurplanetnumeccenxxxxxB}}}
\newcommand{\hatcurSMEzfehshorteccenxxxxxB}{\ifthenelse{\equal{\hatcurSMEversioneccenxxxxxB}{i}}{\hatcurSMEizfehshort{\hatcurplanetnumeccenxxxxxB}}{\hatcurSMEiizfehshort{\hatcurplanetnumeccenxxxxxB}}}
\newcommand{\hatcurSMEloggeccenxxxxxB}{\ifthenelse{\equal{\hatcurSMEversioneccenxxxxxB}{i}}{\hatcurSMEilogg{\hatcurplanetnumeccenxxxxxB}}{\hatcurSMEiilogg{\hatcurplanetnumeccenxxxxxB}}}
\newcommand{\hatcurSMEvsineccenxxxxxB}{\ifthenelse{\equal{\hatcurSMEversioneccenxxxxxB}{i}}{\hatcurSMEivsin{\hatcurplanetnumeccenxxxxxB}}{\hatcurSMEiivsin{\hatcurplanetnumeccenxxxxxB}}}
\newcommand{\hatcurSMEvmaceccenxxxxxB}{\ifthenelse{\equal{\hatcurSMEversioneccenxxxxxB}{i}}{\hatcurSMEivmac{\hatcurplanetnumeccenxxxxxB}}{\hatcurSMEiivmac{\hatcurplanetnumeccenxxxxxB}}}
\newcommand{\hatcurSMEvmiceccenxxxxxB}{\ifthenelse{\equal{\hatcurSMEversioneccenxxxxxB}{i}}{\hatcurSMEivmic{\hatcurplanetnumeccenxxxxxB}}{\hatcurSMEiivmic{\hatcurplanetnumeccenxxxxxB}}}

\newcommand{\hatcur}[1]{\ifnum#1=47 %
\hatcurxxxxxA
\else
\ifnum#1=48 %
\hatcurxxxxxB
\else
??????\fi
\fi
}
\newcommand{\hatcurb}[1]{\ifnum#1=47 %
\hatcurbxxxxxA
\else
\ifnum#1=48 %
\hatcurbxxxxxB
\else
??????\fi
\fi
}
\newcommand{\hatcurc}[1]{\ifnum#1=47 %
\hatcurcxxxxxA
\else
\ifnum#1=48 %
\hatcurcxxxxxB
\else
??????\fi
\fi
}
\newcommand{\hatcurCCtassvi}[1]{\ifnum#1=47 %
\hatcurCCtassvixxxxxA
\else
\ifnum#1=48 %
\hatcurCCtassvixxxxxB
\else
??????\fi
\fi
}
\newcommand{\hatcurCCtassvieccen}[1]{\ifnum#1=47 %
\hatcurCCtassvieccenxxxxxA
\else
\ifnum#1=48 %
\hatcurCCtassvieccenxxxxxB
\else
??????\fi
\fi
}
\newcommand{\hatcurisocite}[1]{\ifnum#1=47 %
\hatcurisocitexxxxxA
\else
\ifnum#1=48 %
\hatcurisocitexxxxxB
\else
??????\fi
\fi
}
\newcommand{\hatcurisofull}[1]{\ifnum#1=47 %
\hatcurisofullxxxxxA
\else
\ifnum#1=48 %
\hatcurisofullxxxxxB
\else
??????\fi
\fi
}
\newcommand{\hatcurisoshort}[1]{\ifnum#1=47 %
\hatcurisoshortxxxxxA
\else
\ifnum#1=48 %
\hatcurisoshortxxxxxB
\else
??????\fi
\fi
}
\newcommand{\hatcurjhkfilset}[1]{\ifnum#1=47 %
\hatcurjhkfilsetxxxxxA
\else
\ifnum#1=48 %
\hatcurjhkfilsetxxxxxB
\else
??????\fi
\fi
}
\newcommand{\hatcurlumind}[1]{\ifnum#1=47 %
\hatcurlumindxxxxxA
\else
\ifnum#1=48 %
\hatcurlumindxxxxxB
\else
??????\fi
\fi
}
\newcommand{\hatcurplanetnum}[1]{\ifnum#1=47 %
\hatcurplanetnumxxxxxA
\else
\ifnum#1=48 %
\hatcurplanetnumxxxxxB
\else
??????\fi
\fi
}
\newcommand{\hatcurplanetnumeccen}[1]{\ifnum#1=47 %
\hatcurplanetnumeccenxxxxxA
\else
\ifnum#1=48 %
\hatcurplanetnumeccenxxxxxB
\else
??????\fi
\fi
}
\newcommand{\hatcurRVgammaabs}[1]{\ifnum#1=47 %
\hatcurRVgammaabsxxxxxA
\else
\ifnum#1=48 %
\hatcurRVgammaabsxxxxxB
\else
??????\fi
\fi
}
\newcommand{\hatcurRVgammarel}[1]{\ifnum#1=47 %
\hatcurRVgammarelxxxxxA
\else
\ifnum#1=48 %
\hatcurRVgammarelxxxxxB
\else
??????\fi
\fi
}
\newcommand{\hatcurSMElogg}[1]{\ifnum#1=47 %
\hatcurSMEloggxxxxxA
\else
\ifnum#1=48 %
\hatcurSMEloggxxxxxB
\else
??????\fi
\fi
}
\newcommand{\hatcurSMEloggeccen}[1]{\ifnum#1=47 %
\hatcurSMEloggeccenxxxxxA
\else
\ifnum#1=48 %
\hatcurSMEloggeccenxxxxxB
\else
??????\fi
\fi
}
\newcommand{\hatcurSMEteff}[1]{\ifnum#1=47 %
\hatcurSMEteffxxxxxA
\else
\ifnum#1=48 %
\hatcurSMEteffxxxxxB
\else
??????\fi
\fi
}
\newcommand{\hatcurSMEteffeccen}[1]{\ifnum#1=47 %
\hatcurSMEteffeccenxxxxxA
\else
\ifnum#1=48 %
\hatcurSMEteffeccenxxxxxB
\else
??????\fi
\fi
}
\newcommand{\hatcurSMEversion}[1]{\ifnum#1=47 %
\hatcurSMEversionxxxxxA
\else
\ifnum#1=48 %
\hatcurSMEversionxxxxxB
\else
??????\fi
\fi
}
\newcommand{\hatcurSMEversioneccen}[1]{\ifnum#1=47 %
\hatcurSMEversioneccenxxxxxA
\else
\ifnum#1=48 %
\hatcurSMEversioneccenxxxxxB
\else
??????\fi
\fi
}
\newcommand{\hatcurSMEvmac}[1]{\ifnum#1=47 %
\hatcurSMEvmacxxxxxA
\else
\ifnum#1=48 %
\hatcurSMEvmacxxxxxB
\else
??????\fi
\fi
}
\newcommand{\hatcurSMEvmaceccen}[1]{\ifnum#1=47 %
\hatcurSMEvmaceccenxxxxxA
\else
\ifnum#1=48 %
\hatcurSMEvmaceccenxxxxxB
\else
??????\fi
\fi
}
\newcommand{\hatcurSMEvmic}[1]{\ifnum#1=47 %
\hatcurSMEvmicxxxxxA
\else
\ifnum#1=48 %
\hatcurSMEvmicxxxxxB
\else
??????\fi
\fi
}
\newcommand{\hatcurSMEvmiceccen}[1]{\ifnum#1=47 %
\hatcurSMEvmiceccenxxxxxA
\else
\ifnum#1=48 %
\hatcurSMEvmiceccenxxxxxB
\else
??????\fi
\fi
}
\newcommand{\hatcurSMEvsin}[1]{\ifnum#1=47 %
\hatcurSMEvsinxxxxxA
\else
\ifnum#1=48 %
\hatcurSMEvsinxxxxxB
\else
??????\fi
\fi
}
\newcommand{\hatcurSMEvsineccen}[1]{\ifnum#1=47 %
\hatcurSMEvsineccenxxxxxA
\else
\ifnum#1=48 %
\hatcurSMEvsineccenxxxxxB
\else
??????\fi
\fi
}
\newcommand{\hatcurSMEzfeh}[1]{\ifnum#1=47 %
\hatcurSMEzfehxxxxxA
\else
\ifnum#1=48 %
\hatcurSMEzfehxxxxxB
\else
??????\fi
\fi
}
\newcommand{\hatcurSMEzfeheccen}[1]{\ifnum#1=47 %
\hatcurSMEzfeheccenxxxxxA
\else
\ifnum#1=48 %
\hatcurSMEzfeheccenxxxxxB
\else
??????\fi
\fi
}
\newcommand{\hatcurSMEzfehshort}[1]{\ifnum#1=47 %
\hatcurSMEzfehshortxxxxxA
\else
\ifnum#1=48 %
\hatcurSMEzfehshortxxxxxB
\else
??????\fi
\fi
}
\newcommand{\hatcurSMEzfehshorteccen}[1]{\ifnum#1=47 %
\hatcurSMEzfehshorteccenxxxxxA
\else
\ifnum#1=48 %
\hatcurSMEzfehshorteccenxxxxxB
\else
??????\fi
\fi
}

\newcounter{planetcounter}

\newboolean{emulateapj}
\setboolean{emulateapj}{true}

\newboolean{rvtablelong}
\setboolean{rvtablelong}{true}

\newboolean{astroph}
\setboolean{astroph}{true}

\shortauthors{Bakos et al.}
\shorttitle{
\setcounter{planetcounter}{1}
\loopand\hatcur{47}\lowercase{b}\loopcommanospace
\setcounter{planetcounter}{2}
\loopand\hatcur{48}\lowercase{b}\loopcommanospace
}
\ifthenelse{\boolean{emulateapj}}{
    \newcommand{\titledag}{$\dagger$}
}{
    \newcommand{\titledag}{\dagger}
}

\begin{document}

\title{
\hatcur{47}\lowercase{b} and \hatcur{48}\lowercase{b}: 
Two Low Density Sub-Saturn-Mass Transiting Planets on the Edge of the 
Period--Mass Desert\altaffilmark{\titledag}
}

\author{
    G.~\'A.~Bakos\altaffilmark{1,$\dagger$},
    J.~D.~Hartman\altaffilmark{1},
    G.~Torres\altaffilmark{2},
    D.~W.~Latham\altaffilmark{2},
    B.~Sato\altaffilmark{3},
    A.~Bieryla\altaffilmark{2},
    A.~Shporer\altaffilmark{4,5},
    A.~W.~Howard\altaffilmark{6},
    B.~J.~Fulton\altaffilmark{6,14},
    L.~A.~Buchhave\altaffilmark{7},
    K.~Penev\altaffilmark{1},
    G.~Kov\'acs\altaffilmark{8},
    T.~Kov\'acs\altaffilmark{8},
    Z.~Csubry\altaffilmark{1},
    G.~A.~Esquerdo\altaffilmark{3},
    M.~Everett\altaffilmark{9},
    T.~Szklen\'ar\altaffilmark{12},
    S.~N.~Quinn\altaffilmark{13},
    B.~B\'eky\altaffilmark{10},
    G.~W.~Marcy\altaffilmark{11},
    R.~W.~Noyes\altaffilmark{2},
    J.~L\'az\'ar\altaffilmark{12},
    I.~Papp\altaffilmark{12},
    P.~S\'ari\altaffilmark{12}
}
\altaffiltext{1}{Department of Astrophysical Sciences, Princeton
  University, Princeton, NJ 08544; email: gbakos@astro.princeton.edu}


\altaffiltext{$\dagger$}{Packard Fellow}

\altaffiltext{2}{Harvard-Smithsonian Center for Astrophysics,
    Cambridge, MA}

\altaffiltext{3}{Department of Earth and Planetary Sciences, Tokyo Institute
    of Technology, 2-12-1 Ookayama, Meguro-ku, Tokyo 152-8551}

\altaffiltext{4}{Jet Propulsion Laboratory, California Institute of Technology, 4800 Oak Grove Drive, Pasadena, CA 91109, USA}

\altaffiltext{5}{NASA Sagan Fellow}

\altaffiltext{6}{Institute for Astronomy, University of Hawaii, 2680 Woodlawn Drive, Honolulu, HI 96822}

\altaffiltext{7}{Niels Bohr Institute, University of Copenhagen, DK-2100, Denmark, and Centre for Star and Planet Formation, Natural History Museum of Denmark, DK-1350 Copenhagen}

\altaffiltext{8,15}{Konkoly Observatory of the Hungarian Academy of
    Sciences, Budapest, Hungary}

\altaffiltext{9}{Steward Observatory, University of Arizona, Tucson, AZ}

\altaffiltext{10}{Google}

\altaffiltext{11}{Department of Astronomy, University of California,
    Berkeley, CA}

\altaffiltext{12}{Hungarian Astronomical Association, Budapest, 
    Hungary}

\altaffiltext{13}{Department of Physics and Astronomy, Georgia State University, Atlanta, GA}

\altaffiltext{14}{
    NSF Graduate Research Fellow}

\altaffiltext{15}{
    Institute of Theoretical Physics, Eötvös University, H-1117 Budapest,
    Hungary
}

\altaffiltext{$\dagger$}{
    Based in part on observations obtained at the W.~M.~Keck
    Observatory, which is operated by the University of California and
    the California Institute of Technology. Keck time has been granted
    by NOAO (A284Hr, A245Hr) and NASA (N108Hr, N154Hr, N130Hr). Based
    in part on data collected at Subaru Telescope (program o11170),
    which is operated by the National Astronomical Observatory of
    Japan. Based in part on observations made with the Nordic Optical
    Telescope, operated on the island of La Palma jointly by Denmark,
    Finland, Iceland, Norway, and Sweden, in the Spanish Observatorio
    del Roque de los Muchachos of the Instituto de Astrofisica de
    Canarias. Based in part on observations obtained with facilities
    of the Las Cumbres Observatory Global Telescope. 
}


\begin{abstract}

\setcounter{footnote}{10}
We report the discovery of two new transiting extrasolar planets
orbiting moderately bright ($V = 10.7$ and $12.2$ mag) F stars (masses
of $\hatcurISOmshort{47}$\,\msun\ and $\hatcurISOmshort{48}$\,\msun,
respectively).  The planets have periods of $P =
\hatcurLCPshort{47}$\,d and $\hatcurLCPshort{48}$\,d, and masses of
$\hatcurPPmshort{47}$\,\mjup\ and $\hatcurPPmshort{48}$\,\mjup\ which
are almost half-way between those of Neptune and Saturn.  With radii of
$\hatcurPPrshort{47}$\,\rjup\ and $\hatcurPPrshort{48}$\,\rjup, these
very low density planets are the two lowest mass planets with radii in
excess that of Jupiter.  Comparing with other recent planet
discoveries, we find that sub-Saturns ($0.18\mjup<\mpl<0.3\mjup$) and
super-Neptunes ($0.05\mjup<\mpl\le0.18\mjup$) exhibit a wide range of
radii, and their radii exhibit a weaker correlation with irradiation
than higher mass planets.  The two planets are both suitable for
measuring the Rossiter-McLaughlin effect and for atmospheric
characterization.  Measuring the former effect would allow an
interesting test of the theory that star--planet tidal interactions are
responsible for the tendency of close-in giant planets around
convective envelope stars to be on low obliquity orbits.  Both planets
fall on the edge of the short period Neptunian desert in the semi-major
axis--mass plane.
\setcounter{footnote}{0}
\end{abstract}

\keywords{
    planetary systems ---
    stars: individual (\hatcur{47}, \hatcurCCgsc{47}, \hatcur{48}, 
    \hatcurCCgsc{48}) ---
    techniques: spectroscopic, photometric
}


\section{Introduction}
\label{sec:introduction}

One of the startling discoveries in the field of exoplanets is that at
fixed mass planets show a very wide spread in radii (and bulk density). 
For example, within the narrow mass range of 0.85\,\mjup\ to
0.9\,\mjup, planets have been found with radii ranging from
0.78\,\rjup\ \citep[WASP-59b;][]{hebrard:2013} up to 2.1\,\rjup\
\citep[WASP-79b;][]{smalley:2012}.  While some variance is expected due
to differences in composition, age, and irradiation
\citep[e.g.,][]{burrows:2007}, explaining the very large radius planets
remains a puzzle \citep{spiegel:2013}.  Observationally the radii of
giant planets have been found to correlate tightly with the degree of
stellar irradiation, with more highly irradiated planets having larger
radii \citep{kovacs:2010:hat15}.  While it is unclear to what extent
selection effects are responsible for this correlation, most
theoretical models do predict larger radii with increased irradiation.

There is some evidence that the influence of irradiation on the
planetary radii depends on planetary mass, with super-Jupiters spanning
a smaller range of radii than sub-Jupiters
\citep[e.g.,][]{huang:2015:hat56}.  On the other hand, planet mass and
radius are essentially uncorrelated for planets with masses greater
than Saturn.  Below Saturn-mass the radii are observed to decrease,
with all planets discovered to date below the mass of Neptune having
radii less than 0.85\rjup, excluding the Kepler-51 system
\citep{kento:2014} which has unusually bloated planets (with large
uncertainties on their masses).  However, due to the small number of
low-mass transiting planets with precisely measured masses and radii,
it has been difficult to draw firm conclusions about the spread in the
mass--radius relation below the mass of Saturn or the influence of
other parameters such as irradiation or metallicity on this relation.

In this paper we present the discovery and characterization of two new
low-mass transiting planets by the HATNet project
\citep{bakos:2004:hatnet}, called \hatcurb{47} and \hatcurb{48}.  In
\refsecl{obs} we summarize the detection of the photometric transit
signal and the subsequent spectroscopic and photometric observations of
each star to confirm the planets.  In \refsecl{analysis} we analyze the
data to rule out false positive scenarios, and to determine the stellar
and planetary parameters.  Our findings are briefly discussed in
\refsecl{discussion}.

\section{Observations}
\label{sec:obs}

\hatcurb{47} and \hatcurb{48} were discovered through a combination of
photometric and spectroscopic observations.  See previous HATNet
transiting exoplanet (TEP) discovery papers for a general description
of the method \citep[e.g.][]{bakos:2010:hat11,latham:2009:hat8}.  Below
we summarize the observations that led to the two discoveries presented
here.  Identifying information for these stars is provided later in the
paper (\reftabl{stellar}).

\subsection{Photometric detection}
\label{sec:detection}

\begin{figure}[]
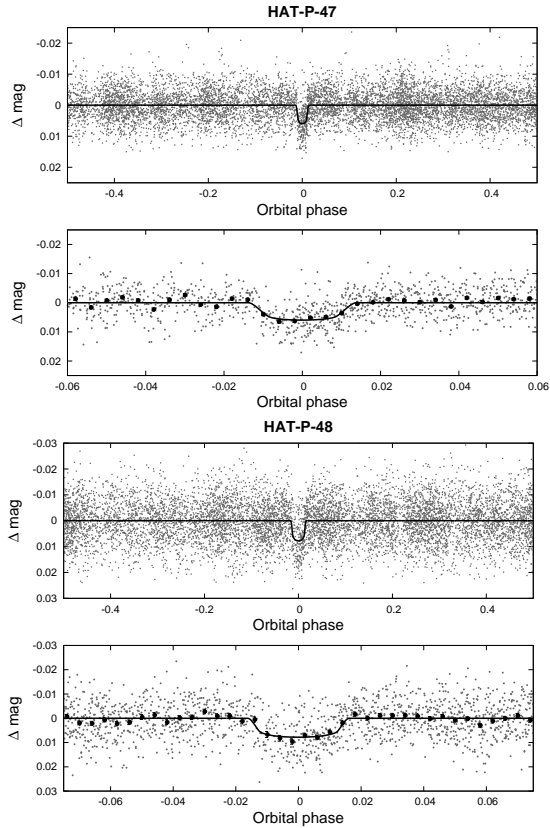

\plotone{\hatcurhtr{47}-hatnet.eps}
\plotone{\hatcurhtr{48}-hatnet.eps}
\caption[]{
    HATNet \lcs{} of \hatcur{47} (top) and \hatcur{48} (bottom) phase
    folded with the transit period.  In each case we show two panels:
    the top shows the unbinned light curve, while the bottom shows the
    region zoomed-in on the transit, with dark filled circles for the
    light curve binned in phase with a binsize of 0.004.  The solid
    line shows the model fit to the light curve.
\label{fig:hatnet}}
\end{figure}

\reftabl{photobs} summarizes the photometric observations that we
performed of \hatcur{47} and \hatcur{48}.  These include discovery
light curves obtained with the fully automated HATNet system, which
were generated and filtered for noise following
\cite{bakos:2010:hat11}.  These \lcs{} were searched for periodic
transit signals using the Box Least-Squares \citep[BLS;
see][]{kovacs:2002:BLS} method, leading to the identification of
\hatcur{47} and \hatcur{48} as candidate TEP systems (\reffigl{hatnet})
with the following properties:

\begin{itemize}
\item {\em \hatcur{47}} -- \hatcurCCgsc{47} (also known as
  \hatcurCCtwomass{47}; $\alpha = \hatcurCCra{47}$, $\delta =
  \hatcurCCdec{47}$; J2000; $V=\hatcurCCtassmv{47}$, \citealp{droege:2006}). 
  A signal was detected for this star with an apparent depth of
  $\sim$\hatcurLCdip{47}\,mmag, and a period of
  $P=\hatcurLCPshort{47}$\,days.  Discrete Fourier Transformation (DFT)
  frequency spectrum did not detect any component in excess of 0.5\,mmag
  amplitude in the [0,50] cycles/day frequency range.  No significant
  second frequency peak was seen in the BLS spectrum.
\item {\em \hatcur{48}} -- \hatcurCCgsc{48} (also known as
  \hatcurCCtwomass{48}; $\alpha = \hatcurCCra{48}$, $\delta =
  \hatcurCCdec{48}$; J2000; $V=\hatcurCCtassmv{48}$, \citealp{droege:2006}). 
  A signal was detected for this star with an apparent depth of
  $\sim$\hatcurLCdip{48}\,mmag, and a period of
  $P=\hatcurLCPshort{48}$\,days.  The DFT frequency spectrum showed no
  significant peak in excess of 0.6\,mmag in the range of [0,50] 
  cycles/day.
\end{itemize}

\ifthenelse{\boolean{emulateapj}}{
    \begin{deluxetable*}{llrrr}
}{
    \begin{deluxetable}{llrrr}
}
\tablewidth{0pc}
\tabletypesize{\scriptsize}
\tablecaption{
    Summary of photometric observations
    \label{tab:photobs}
}
\tablehead{
    \colhead{~~~~~~~~Instrument/Field~~~~~~~~}  &
    \colhead{Date(s)} &
    \colhead{Number of Images} &
    \colhead{Mode Cadence (min)} &
    \colhead{Filter}
}
\startdata
\sidehead{\textbf{\hatcur{47}}}
~~~~HAT-5/G212 & 2010 Sep--2010 Nov & 2200 & $5.5$ & Sloan~\band{r} \\
~~~~HAT-8/G212 & 2010 Aug--2010 Nov & 6000 & $3.5$ & Sloan~\band{r} \\
~~~~KeplerCam & 2010 Dec 26 & 137 & $1.2$ & Sloan~\band{i}
\\
~~~~KeplerCam & 2011 Jan 09 & 466 & $0.7$ & Sloan~\band{i}
\\
~~~~BOS       & 2011 Aug 24 &  94 & $1.7$ & Sloan~\band{i}
\\
~~~~BOS       & 2011 Oct 01 & 307 & $0.7$ & Sloan~\band{i}
\\
~~~~FTN       & 2011 Oct 15 & 266 & $0.7$ & Sloan~\band{i}
\\
~~~~FTN       & 2012 Jan 13 & 345 & $0.6$ & Sloan~\band{i}
\\
\sidehead{\textbf{\hatcur{48}}}
~~~~HAT-5/G212 & 2010 Sep--2010 Nov & 2200 & $5.5$ & Sloan~\band{r} \\
~~~~HAT-8/G212 & 2010 Aug--2010 Nov & 6000 & $3.5$ & Sloan~\band{r} \\
~~~~KeplerCam         & 2011 Jan 12        & 328 & $0.9$ & Sloan~\band{i} \\
~~~~KeplerCam         & 2011 Sep 20        & 231 & $0.9$ & Sloan~\band{i} \\
~~~~KeplerCam         & 2011 Oct 30        & 137 & $1.5$ & Sloan~\band{i} \\
~~~~KeplerCam         & 2011 Nov 21        & 201 & $1.5$ & Sloan~\band{i} \\
~~~~FTN               & 2012 Jan 26        &  99 & $1.1$ & Sloan~\band{i} \\
\enddata
\ifthenelse{\boolean{emulateapj}}{
    \end{deluxetable*}
}{
    \end{deluxetable}
}

\subsection{Reconnaissance Spectroscopy}
\label{sec:recspec}

High-resolution, low-S/N ``reconnaissance'' spectra were obtained for
\hatcur{47} and \hatcur{48} using the Tillinghast Reflector Echelle
Spectrograph \citep[TRES;][]{furesz:2008} on the 1.5\,m Tillinghast
Reflector at the Fred Lawrence Whipple Observatory (FLWO) in AZ.  The
reconnaissance spectroscopic observations and results for each system
are summarized in \reftabl{reconspecobs}.  The observations were
reduced and analyzed following the procedure described by
\cite{quinn:2012:hat25} and \cite{buchhave:2010:hat16}.

Based on the observations summarized in \reftabl{reconspecobs} we find
that both targets are dwarf stars with radial velocity (RV) root mean
square (rms) residuals consistent with no detectable RV variation
within the $\lesssim 200$\,\ms\ precision of the measurements.  All
spectra were single-lined, i.e., there is no evidence that any of these
targets consist of more than one star.

\ifthenelse{\boolean{emulateapj}}{
    \begin{deluxetable*}{llrrrrr}
}{
    \begin{deluxetable}{llrrrrr}
}
\tablewidth{0pc}
\tabletypesize{\scriptsize}
\tablecaption{
    Summary of reconnaissance spectroscopy observations.
    \label{tab:reconspecobs}
}
\tablehead{
    \multicolumn{1}{c}{Instrument}          &
    \multicolumn{1}{c}{$HJD - 2400000$}             &
    \multicolumn{1}{c}{$\teffstar$}         &
    \multicolumn{1}{c}{$\loggstar$}         &
    \multicolumn{1}{c}{$\vsini$}            &
    \multicolumn{1}{c}{$\gamma_{\rm RV}$\tablenotemark{a}} \\
    &
    &
    \multicolumn{1}{c}{(K)}                 &
    \multicolumn{1}{c}{(cgs)}               &
    \multicolumn{1}{c}{(\kms)}              &
    \multicolumn{1}{c}{(\kms)}
}
\startdata
\sidehead{\textbf{\hatcur{47}}}
~~~~TRES              & $55544.64583$ & $6700$  & $4.1$ & $14$ & $2.72$ \\
~~~~TRES              & $55546.65187$ & $6650$  & $4.2$ & $14$ & $2.69$ \\
\sidehead{\textbf{\hatcur{48}}}
~~~~TRES              & $55546.76439$ & $5850 \pm 50$ & $4.15 \pm 0.10$ & $3.0 \pm 0.5$ &  $18.31$ \\
~~~~TRES              & $55549.69608$ & $6050 \pm 100$ & $4.53 \pm 0.18$ & $4.0 \pm 0.8$ & $18.26$ \\
\enddata 
\tablenotetext{a}{
    The heliocentric RV of the target on the IAU system, with a
    systematic uncertainty of approximately $0.1$\,\kms\ mostly
    limited by how well the velocities of the standard stars have been
    established.
}
\ifthenelse{\boolean{emulateapj}}{
    \end{deluxetable*}
}{
    \end{deluxetable}
}

\subsection{High resolution, high S/N spectroscopy}
\label{sec:hispec}

We proceeded with the follow-up of each candidate by obtaining
high-resolution, high-S/N spectra to characterize the RV variations,
and to refine the determination of the stellar parameters.  The
observations were made with HIRES \citep{vogt:1994} on the Keck-I
telescope in Hawaii, with FIES on the Nordic Optical Telescope on the
island of La Palma, Spain \citep{djupvik:2010}, and with the
High-Dispersion Spectrograph \citep[HDS;][]{noguchi:2002} on the 8.2\,m
Subaru telescope in Hawaii.  For HIRES and HDS we used an I$_{2}$
absorption cell, while for FIES we relied on simultaneous ThAr spectra
to determine the wavelength correction.  All but one of the HIRES
observations were obtained with the C2 decker which provides a
resolution of $R=45,\!000$, with the remaining observation made with
the B5 decker on the night of 01 Sep 2011.  This latter decker provides
the same resolution as C2, but with a shorter slit length.  For the
FIES observations we used the high-resolution fiber yielding a spectral
resolution of $R=67,\!000$, while for HDS we used the $0\farcs6 \times
2\farcs0$ slit yielding a spectral resolution of $R=60,\!000$.  The
HIRES observations were reduced to radial velocities in the barycentric
frame following the procedure described by \cite{butler:1996}; the FIES
observations were reduced following \cite{buchhave:2010:hat16}; and the
HDS observations were reduced following \cite{sato:2002,sato:2012}. 
The RV measurements and uncertainties are given in
Tables~\ref{tab:rvs47} and~\ref{tab:rvs48} for \hatcur{47} and
\hatcur{48}, respectively.  The period-folded data, along with our best
fit described below in \refsecl{analysis} are displayed in
Figures~\ref{fig:rvbis47} and~\ref{fig:rvbis48}.

\setcounter{planetcounter}{1}
%
\begin{figure} []
\plotone{\hatcurhtr{47}-rv.eps}
\ifthenelse{\value{planetcounter}=1}{
\caption{
    {\em Top panel:} RV measurements from Keck/HIRES (filled circles)
    and Subaru/HDS (open triangles) for \hbox{\hatcur{47}{}} shown as a
    function of orbital phase, along with our best-fit circular model
    (solid line; see \reftabl{planetparam}), and our best-fit eccentric
    model (dashed line).  Zero phase corresponds to the time of
    mid-transit.  The center-of-mass velocity has been subtracted.
    {\em Second panel:} Velocity $O\!-\!C$ residuals from the best fit. 
    The error bars for Keck/HIRES include a component from
    astrophysical jitter (\hatcurRVjitterA{47}\,\ms) added in
    quadrature to the formal errors (see \refsecl{globmod}).
    {\em Third panel:} Bisector spans (BS), with the mean value
    subtracted.  The measurement from the template spectrum is
    included.  The BS uncertainties are internal errors determined for
    each spectrum from the scatter of the individual BS values measured
    on separate orders of the spectrum; they do not include the unknown
    contribution from stellar jitter.
    {\em Bottom panel:} Chromospheric activity index $S$.
}}{
\caption{
    Keck/HIRES observations of \hatcur{47}. The panels are as in
    \reffigl{rvbis47}.  The parameters used in the best-fit model are
    given in \reftabl{planetparam}.
}}
\label{fig:rvbis47}
\end{figure}
\setcounter{planetcounter}{2}
%
\begin{figure} []
\plotone{\hatcurhtr{48}-rv.eps}
\ifthenelse{\value{planetcounter}=1}{
\caption{
    {\em Top panel:} Keck/HIRES RV measurements for
        \hbox{\hatcur{48}{}} shown as a function of orbital phase,
        along with our best-fit model (see \reftabl{planetparam}). 
        Zero phase corresponds to the time of mid-transit.  The
        center-of-mass velocity has been subtracted.
    {\em Second panel:} Velocity $O\!-\!C$ residuals from the best fit. 
        The error bars include a component from astrophysical jitter
        (\hatcurRVjitterA{48}\,\ms) added in quadrature to the formal
        errors (see \refsecl{globmod}).
    {\em Third panel:} Bisector spans (BS), with the mean value
        subtracted.  The measurement from the template spectrum is
        included (see \refsecl{blend}).
    {\em Bottom panel:} Chromospheric activity index $S$
        measured from the Keck spectra.
    Note the different vertical scales of the panels. Observations
    shown twice are represented with open symbols.
}}{
\caption{
    Keck/HIRES, Subaru/HDS, and NOT/FIES observations of \hatcur{48}. 
    The panels are as in \reffigl{rvbis47}.  Here we use filled circles
    to show Keck/HIRES observations, open triangles to show Subaru/HDS
    observations, and filled triangles to show NOT/FIES observations. 
    BS measurements are only available for the Keck/HIRES and
    Subaru/HDS observations, while S index measurements are only
    available for the Keck/HIRES observations.  The parameters used in
    the best-fit model are given in \reftabl{planetparam}.
}}
\label{fig:rvbis48}
\end{figure}

In each figure we show also the spectral line bisector spans (BSs)
computed from the Keck/HIRES spectra following \cite{torres:2007:hat3}
and the $S$ activity index calculated following \citet{isaacson:2010}.

\ifthenelse{\boolean{emulateapj}}{
    \begin{deluxetable*}{lrrrrrrr}
}{
    \begin{deluxetable}{lrrrrrrr}
}
\tablewidth{0pc}
\tablecaption{
    Relative radial velocities, bisector spans, and activity index
    measurements of \hatcur{47}.
    \label{tab:rvs47}
}
\tablehead{
    \colhead{BJD\tablenotemark{a}} &
    \colhead{RV\tablenotemark{b}} &
    \colhead{\ensuremath{\sigma_{\rm RV}}\tablenotemark{c}} &
    \colhead{BS} &
    \colhead{\ensuremath{\sigma_{\rm BS}}} &
    \colhead{S\tablenotemark{d}} &
    \colhead{Phase} &
    \colhead{Instrument}\\
    \colhead{\hbox{(2,454,000$+$)}} &
    \colhead{(\ms)} &
    \colhead{(\ms)} &
    \colhead{(\ms)} &
    \colhead{(\ms)} &
    \colhead{} &
    \colhead{} &
    \colhead{}
}
\startdata
\input{\hatcurhtr{47}_rvtable.tex}
\enddata
\tablenotetext{a}{
    Barycentric Julian Date calculated directly from UTC, {\em
      without} correction for leap seconds.
}
\tablenotetext{b}{
    The zero-point of these velocities is arbitrary. An overall offset
    $\gamma_{\rm rel}$ fitted independently to the velocities for each
    instrument has been subtracted.
}
\tablenotetext{c}{
    Internal errors excluding the component of astrophysical jitter
    considered in \refsecl{globmod}.
}
\tablenotetext{d}{
    Chromospheric activity index.
}
\ifthenelse{\boolean{rvtablelong}}{
    \tablecomments{
        Note that for the iodine-free Keck/HIRES template exposures we
        do not measure the RV but do measure the BS and S index.  Such
        template exposures can be distinguished by the missing RV
        value.  For the Subaru iodine-free template we did not measure
        the BS and S index (consequently, it is missing from the
        table).  We exclude BS measurements for a handful of
        measurements which were heavily affected by contamination from
        scattered moonlight.
    }
}{
    \tablecomments{
    Note that for the iodine-free Keck/HIRES template exposures we do
    not measure the RV but do measure the BS and S index.  Such
    template exposures can be distinguished by the missing RV value. 
    For the Subaru iodine-free template we did not measure the BS and S
    index (consequently, it is missing from the table).  We exclude BS
    measurements for a handful of measurements which were heavily
    affected by contamination from scattered moonlight.  This table is
    presented in its entirety in the electronic edition of the
    Astrophysical Journal.  A portion is shown here for guidance
    regarding its form and content.
    }
} 
\ifthenelse{\boolean{emulateapj}}{
    \end{deluxetable*}
}{
    \end{deluxetable}
}
%
\ifthenelse{\boolean{emulateapj}}{
    \begin{deluxetable*}{lrrrrrrr}
}{
    \begin{deluxetable}{lrrrrrrr}
}
\tablewidth{0pc}
\tablecaption{
    Relative radial velocities, bisector spans, and activity index
    measurements of \hatcur{48}.
    \label{tab:rvs48}
}
\tablehead{
    \colhead{BJD\tablenotemark{a}} &
    \colhead{RV\tablenotemark{b}} &
    \colhead{\ensuremath{\sigma_{\rm RV}}\tablenotemark{c}} &
    \colhead{BS} &
    \colhead{\ensuremath{\sigma_{\rm BS}}} &
    \colhead{S\tablenotemark{d}} &
    \colhead{Phase} &
    \colhead{Instrument}\\
    \colhead{\hbox{(2,454,000$+$)}} &
    \colhead{(\ms)} &
    \colhead{(\ms)} &
    \colhead{(\ms)} &
    \colhead{(\ms)} &
    \colhead{} &
    \colhead{} &
    \colhead{}
}
\startdata
\input{\hatcurhtr{48}_rvtable.tex}
\enddata
\tablenotetext{a}{
    Barycentric Julian Date calculated directly from UTC, {\em
      without} correction for leap seconds.
}
\tablenotetext{b}{
    The zero-point of these velocities is arbitrary. An overall offset
    $\gamma_{\rm rel}$ fitted independently to the velocities for each
    instrument has been subtracted.
}
\tablenotetext{c}{
    Internal errors excluding the component of astrophysical jitter
    considered in \refsecl{globmod}.
}
\tablenotetext{d}{
    Chromospheric activity index.
}
\ifthenelse{\boolean{rvtablelong}}{
    \tablecomments{
        Note that for the iodine-free Keck/HIRES template exposures we
        do not measure the RV but do measure the BS as well as the S
        index (in the case of HIRES).  Such template exposures for
        HIRES and HDS can be distinguished by the missing RV value.
    }
}{
    \tablecomments{
        Note that for the iodine-free Keck/HIRES template exposures we
        do not measure the RV but do measure the BS and S index.  Such
        template exposures can be distinguished by the missing RV
        value.  This table is presented in its entirety in the
        electronic edition of the Astrophysical Journal.  A portion is
        shown here for guidance regarding its form and content.
    }
} 
\ifthenelse{\boolean{emulateapj}}{
    \end{deluxetable*}
}{
    \end{deluxetable}
}

\subsection{Photometric follow-up observations}
\label{sec:phot}

%
\setcounter{planetcounter}{1}
%
\begin{figure}[]
\plotone{\hatcurhtr{47}-lc.eps}
\ifthenelse{\value{planetcounter}=1}{
\caption{
    Unbinned transit \lcs{} for \hatcur{47}, acquired with KeplerCam at
    the \flwof{} telescope.  The light curves have been EPD- and
    TFA-processed, as described in \refsec{globmod}.  The dates of the
    events are indicated.  Curves after the first are displaced
    vertically for clarity.  Our best fit from the global modeling
    described in \refsecl{globmod} is shown by the solid lines. 
    Residuals from the fits are displayed at the bottom, in the same
    order as the top curves.  The error bars represent the photon and
    background shot noise, plus the readout noise.
}}{
\caption{
    Similar to \reffigl{lc47}; here we show the follow-up
    \lcs{} for \hatcur{47}.
}}
\label{fig:lc47}
\end{figure}
\setcounter{planetcounter}{2}
%
\begin{figure}[]
\plotone{\hatcurhtr{48}-lc.eps}
\ifthenelse{\value{planetcounter}=1}{
\caption{
    Unbinned transit \lcs{} for \hatcur{48}, acquired with KeplerCam at
    the \flwof{} telescope.  The light curves have been EPD and TFA
    processed, as described in \refsec{globmod}.  The dates of the
    events are indicated.  Curves after the first are displaced
    vertically for clarity.  Our best fit from the global modeling
    described in \refsecl{globmod} is shown by the solid lines. 
    Residuals from the fits are displayed at the bottom, in the same
    order as the top curves.  The error bars represent the photon and
    background shot noise, plus the readout noise.
}}{
\caption{
    Similar to \reffigl{lc47}; here we show the follow-up \lcs{} for
    \hatcur{48}.  The facility used for each each light curve is
    indicated next to the date of the event.
}}
\label{fig:lc48}
\end{figure}

We conducted additional photometric observations of both stars with the
KeplerCam CCD camera on the \flwof{} telescope.  For \hatcur{47} we
also made use of the Spectral CCD on the 2.0\,m Faulkes Telescope North
(FTN) at Haleakala Observatory in Hawaii, and the SBIG CCD imager on
the Byrne Observatory at Sedgwick (BOS) 0.8\,m telescope, at Sedgwick
Reserve in the Santa Ynez Valley, CA.  Both FTN and BOS are operated by
the Las Cumbres Observatory Global Telescope\footnote{http://lcogt.net}
\citep[][]{brown:2013}.  The observations for each target are
summarized in \reftabl{photobs}.

The reduction of the KeplerCam images to light curves was performed as
described by \citet{bakos:2010:hat11}.  The FTN and BOS images were
reduced in a similar manner.  We performed external parameter
decorrelation (EPD) and the trend filtering algorithm (TFA) to remove
trends simultaneously with the light curve modeling (for more details,
see \citet{bakos:2010:hat11}).  The final time series, together with
our best-fit transit \lc{} model, are shown in the top portion of
Figures~\ref{fig:lc47} and~\ref{fig:lc48}, while the individual
measurements are reported in Tables~\ref{tab:phfu47}
and~\ref{tab:phfu48}.  All relevant data (discovery, follow-up), just
like for other HATNet discoveries, are also reported at the HATNet
website\footnote{\url{www.hatnet.org}}.

\begin{deluxetable}{lrrrr}
\tablewidth{0pc}
\tablecaption{
    High-precision differential photometry of
    \hatcur{47}\label{tab:phfu47}.
}
\tablehead{
    \colhead{BJD\tablenotemark{a}} & 
    \colhead{Mag\tablenotemark{b}} & 
    \colhead{\ensuremath{\sigma_{\rm Mag}}} &
    \colhead{Mag(orig)\tablenotemark{c}} & 
    \colhead{Filter} \\
    \colhead{\hbox{~~~~(2,400,000$+$)~~~~}} & 
    \colhead{} & 
    \colhead{} &
    \colhead{} & 
    \colhead{}
}
\startdata
\input{\hatcurhtr{47}_phfu_tab_short.tex}
\enddata
\tablenotetext{a}{
    Barycentric Julian Date calculated directly from UTC, {\em
      without} correction for leap seconds.
}
\tablenotetext{b}{
    The out-of-transit level has been subtracted. These magnitudes have
    been subjected to the EPD and TFA procedures, carried out
    simultaneously with the transit fit.
}
\tablenotetext{c}{
    Raw magnitude values without application of the EPD and TFA
    procedures.
}
\tablecomments{
    This table is available in a machine-readable form in the online
    journal.  A portion is shown here for guidance regarding its form
    and content.
}
\end{deluxetable}
%
\begin{deluxetable}{lrrrr}
\tablewidth{0pc}
\tablecaption{
    High-precision differential photometry of
    \hatcur{48}\label{tab:phfu48}.
}
\tablehead{
    \colhead{BJD\tablenotemark{a}} & 
    \colhead{Mag\tablenotemark{b}} & 
    \colhead{\ensuremath{\sigma_{\rm Mag}}} &
    \colhead{Mag(orig)\tablenotemark{c}} & 
    \colhead{Filter} \\
    \colhead{\hbox{~~~~(2,400,000$+$)~~~~}} & 
    \colhead{} & 
    \colhead{} &
    \colhead{} & 
    \colhead{}
}
\startdata
\input{\hatcurhtr{48}_phfu_tab_short.tex}
\enddata
\tablenotetext{a}{
    Barycentric Julian Date calculated directly from UTC, {\em
      without} correction for leap seconds.
}
\tablenotetext{b}{
    The out-of-transit level has been subtracted. These magnitudes have
    been subjected to the EPD and TFA procedures, carried out
    simultaneously with the transit fit.
}
\tablenotetext{c}{
    Raw magnitude values without application of the EPD and TFA
    procedures.
}
\tablecomments{
    This table is available in a machine-readable form in the online
    journal.  A portion is shown here for guidance regarding its form
    and content.
}
\end{deluxetable}

\section{Analysis}
\label{sec:analysis}

\subsection{Excluding blend scenarios}

The analyses of our reconnaissance spectroscopic observations discussed
in \refsecl{recspec} rule out many of the astrophysical false positive
scenarios for \hatcur{47} and \hatcur{48}.  To rule out remaining
scenarios we conduct an analysis similar to that done in
\cite{hartman:2011:hat32hat33,hartman:2012:hat39hat41}.  This involves
modeling the available light curves, absolute photometry, and stellar
atmospheric parameters as a combination of three stars (either a
hierarchical triple, or an unresolved blend between a foreground star
and a background eclipsing binary system) using the Padova isochrones
\citep{girardi:2002} to constrain the properties of the stars in the
simulated systems.  For each simulation we also predict the RVs and BS
values that would have been measured with Keck/HIRES at the times of
observation, and we compare them with the actual observations.

We find that for \hatcur{47} the photometry and measured stellar
atmospheric parameters rule out hierarchical triple eclipsing stellar
binary systems with the more massive of the eclipsing stars having $M <
0.89\,\msun$, or blended eclipsing binary systems where the background
eclipsing binary has a distance modulus that is more than $2.25$\,mag
larger than the distance modulus to the foreground star.  In both cases
these are $5\sigma$ limits based on Monte Carlo simulations which allow
for the possibility of time-correlated noise in the photometry.  We
also find that the simulated RV and BS measurements exclude
hierarchical triple systems where the more massive of the eclipsing
stars has $M > 0.69\,\msun$, or blended systems where the difference in
distance moduli is less than $3.35$\,mag.  Systems excluded by these
limits would show RV and/or BS scatter (RMS) that is at least 5 times
greater than what was observed.  Combining these constraints we
conclude that \hatcur{47} cannot be a hierarchical triple eclipsing
stellar binary system, or a blend between a foreground star and a
background eclipsing binary.

The analysis for \hatcur{48} yields a similar result. Here the
photometry rules out hierarchical triples with the more massive of the
eclipsing stars having $M < 0.88\,\msun$ or blended eclipsing binary
systems with distance moduli differences $> 1$\,mag, while the
simulated RV and BS measurements exclude hierarchical triples with $M
> 0.69$\,\msun, or blended eclipsing binary systems with distance
moduli differences $< 2.35$\,mag. As for \hatcur{47} we conclude that
\hatcur{48} cannot be a hierarchical triple eclipsing stellar binary
system, or a blend between a foreground star and a background
eclipsing binary.

While we exclude the possibility that either object is solely a
combination of stellar mass components, we cannot rule out the
possibility that either system is a combination of two stars
(physically associated, or aligned on the sky by chance), one of which
hosts a planet.  However, given the absence of evidence that either
object is composed of more than one star, we proceed by analyzing both
objects as single stars orbited by transiting planets.

\subsection{Global modeling of the data}
\label{sec:globmod}

We analyzed both systems following the procedure of
\cite{bakos:2010:hat11} as amended by \cite{hartman:2012:hat39hat41}. 
To summarize: (1) we determine stellar atmospheric parameters for each
star by applying the Stellar Parameter Classification method
\citep{buchhave:2012} to the Keck/HIRES iodine-free template spectra;
(2) we then conduct a Markov-Chain Monte Carlo (MCMC)-based modeling of
the available light curves and RVs, which, among others, results in a
posterior distribution for the mean stellar density.  We fix the limb
darkening coefficients to values taken from \cite{claret:2004} for the
measured atmospheric parameters; (3) we use the effective temperatures
and metalicities of the stars measured from the spectra, together with
the above determined stellar densities to derive the stellar properties
based on the Yonsei-Yale (YY) theoretical stellar evolution models
\citep{yi:2001}.  The stellar properties so-determined include the
masses, radii and ages.  We also determine the planetary parameters
(e.g.~mass and radius) which depend on these values; (4) we re-analyze
the Keck/HIRES spectra fixing the stellar surface gravities to the
values found in (3), and we go back to steps (2) and (3).

For both systems we conducted the analysis twice: fixing the
eccentricity to zero, and allowing it to vary.  For each system we find
that the eccentricity is consistent with zero, but with a poor
constraint (the 95\% upper limits on the eccentricity are $e < 0.31$
for \hatcurb{47} and $e < 0.46$ for \hatcurb{48}).  Following
\citet{anderson:2012} we adopt the parameter values associated with the
fixed circular orbits.  The adopted stellar parameters are given in
\reftabl{stellar} while the adopted planetary parameters are given in
\reftabl{planetparam}.  We find that \hatcur{47} is a
\hatcurISOmlong{47}\,\msun\ mass star with a radius of
\hatcurISOrlong{47}\,\rsun, and is located at a reddening-corrected
distance of \hatcurXdistred{47}\,pc, while \hatcur{48} is a
\hatcurISOmlong{48}\,\msun\ mass star with a radius of
\hatcurISOrlong{48}\,\rsun, and is located at a reddening-corrected
distance of \hatcurXdistred{48}\,pc.  The respective planets have
masses of \hatcurPPmlong{47}\,\mjup\ and \hatcurPPmlong{48}\,\mjup, and
radii of \hatcurPPrlong{47}\,\rjup\ and \hatcurPPrlong{48}\,\rjup.  The
parameters which result when the eccentricities are allowed to vary are
listed in Tables~\ref{tab:stellareccen} and~\ref{tab:planetparameccen}.

\ifthenelse{\boolean{emulateapj}}{
  \begin{deluxetable*}{lccr}
}{
  \begin{deluxetable}{lccr}
}
\tablewidth{0pc}
\tabletypesize{\scriptsize}
\tablecaption{
    Adopted stellar parameters for \hatcur{47}--\hatcur{48} assuming circular orbits
    \label{tab:stellar}
}
\tablehead{
    \multicolumn{1}{c}{} &
    \multicolumn{1}{c}{{\bf HAT-P-47}} &
    \multicolumn{1}{c}{{\bf HAT-P-48}} &
    \multicolumn{1}{c}{} \\
    \multicolumn{1}{c}{~~~~~~~~Parameter~~~~~~~~} &
    \multicolumn{1}{c}{Value}                     &
    \multicolumn{1}{c}{Value}                     &
    \multicolumn{1}{c}{Source}                    
}
\startdata
\noalign{\vskip -3pt}
\sidehead{Identifying Information}
~~~~R.A.                              &  \hatcurCCra{47}      &  \hatcurCCra{48}      & 2MASS\\
~~~~Dec.                              &  \hatcurCCdec{47}     &  \hatcurCCdec{48}     & 2MASS\\
~~~~GSC ID                            &  \hatcurCCgsc{47}     &  \hatcurCCgsc{48}     & GSC\\
~~~~2MASS ID                          &  \hatcurCCtwomass{47} &  \hatcurCCtwomass{48} & 2MASS\\
\sidehead{Spectroscopic properties}
~~~~$\teffstar$ (K)\dotfill          &  \hatcurSMEteff{47}   &  \hatcurSMEteff{48}   & SPC\tablenotemark{a}\\
~~~~$\feh$\dotfill                   &  \hatcurSMEzfeh{47}   &  \hatcurSMEzfeh{48}   & SPC                 \\
~~~~$\vsini$ (\kms)\dotfill          &  \hatcurSMEvsin{47}   &  \hatcurSMEvsin{48}   & SPC                 \\
~~~~$\gamma_{\rm RV}$ (\kms) \dotfill&  $2.70 \pm 0.1$       &  $18.29\pm0.1$        & TRES                \\
\sidehead{Photometric properties}
~~~~$V$ (mag)\dotfill               &  \hatcurCCtassmv{47}  &  \hatcurCCtassmv{48}  & TASS                \\
~~~~$I_{C}$ (mag)\dotfill            &  \hatcurCCtassmI{47}  &  \hatcurCCtassmI{48}  & TASS               \\
~~~~$J$ (mag)\dotfill               &  \hatcurCCtwomassJmag{47} &  \hatcurCCtwomassJmag{48} & 2MASS       \\
~~~~$H$ (mag)\dotfill               &  \hatcurCCtwomassHmag{47} &  \hatcurCCtwomassHmag{48} & 2MASS       \\
~~~~$K_s$ (mag)\dotfill             &  \hatcurCCtwomassKmag{47} &  \hatcurCCtwomassKmag{48} & 2MASS       \\
\sidehead{Derived properties}
~~~~$\mstar$ ($\msun$)\dotfill      &  \hatcurISOmlong{47}   &  \hatcurISOmlong{48}   & \hatcurisoshort{47}+\hatcurlumind{47}+SPC\tablenotemark{b}\\
~~~~$\rstar$ ($\rsun$)\dotfill      &  \hatcurISOrlong{47}   &  \hatcurISOrlong{48}   & \hatcurisoshort{47}+\hatcurlumind{47}+SPC         \\
~~~~$\loggstar$ (cgs)\dotfill       &  \hatcurISOlogg{47}    &  \hatcurISOlogg{48}    & \hatcurisoshort{47}+\hatcurlumind{47}+SPC         \\
~~~~$\lstar$ ($\lsun$)\dotfill      &  \hatcurISOlum{47}     &  \hatcurISOlum{48}     & \hatcurisoshort{47}+\hatcurlumind{47}+SPC         \\
~~~~$M_V$ (mag)\dotfill             &  \hatcurISOmv{47}      &  \hatcurISOmv{48}      & \hatcurisoshort{47}+\hatcurlumind{47}+SPC         \\
~~~~$M_K$ (mag,\hatcurjhkfilset{47})\dotfill &  \hatcurISOMK{47} &  \hatcurISOMK{48} & \hatcurisoshort{47}+\hatcurlumind{47}+SPC         \\
~~~~Age (Gyr)\dotfill               &  \hatcurISOage{47}     &  \hatcurISOage{48}     & \hatcurisoshort{47}+\hatcurlumind{47}+SPC         \\
~~~~$A_{V}$ (mag)\tablenotemark{c}\dotfill           &  \hatcurXAv{47}\phn  &  \hatcurXAv{48}    & \hatcurisoshort{47}+\hatcurlumind{47}+SPC\\
~~~~Distance (pc)\dotfill           &  \hatcurXdistred{47}\phn  &  \hatcurXdistred{48}\phn  & \hatcurisoshort{47}+\hatcurlumind{47}+SPC\\
~~~~$\log R^{\prime}_{\rm HK}$\tablenotemark{d}\dotfill & $-5.125 \pm 0.015$ & $-5.203 \pm 0.029$ & Keck/HIRES\\ [-1.5ex]
\enddata
\tablenotetext{a}{
    SPC = ``Stellar Parameter Classification'' method based on
    cross-correlating high-resolution spectra against synthetic
    templates \citep{buchhave:2012}. These parameters rely primarily
    on SPC, but have a small dependence also on the iterative analysis
    incorporating the isochrone search and global modeling of the
    data, as described in the text.  } \tablenotetext{b}{
    \hatcurisoshort{47}+\hatcurlumind{47}+SPC = Based on the \hatcurisoshort{47}\
    isochrones \citep{\hatcurisocite{47}}, \hatcurlumind{47}\ as a luminosity
    indicator, and the SPC results.
}
\tablenotetext{c}{
  \band{V} extinction determined by comparing the measured 2MASS and
  TASS photometry for the star to the expected magnitudes from the
  \hatcurisoshort{47}+\hatcurlumind{47}+SPC model for the star. We use
  the \citet{cardelli:1989} extinction law.  
}
\tablenotetext{d}{
  Chromospheric activity index defined in \citet{noyes:1984}
  determined from the Keck/HIRES spectra following
  \cite{isaacson:2010}. In each case we give the average value and the
  standard deviation from the individual spectra.
}
\ifthenelse{\boolean{emulateapj}}{
  \end{deluxetable*}
}{
  \end{deluxetable}
}

\ifthenelse{\boolean{emulateapj}}{
  \begin{deluxetable*}{lccr}
}{
  \begin{deluxetable}{lccr}
}
\tablewidth{0pc}
\tabletypesize{\scriptsize}
\tablecaption{
    Derived stellar parameters for \hatcur{47}--\hatcur{48} allowing eccentric orbits\tablenotemark{a}
    \label{tab:stellareccen}
}
\tablehead{
    \multicolumn{1}{c}{} &
    \multicolumn{1}{c}{{\bf HAT-P-47}} &
    \multicolumn{1}{c}{{\bf HAT-P-48}} &
    \multicolumn{1}{c}{} \\
    \multicolumn{1}{c}{~~~~~~~~Parameter~~~~~~~~} &
    \multicolumn{1}{c}{Value}                     &
    \multicolumn{1}{c}{Value}                     &
    \multicolumn{1}{c}{Source}                    
}
\startdata
\noalign{\vskip -3pt}
~~~~$\mstar$ ($\msun$)\dotfill      &  \hatcurISOmlongeccen{47}   &  \hatcurISOmlongeccen{48}   & \hatcurisoshort{47}+\hatcurlumind{47}+SPC \\
~~~~$\rstar$ ($\rsun$)\dotfill      &  \hatcurISOrlongeccen{47}   &  \hatcurISOrlongeccen{48}   & \hatcurisoshort{47}+\hatcurlumind{47}+SPC         \\
~~~~$\loggstar$ (cgs)\dotfill       &  \hatcurISOloggeccen{47}    &  \hatcurISOloggeccen{48}    & \hatcurisoshort{47}+\hatcurlumind{47}+SPC         \\
~~~~$\lstar$ ($\lsun$)\dotfill      &  \hatcurISOlumeccen{47}     &  \hatcurISOlumeccen{48}     & \hatcurisoshort{47}+\hatcurlumind{47}+SPC         \\
~~~~$M_V$ (mag)\dotfill             &  \hatcurISOmveccen{47}      &  \hatcurISOmveccen{48}      & \hatcurisoshort{47}+\hatcurlumind{47}+SPC         \\
~~~~$M_K$ (mag,\hatcurjhkfilset{47})\dotfill &  \hatcurISOMKeccen{47} &  \hatcurISOMKeccen{48} & \hatcurisoshort{47}+\hatcurlumind{47}+SPC         \\
~~~~Age (Gyr)\dotfill               &  \hatcurISOageeccen{47}     &  \hatcurISOageeccen{48}     & \hatcurisoshort{47}+\hatcurlumind{47}+SPC         \\
~~~~$A_{V}$ (mag)\dotfill           &  \hatcurXAveccen{47}\phn  &  \hatcurXAveccen{48}    & \hatcurisoshort{47}+\hatcurlumind{47}+SPC\\
~~~~Distance (pc)\dotfill           &  \hatcurXdistredeccen{47}\phn  &  \hatcurXdistredeccen{48}\phn  & \hatcurisoshort{47}+\hatcurlumind{47}+SPC\\
\enddata
\tablenotetext{a}{
    Quantities and abbreviations are as in \reftabl{stellar}, which gives our adopted values, determined assuming circular orbits. We do not list parameters that are independent of the eccentricity.
}
\ifthenelse{\boolean{emulateapj}}{
  \end{deluxetable*}
}{
  \end{deluxetable}
}

\ifthenelse{\boolean{emulateapj}}{
  \begin{deluxetable*}{lcc}
}{
  \begin{deluxetable}{lcc}
}
\tabletypesize{\scriptsize}
\tablecaption{Adopted orbital and planetary parameters 
assuming circular orbits\label{tab:planetparam}}
\tablehead{
    \multicolumn{1}{c}{} &
    \multicolumn{1}{c}{{\bf HAT-P-47b}} &
    \multicolumn{1}{c}{{\bf HAT-P-48b}} \\
    \multicolumn{1}{c}{~~~~~~~~Parameter~~~~~~~~} &
    \multicolumn{1}{c}{Value}                     &
    \multicolumn{1}{c}{Value}                     
}
\startdata
\noalign{\vskip -3pt}
\sidehead{\Lc{} parameters}
~~~$P$ (days)             \dotfill    & $\hatcurLCP{47}$              & $\hatcurLCP{48}$              \\
~~~$T_c$ (${\rm BJD}$)    
      \tablenotemark{a}   \dotfill    & $\hatcurLCT{47}$              & $\hatcurLCT{48}$              \\
~~~$T_{14}$ (days)
      \tablenotemark{a}   \dotfill    & $\hatcurLCdur{47}$            & $\hatcurLCdur{48}$            \\
~~~$T_{12} = T_{34}$ (days)
      \tablenotemark{a}   \dotfill    & $\hatcurLCingdur{47}$         & $\hatcurLCingdur{48}$         \\
~~~$\arstar$              \dotfill    & $\hatcurPPar{47}$             & $\hatcurPPar{48}$             \\
~~~$\zrstar$\tablenotemark{b}              \dotfill    & $\hatcurLCzeta{47}$\phn       & $\hatcurLCzeta{48}$\phn       \\
~~~$\rpl/\rstar$          \dotfill    & $\hatcurLCrprstar{47}$        & $\hatcurLCrprstar{48}$        \\
~~~$b^2$                  \dotfill    & $\hatcurLCbsq{47}$            & $\hatcurLCbsq{48}$            \\
~~~$b \equiv a \cos i/\rstar$
                          \dotfill    & $\hatcurLCimp{47}$            & $\hatcurLCimp{48}$            \\
~~~$i$ (deg)              \dotfill    & $\hatcurPPi{47}$\phn          & $\hatcurPPi{48}$\phn          \\

\sidehead{Limb-darkening coefficients \tablenotemark{c}}
~~~$c_1,i$ (linear term)  \dotfill    & $\hatcurLBii{47}$             & $\hatcurLBii{48}$             \\
~~~$c_2,i$ (quadratic term) \dotfill  & $\hatcurLBiii{47}$            & $\hatcurLBiii{48}$            \\
~~~$c_1,r$               \dotfill     & $\hatcurLBir{47}$             & $\hatcurLBir{48}$             \\
~~~$c_2,r$               \dotfill     & $\hatcurLBiir{47}$            & $\hatcurLBiir{48}$            \\

\sidehead{RV parameters}
~~~$K$ (\ms)              \dotfill    & $\hatcurRVK{47}$\phn\phn      & $\hatcurRVK{48}$\phn\phn      \\
~~~$e$                    \dotfill    & $0$ (fixed)          & $0$ (fixed)          \\
~~~RV jitter Keck/HIRES (\ms)\tablenotemark{d}        \dotfill    & \hatcurRVjitterA{47}           & \hatcurRVjitterA{48}           \\

~~~RV jitter Subaru/HDS (\ms)\tablenotemark{d}        \dotfill    & \hatcurRVjitterB{47}           & \hatcurRVjitterB{48}           \\

~~~RV jitter NOT/FIES (\ms)\tablenotemark{d}        \dotfill    & $\cdots$           & \hatcurRVjitterC{48}           \\

\sidehead{Planetary parameters}
~~~$\mpl$ ($\mjup$)       \dotfill    & $\hatcurPPmlong{47}$          & $\hatcurPPmlong{48}$          \\
~~~$\rpl$ ($\rjup$)       \dotfill    & $\hatcurPPrlong{47}$          & $\hatcurPPrlong{48}$          \\
~~~$C(\mpl,\rpl)$
    \tablenotemark{e}     \dotfill    & $\hatcurPPmrcorr{47}$         & $\hatcurPPmrcorr{48}$         \\
~~~$\rhopl$ (\gcmc)       \dotfill    & $\hatcurPPrho{47}$            & $\hatcurPPrho{48}$            \\
~~~$\log g_p$ (cgs)       \dotfill    & $\hatcurPPlogg{47}$           & $\hatcurPPlogg{48}$           \\
~~~$a$ (AU)               \dotfill    & $\hatcurPParel{47}$           & $\hatcurPParel{48}$           \\
~~~$T_{\rm eq}$ (K)\tablenotemark{f}        \dotfill   & $\hatcurPPteff{47}$           & $\hatcurPPteff{48}$           \\
~~~$\Theta$\tablenotemark{g} \dotfill & $\hatcurPPtheta{47}$          & $\hatcurPPtheta{48}$          \\
~~~$\langle F \rangle$ ($10^{8}$\ergscmsq) \tablenotemark{h}
                          \dotfill    & $\hatcurPPfluxavg{47}$        & $\hatcurPPfluxavg{48}$        \\ [-1.5ex]
\enddata
\tablenotetext{a}{
    Reported times are in Barycentric Julian Date calculated directly
    from UTC, {\em without} correction for leap seconds.
    \ensuremath{T_c}: Reference epoch of mid transit that
    minimizes the correlation with the orbital period.
    \ensuremath{T_{14}}: total transit duration, time
    between first to last contact;
    \ensuremath{T_{12}=T_{34}}: ingress/egress time, time between first
    and second, or third and fourth contact.
}
\tablenotetext{b}{
    Reciprocal of the half duration of the transit used as a jump
    parameter in our MCMC analysis in place of $\arstar$. It is
    related to $\arstar$ by the expression $\zrstar = \arstar
    (2\pi(1+e\sin \omega))/(P \sqrt{1 - b^{2}}\sqrt{1-e^{2}})$
    \citep{bakos:2010:hat11}.
}
\tablenotetext{c}{
    Values for a quadratic law, adopted from the tabulations by
    \cite{claret:2004} according to the spectroscopic (SPC) parameters
    listed in \reftabl{stellar}.
}
\tablenotetext{d}{
    Error term, either astrophysical or instrumental in origin, added
    in quadrature to the formal RV errors for the listed instrument
    such that $\chi^{2}$ per degree of freedom is unity. For both
    \hatcur{47} and \hatcur{48} we did not add jitter to the
    Subaru/HDS RV errors because the formal errors for these
    observations exceeded the scatter in the RV residuals.
}
\tablenotetext{e}{
    Correlation coefficient between the planetary mass \mpl\ and radius
    \rpl.
}
\tablenotetext{f}{
    Planet equilibrium temperature averaged over the orbit, calculated
    assuming a Bond albedo of zero, and that flux is reradiated from
    the full planet surface.
}
\tablenotetext{g}{
    The Safronov number is given by $\Theta = \frac{1}{2}(V_{\rm
    esc}/V_{\rm orb})^2 = (a/\rpl)(\mpl / \mstar )$
    \citep[see][]{hansen:2007}.
}
\tablenotetext{h}{
    Incoming flux per unit surface area, averaged over the orbit.
}
\ifthenelse{\boolean{emulateapj}}{
  \end{deluxetable*}
}{
  \end{deluxetable}
}
%

\ifthenelse{\boolean{emulateapj}}{
  \begin{deluxetable*}{lcc}
}{
  \begin{deluxetable}{lcc}
}
\tabletypesize{\scriptsize}
\tablecaption{Orbital and planetary parameters 
%
%
allowing eccentric orbits\tablenotemark{a}\label{tab:planetparameccen}}
\tablehead{
    \multicolumn{1}{c}{} &
    \multicolumn{1}{c}{{\bf HAT-P-47b}} &
    \multicolumn{1}{c}{{\bf HAT-P-48b}} \\
    \multicolumn{1}{c}{~~~~~~~~Parameter~~~~~~~~} &
    \multicolumn{1}{c}{Value}                     &
    \multicolumn{1}{c}{Value}                     
}
\startdata
\noalign{\vskip -3pt}
\sidehead{\Lc{} parameters}
~~~$\arstar$              \dotfill    & $\hatcurPPareccen{47}$             & $\hatcurPPareccen{48}$             \\
~~~$\zrstar$              \dotfill    & $\hatcurLCzetaeccen{47}$\phn       & $\hatcurLCzetaeccen{48}$\phn       \\
~~~$i$ (deg)              \dotfill    & $\hatcurPPieccen{47}$\phn          & $\hatcurPPieccen{48}$\phn          \\


\sidehead{RV parameters}
~~~$K$ (\ms)              \dotfill    & $\hatcurRVKeccen{47}$\phn\phn      & $\hatcurRVKeccen{48}$\phn\phn      \\
~~~$\sqrt{e} \cos \omega$ 
                          \dotfill    & $\hatcurRVrkeccen{47}$\phs          & $\hatcurRVrkeccen{48}$\phs          \\
~~~$\sqrt{e} \sin \omega$
                          \dotfill    & $\hatcurRVrheccen{47}$              & $\hatcurRVrheccen{48}$              \\
~~~$e \cos \omega$ 
                          \dotfill    & $\hatcurRVkeccen{47}$\phs          & $\hatcurRVkeccen{48}$\phs          \\
~~~$e \sin \omega$
                          \dotfill    & $\hatcurRVheccen{47}$              & $\hatcurRVheccen{48}$              \\
~~~$e$                    \dotfill    & $\hatcurRVecceneccen{47}$          & $\hatcurRVecceneccen{48}$          \\
~~~$\omega$ (deg)         \dotfill    & $\hatcurRVomegaeccen{47}$\phn      & $\hatcurRVomegaeccen{48}$\phn      \\

\sidehead{Secondary eclipse parameters}
~~~$T_s$ (BJD)            \dotfill    & $\hatcurXsecondaryeccen{47}$       & $\hatcurXsecondaryeccen{48}$       \\
~~~$T_{s,14}$              \dotfill   & $\hatcurXsecdureccen{47}$          & $\hatcurXsecdureccen{48}$          \\
~~~$T_{s,12}$              \dotfill   & $\hatcurXsecingdureccen{47}$       & $\hatcurXsecingdureccen{48}$       \\

\sidehead{Planetary parameters}
~~~$\mpl$ ($\mjup$)       \dotfill    & $\hatcurPPmlongeccen{47}$          & $\hatcurPPmlongeccen{48}$          \\
~~~$\rpl$ ($\rjup$)       \dotfill    & $\hatcurPPrlongeccen{47}$          & $\hatcurPPrlongeccen{48}$          \\
~~~$C(\mpl,\rpl)$
         \dotfill    & $\hatcurPPmrcorreccen{47}$         & $\hatcurPPmrcorreccen{48}$         \\
~~~$\rhopl$ (\gcmc)       \dotfill    & $\hatcurPPrhoeccen{47}$            & $\hatcurPPrhoeccen{48}$            \\
~~~$\log g_p$ (cgs)       \dotfill    & $\hatcurPPloggeccen{47}$           & $\hatcurPPloggeccen{48}$           \\
~~~$a$ (AU)               \dotfill    & $\hatcurPPareleccen{47}$           & $\hatcurPPareleccen{48}$           \\
~~~$T_{\rm eq}$ (K)        \dotfill   & $\hatcurPPteffeccen{47}$           & $\hatcurPPteffeccen{48}$           \\
~~~$\Theta$ \dotfill & $\hatcurPPthetaeccen{47}$          & $\hatcurPPthetaeccen{48}$          \\
~~~$\langle F \rangle$ ($10^{8}$\ergscmsq)
                          \dotfill    & $\hatcurPPfluxavgeccen{47}$        & $\hatcurPPfluxavgeccen{48}$        \\ [-1.5ex]
\enddata
\tablenotetext{a}{
    Quantities and definitions are as in \reftabl{planetparam}, which
    gives our adopted values, determined assuming circular orbits. 
    Here we do not list parameters that are effectively independent of
    the eccentricity.
}
\ifthenelse{\boolean{emulateapj}}{
  \end{deluxetable*}
}{
  \end{deluxetable}
}
%



\section{Discussion}
\label{sec:discussion}

\begin{figure*}[]
\begin{center}
\includegraphics[width=150mm]{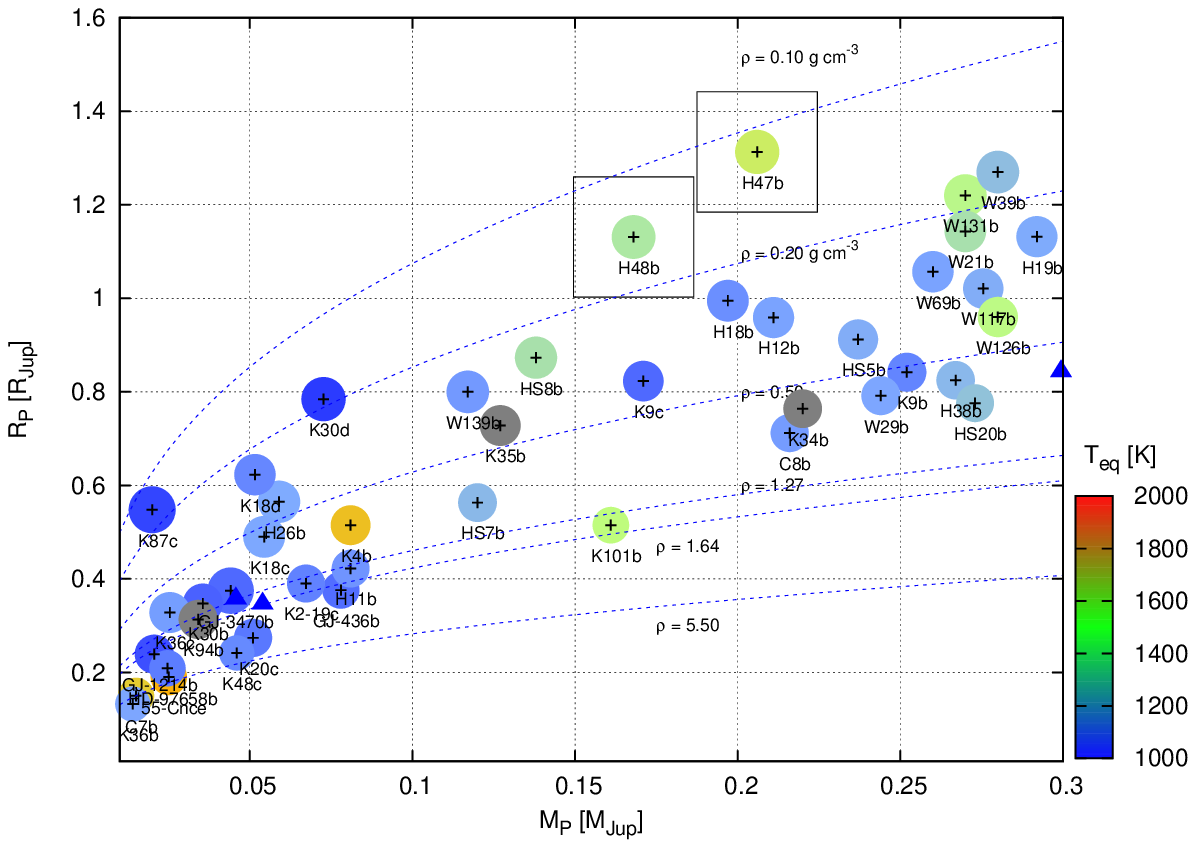}
\caption{
    Mass--radius diagram of sub-Saturn mass TEPs ($\mpl<0.3\mjup$),
    where the relative error of the mass determination is
    $\delta(\mpl)<20$\%.  \hatcurb{47} and \hatcurb{48} are highlighted
    by large boxes (these are not error-bars).  The color-bar indicates
    equilibrium temperature (with a palette of R,G,B = 2000, 1500,
    1000\,K).  Solar System planets are indicated by blue triangles. 
    Both \hatcurb{47} and \hatcurb{48} stand out by their very low
    density.
\label{fig:exomr}}
\end{center}
\end{figure*}

We show the location of \hatcurb{47} and \hatcurb{48} on a mass--radius
diagram in \reffig{exomr}.  In this same Figure we show and label other
transiting exoplanets with $\mpl < 0.3$\,\mjup\ (sub-Saturn mass), and
with masses having relative error $\delta(\mpl) < 20$\%.  The values
are taken from our privately maintained database of up-to-date
exoplanet parameters, which is broadly consistent with the NASA
exoplanet
archive\footnote{\url{http://exoplanetarchive.ipac.caltech.edu/}}. 
\hatcurb{47} and \hatcurb{48} are the two lowest density sub-Saturn
planets, and they are the two lowest mass planets discovered to date
with radii larger than Jupiter.  If all planets are considered with
$\delta(\mpl)<20$\% (314 of them, as of 2016 April 20), \hatcurb{47} is
the 4th lowest density, and \hatcurb{48} is the 11th lowest density
exoplanet.  While the orbital solution for both planets is broadly
consistent with circular orbits, it is possible that they are on
slightly eccentric orbits, which could contribute to their large radii
via tidal heating.  Carrying out measurements of the occultations
(secondary eclipses) of these planets could help in constraining their
eccentricities.

\begin{figure*}[]
\begin{center}
\plotone{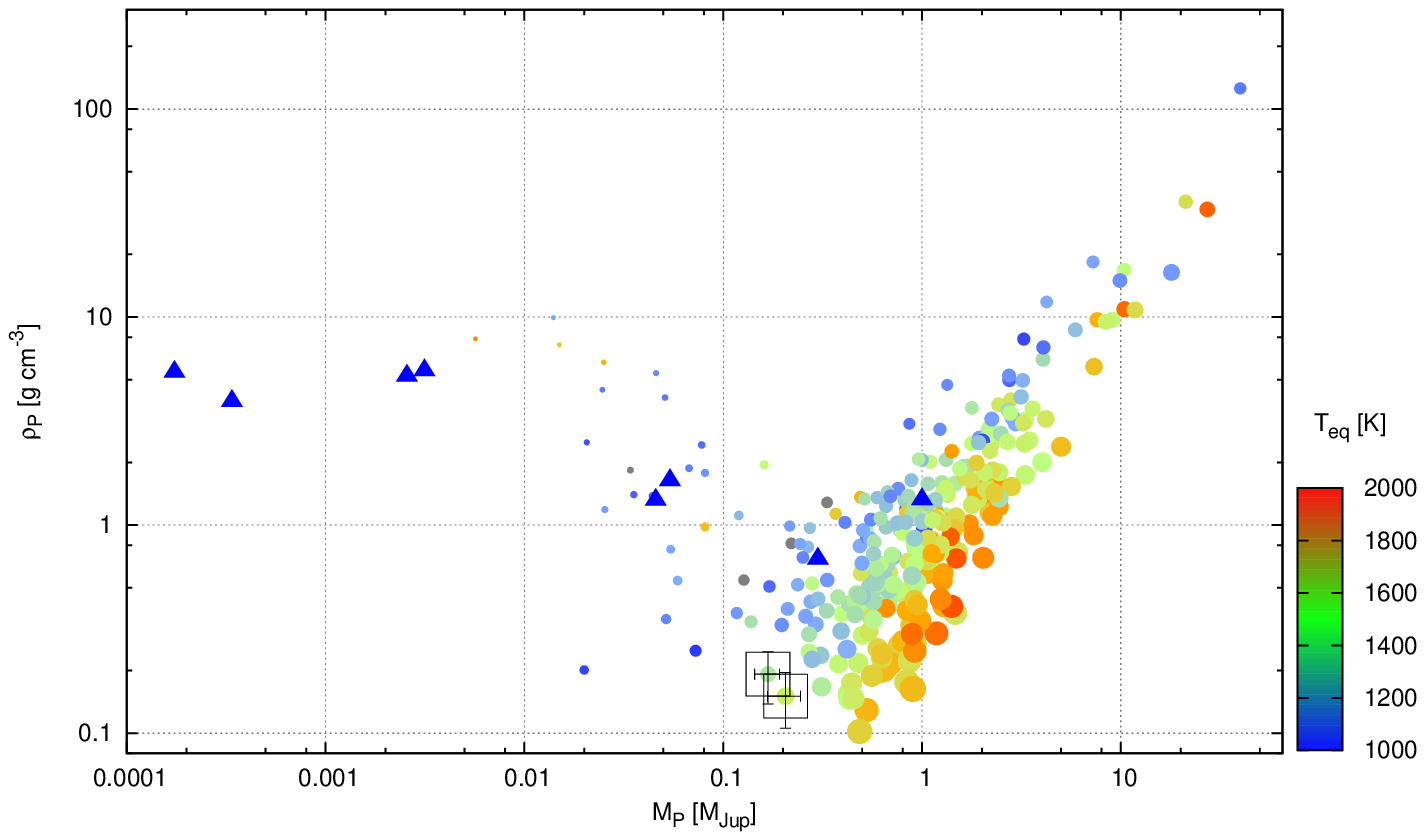}
\caption[]{
    Planetary mean density vs.~mass for TEPs with mass measured at
    $\delta(\mpl)<20$\% precision.  \hatcurb{47} and \hatcurb{48} are
    highlighted by large boxes.  The size of the points scales with
    planetary radius, while the color indicates equilibrium temperature
    (as in \reffig{exomr}).  \hatcurb{47} and \hatcurb{48} fall in a
    so-far unpopulated region of the parameter space; they are the
    lowest density sub-Saturn objects.  Solar system planets are marked
    with blue triangles.
\label{fig:rho}}
\end{center}
\end{figure*}

We also plot the mass--density diagram, this time not limiting
ourselves to $\mpl<0.3\,\mjup$, but considering {\em all transiting
sub-stellar} objects with relative error on mass $<20$\%
(\reffig{rho}).  \hatcurb{47} and \hatcurb{48} clearly fall in a so-far
unpopulated region of the parameter space: they are somewhat detached
from the rest of the population in being light and very low density. 
Both objects fall close to, but below the approximate locus of
Neptunian-to-Jovian class transition point of $\sim0.4\,\mjup$
\citep{chen:2016}.

\begin{figure*}[]
\plotone{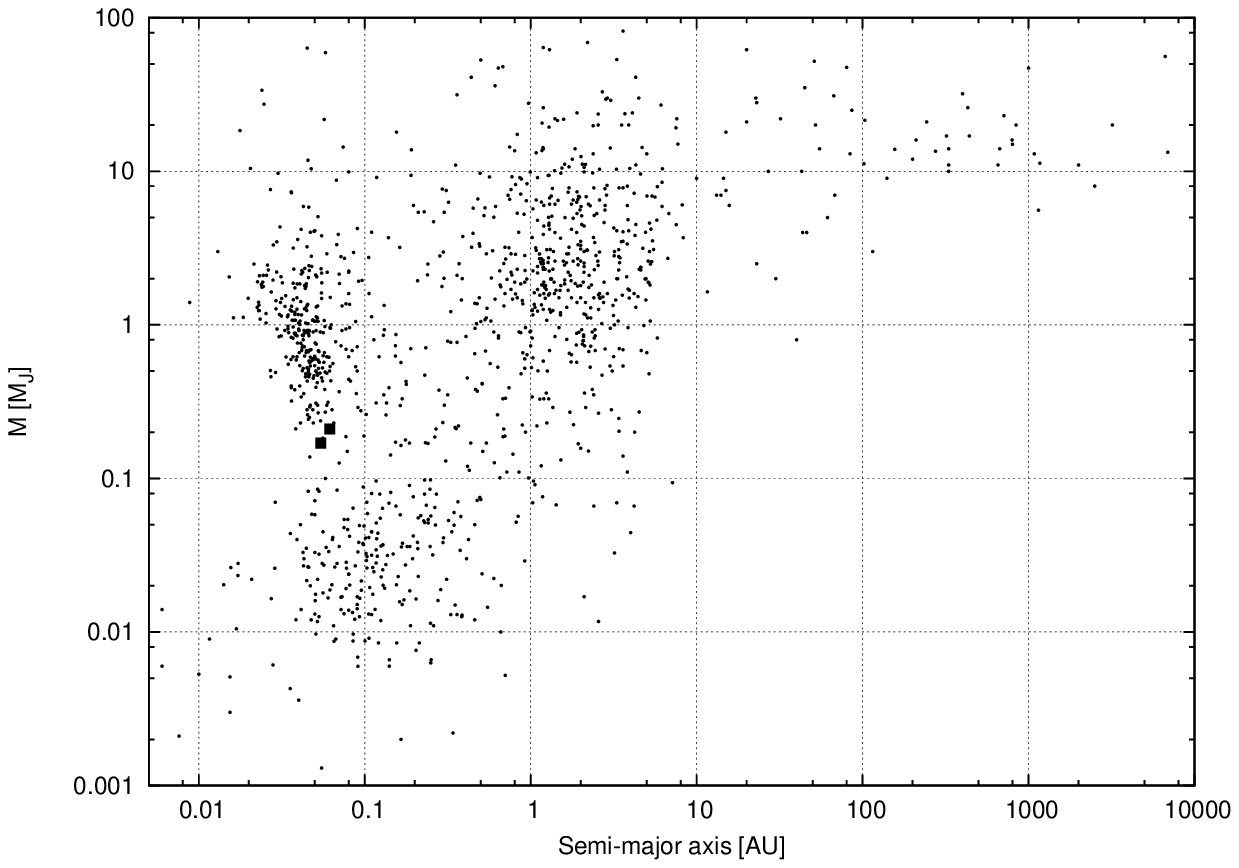}
\caption[]{
    Mass vs.~semi-major axis of exoplanets from
    exoplanet.eu\footnote{\url{http://exoplanetarchive.ipac.caltech.edu/}}. 
    For non-transiting systems, $\mpl \sin i$ is used instead of the
    true mass.  \hatcurb{47} and \hatcurb{48} are indicated with black
    filled boxes, both falling in the sparse region between hot
    Jupiters and hot super-Earths.
\label{fig:exoam}}
\end{figure*}

When plotting the mass vs.~the semi-major axis of the planets
($\mpl\sin i$ for non-transiting radial velocity detections),
\hatcurb{47} and \hatcurb{48} fall in the ``desert'' between hot (small
semi-major axis) Jupiters and hot super Earths (\reffig{exoam}).  This
area of the parameter space has been coined the short-period Neptunian
``desert'' \citep[][and references therein]{szabo:2011,mazeh:2016}.  A
similar desert has been identified for planets around M-dwarf host
stars \citep{gaidos:2016}.  The desert is especially pronounced for
$P<5$\,days, so with their respective periods of \hatcurLCPshort{47}\,d
and \hatcurLCPshort{48}\,d, both objects are on the ``edge'' of the
desert.  One possible explanation for the lack of planets in this
domain is that small gaseous objects can not survive the proximity of
the star, unlike large gaseous objects (hot Jupiters), or small, dense,
rocky planets.

Measuring the Rossiter-McLaughlin effect
\citep[R-M;][]{queloz:2000,winn:2005} for \hatcurb{47} and \hatcurb{48}
would provide an interesting test of the hypothesis that the
obliquities of close-in giant planets were initially nearly random, and
that subsequent star--planet tidal interactions act to reduce the
obliquities \citep[e.g.,][]{albrecht:2012}.  With effective
temperatures of \hatcurSMEteff{47}\,K and \hatcurSMEteff{48}\,K,
\hatcur{47} and \hatcur{48} are expected to have radiative and
convective envelopes, respectively.  While most close-in transiting
planets around stars with convective envelopes have been found to be on
low-obliquities orbits, those around radiative envelope stars have been
found to have a wide range of obliquities \citep{winn:2010}.  If this
difference is the result of a dependence on stellar mass of the
dominant planet migration pathway, then we might expect \hatcurb{48} to
be on a low-obliquity orbit while \hatcurb{47} would likely be
misaligned.  If, however, the low obliquity orbits are the result of
tidal interactions, then based on Figure~24 of \citet{albrecht:2012} we
would expect both \hatcurb{47} and \hatcurb{48} to be misaligned.  This
is due to the low planet masses, and the resulting long tidal
interaction timescales (Equations~2 and~3 from \citealp{albrecht:2012},
which in turn are taken from \citealp{zahn:1977}), even for \hatcur{48}
with its convective envelope.  \hatcurb{48} is at an interesting mass
where we expect the tidal interaction to be comparable to WASP-8b, a
misaligned planet around a convective envelope star which has a mass of
2.2\,\mjup, but a long orbital period of $8.16$\,days
\citep{queloz:2010}.  Measuring the R-M effect is feasible for both
systems.  Assuming aligned orbits, we expect \hatcurb{47} to have an
R-M amplitude of 68\,\ms, while \hatcurb{48} would have an amplitude of
20\,\ms.  Which, given the magnitudes and RV jitter, should be
detectable from single transits with Keck/HIRES.

\begin{figure*}[]
\plotone{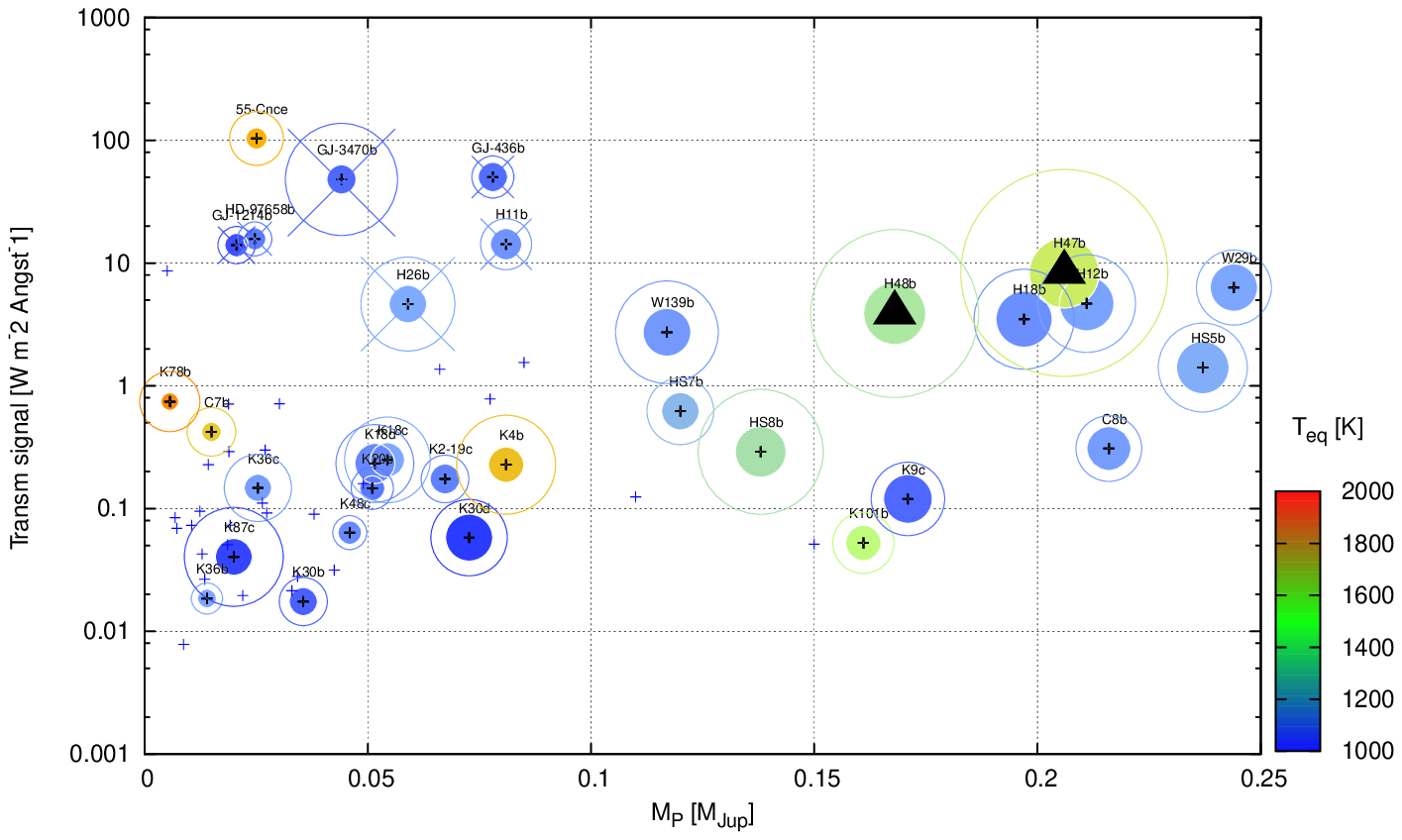}
\caption[]{
    Approximate detectability of a planet's atmosphere in transmission,
    as a function of planetary mass for planets with $\mpl \le
    0.25\,\mjup$.  The size of the filled circles scales with the
    radius of the planet (arbitrary scale), and the radius of the open
    circles scales with the scale height of the atmosphere.  Small plus
    symbols denote planets with uncertain mass or radius measurements
    (error $>$20\%).  The color-bar is the same as for \reffig{rho}. 
    Transmission spectroscopy has been carried out for planets with
    diagonal crosses.  \hatcurb{47} and \hatcurb{48} are marked with
    black triangles.  Abbreviations are: K: Kepler, H: HAT, HS:
    HATSouth, C: Corot, W: WASP.
\label{fig:trans}}
\end{figure*}

Finally, we calculate the expected transmission spectroscopy signature
for both planets as $\delta = 5 \times 2 \rpl H / \rstar^2$
\citep{perryman:2014}, where $H$ is the scale height of the atmosphere. 
In calculating the latter quantity, we assume the molecular weight of
pure molecular hydrogen, and the equilibrium temperature and surface
gravity of the planet as determined from our analysis
(\reftab{planetparam}).  Once $\delta$ is known, we then calculate the
K-band flux of the star, and multiply by $\delta$, to come up with an
approximate measure of the transmission signal.  This quantity does not
take into account the detailed (expected) spectrum of the planetary
atmosphere, but is an order of magnitude estimate of the signature. 
The results are plotted in \reffig{trans} for planets with $\mpl \le
0.25\,\mjup$.  For both \hatcurb{47} and \hatcurb{48} the expected
transmission signature is amongst the largest for sub-Saturn objects,
making these new discoveries especially valuable.

\acknowledgements 

\paragraph{Acknowledgements}
HATNet operations have been funded by NASA grants NNG04GN74G,
NNX08AF23G, NNX13AJ15G and SAO IR\&D grants.  We acknowledge partial
support also from the Kepler Mission under NASA Cooperative Agreement
NCC2-1390 (D.W.L., PI).  This research has made use of Keck telescope
time granted through NOAO (A284Hr, A245Hr) and NASA (N108Hr, N154Hr,
N130Hr).  Based in part on data collected at Subaru Telescope (program
o11170), which is operated by the National Astronomical Observatory of
Japan.  This paper presents observations made with the Nordic Optical
Telescope, operated on the island of La Palma jointly by Denmark,
Finland, Iceland, Norway, and Sweden, in the Spanish Observatorio del
Roque de los Muchachos of the Instituto de Astrofisica de Canarias. 
This paper uses observations obtained with facilities of the Las
Cumbres Observatory Global Telescope.  The Byrne Observatory at
Sedgwick (BOS) is operated by the Las Cumbres Observatory Global
Telescope Network and is located at the Sedgwick Reserve, a part of the
University of California Natural Reserve System.  B.J.F.~notes that
this material is based upon work supported by the National Science
Foundation Graduate Research Fellowship under grant No.~2014184874.
Data presented in this paper are based on observations obtained at the
HAT station at thex Submillimeter Array of SAO, and the HAT station at
the Fred Lawrence Whipple Observatory of SAO.  We wish to thank J.~Johnson
his contribution to the Keck/HIRES radial velocity observations. 
The authors wish to
recognize and acknowledge the very significant cultural role and
reverence that the summit of Mauna Kea has always had within the
indigenous Hawaiian community.  This research has made use of the NASA
Exoplanet Archive, which is operated by the California Institute of
Technology, under contract with the National Aeronautics and Space
Administration under the Exoplanet Exploration Program.

\clearpage
\bibliographystyle{apj}
\bibliography{htrbib.bib}

\end{document}